\documentclass[%
 aip,
 amsmath,amssymb,
 reprint,%
]{revtex4-1}
\usepackage[english]{babel}

\usepackage{graphicx}
\usepackage{dcolumn}
\usepackage{bm}

\usepackage[utf8]{inputenc}
\usepackage[T1]{fontenc}
\usepackage{mathptmx}
\usepackage{etoolbox}
\usepackage{makecell} 
\usepackage{siunitx}
\usepackage{gensymb}
\usepackage[colorlinks=true,linkcolor=blue,citecolor=blue,urlcolor=blue]{hyperref}
\usepackage{multirow}

\makeatletter
\def\@email#1#2{%
 \endgroup
 \patchcmd{\titleblock@produce}
  {\frontmatter@RRAPformat}
  {\frontmatter@RRAPformat{\produce@RRAP{*#1\href{mailto:#2}{#2}}}\frontmatter@RRAPformat}
  {}{}
}%
\makeatother
\begin{document}

\preprint{AIP/123-QED}

\title{Integrated Photonic Devices in Thin-Film Barium Titanate: Opportunities and Challenges}

\author{Hong-Lin Lin}
\thanks{These authors contributed equally to this work.}
\affiliation{Department of Electrical and Computer Engineering, National University of Singapore, 4 Engineering Drive 3, 117583, Singapore}

\author{Minghao Shang}
\thanks{These authors contributed equally to this work.}
\affiliation{Department of Electrical and Computer Engineering, National University of Singapore, 4 Engineering Drive 3, 117583, Singapore}
\affiliation{National Laboratory of Solid State Microstructures, School of Physics, School of Electronic Science and Engineering, National Key Laboratory of Microwave Photonics and Collaborative Innovation Center of Advanced Microstructures, Nanjing University, Nanjing 210093, China}

\author{Yuhui Yin}
\affiliation{Department of Electrical and Computer Engineering, National University of Singapore, 4 Engineering Drive 3, 117583, Singapore}

\author{Wujie Fu}
\affiliation{Department of Electrical and Computer Engineering, National University of Singapore, 4 Engineering Drive 3, 117583, Singapore}

\author{Luo Qi}
\affiliation{Department of Electrical and Computer Engineering, National University of Singapore, 4 Engineering Drive 3, 117583, Singapore}

\author{Aaron J.~Danner}
\email{adanner@nus.edu.sg}
\affiliation{Department of Electrical and Computer Engineering, National University of Singapore, 4 Engineering Drive 3, 117583, Singapore}

\date{\today}

\begin{abstract}
Thin-film barium titanate (BaTiO$_3$, or BTO) has emerged as a promising electro-optic (EO) material for integrated photonics due to its exceptionally large Pockels coefficients, low optical loss, and compatibility with heterogeneous integration. Recent advances in epitaxial thin-film growth, crystal orientation control, ferroelectric domain engineering, and scalable integration have enabled EO performance in thin-film BTO approaching bulk-like properties on photonic platforms. This review provides a comprehensive overview of integrated photonic devices based on thin-film BTO, including fundamental material properties, thin-film growth and integration strategies, passive waveguide configurations, EO response at both the film and device levels, and relevant optical applications. We review the role of film orientation, domain structure, and device geometry in determining accessible EO coefficients and modulation mechanisms, as well as the distinctions between intrinsic and effective EO responses demonstrated at the film and device levels. Representative studies are compared in terms of film-quality-related metrics and device performance figures of merit, alongside emerging applications realized in BTO-based platforms. Finally, we discuss the remaining challenges and outline future development directions toward high-performance BTO-based integrated photonic systems.
\end{abstract}

\maketitle

\section{Introduction}
Integrated photonics has become a key technology for modern optical communication\cite{kikuchi2015fundamentals,kaushal2016optical,nakajima2017multi}, signal processing\cite{zhou2024silicon,teng2020miniaturized,liu2016fully}, sensing\cite{kumar2025high,kazanskiy2022advancement}, and emerging applications such as artificial intelligence\cite{ahmed2025universal,yao2019intelligent}, data-center interconnects\cite{thraskias2018survey,zhang2018silicon}, quantum and neuromorphic optical computing\cite{shastri2021photonics,wang2025neuro,chrysostomidis2025ultra}. As global data traffic continues to grow rapidly and energy efficiency becomes a critical concern, integrated photonic systems offer a scalable solution to overcome the bandwidth, latency, and power-consumption limitations of conventional electronic interconnects. By enabling dense integration of optical components on chip, integrated photonics provides a promising route toward low-power and high-speed information processing platforms.

Early demonstrations of integrated photonic systems were primarily realized on the silicon-on-insulator (SOI) platform, owning to its compatibility with complementary metal oxide semiconductor (CMOS) process and scalability\cite{izhaky2006development,stojanovic2018monolithic}. However, optical modulation on SOI platforms relies mainly on thermo-optic (TO) effects \cite{chung2019low-power,seo2025efficient,parra2024silicon} or free-carrier plasma-dispersion effects\cite{perez-galacho2016simplified,chen2020modeling}. While these mechanisms are well established, they inherently suffer from high power consumption, optical loss, and limited high-speed performance. These limitations motivated the exploration of alternative modulation mechanisms in optical thin films, which can provide higher efficiency, lower power consumption, and broader bandwidths to meet the increasing demands of next-generation optical communication systems.

Ferroelectric perovskite oxide materials exhibiting a strong electro-optic (EO) effect or Pockels effect have emerged as one of the most promising candidates for integrated photonic modulators\cite{thomaschewski2022pockels,yan2025recent}. Compared to carrier-based modulation in Si, EO modulation enables intrinsically broad bandwidth with negligible optical loss, making it an attractive mechanism for high-speed, low-loss integrated photonic devices. Among EO materials, lithium niobate (LiNbO$_3$, or LN) has been extensively studied for decades due to its relatively large EO coefficient (e.g., $r_{33}$ = 30 pm/V)\cite{zhu2021integrated,qi2020integrated,chen2022advances}. Significant progress in thin-film LN on insulator (LNOI) has been reported in recent years, supported by mature wafer preparation techniques and advanced etching methods. As a result, low-loss waveguides, low driving voltages, and broadband EO modulation, have been demonstrated on the LNOI platform.

In addition to LN, barium titanate (BaTiO$_3$, or BTO) has emerged as one of the most promising materials that could potentially lead to improved device performance compared to the thin-film LN platform\cite{karvounis2020barium,guo2021epitaxial,yan2025recent}. BTO exhibits exceptionally large Pockels coefficients, with $r_{42}$ = 1300 pm/V\cite{zgonik1994dielectric}, which is more than 40 times larger than the largest EO coefficient of LN. This extraordinary EO property makes BTO a promising material for realizing ultra-efficient EO modulation. Beyond its EO properties, BTO is also a well-established ferroelectric material with switchable polarization. For instance, recent studies have demonstrated low-voltage switching in thin-film BTO for electronic applications, showing the multifunctional nature of the material\cite{jiang2022enabling,zhang2025single}.

However, integrated photonic devices in BTO thin film have historically been hindered by challenges in epitaxial growth and patterning/etching methods. To achieve high-performance integrated photonic devices in BTO thin films, it is crucial that the material retain bulk-like BTO properties in thin-film form while simultaneously achieving low optical propagation loss. Recently, substantial progress has been made in overcoming these challenges. Advances in epitaxial growth techniques have enabled high-quality BTO thin films with controlled crystal orientation on a variety of substrates, including oxide insulators and Si-based platforms. Both thin-film BTO on insulator (BTOI) and heterogeneously integrated BTO-on-Si platforms have been widely reported, demonstrating the exceptional EO properties of BTO in the integrated photonic form\cite{petraru2002ferroelectric,cao2023characterization,abel2019large,posadas2023rf,ppsiquantum_team2025manufacturable}. To circumvent dry etching challenges, hybrid waveguide configurations using strip-loaded Si or silicon nitride (SiN) have been widely adopted\cite{eltes2020integrated,xiong2014active,tang2004low}. At the same time, recent demonstration of ultra-low-loss BTO ridge waveguides indicates that, with optimized etching processes, compact and low-loss photonic devices can be realized in BTO thin films\cite{raju2025high,isti2025fabrication}. Leveraging these advances, various high-performance EO devices and emerging applications have been demonstrated on thin-film BTO platforms\cite{kohli2025plasmonic,psiquantum_team2025manufacturable,deng2026self-buffered,abel2019large}.

This review provides a comprehensive overview of integrated photonic devices based on thin-film BTO. Section II introduces the fundamental physical properties of BTO relevant to photonics. Section III reviews thin-film BTO growth techniques, crystal orientation control, and integration strategies. Section IV focuses on integrated photonic devices, including both passive photonics and active EO modulators. We introduce EO characterization methods at the film level and device level, and then clarify the distinction between intrinsic and effective EO coefficients; we then discuss how $a$- and $c$-axis oriented films lead to fundamentally different modulation mechanisms in both monolithic and hybrid configurations. Representative demonstrations of BTO's EO response are summarized to illustrate the evolution of state-of-the-art thin-film BTO photonics. Section V reviews emerging applications of thin-film BTO, including high-speed optical communications, metasurface-based devices, optical sensing, nonlinear photonics, and neuromorphic computing. Finally, Section VI discusses challenges and presents a future outlook, highlighting key directions toward robust, scalable, and high-performance BTO-based integrated photonic systems.

\section{BTO material properties}
The material properties of BTO provide the physical basis for its use in integrated photonics. As a ferroelectric perovskite \cite{karvounis2020barium,deng2024recent}, BTO combines strong structural anisotropy, spontaneous polarization, high dielectric permittivity, and large EO response. In thin-film form, these properties are strongly influenced by substrate-induced strain and clamping effects, crystal orientation, and domain configuration. This distinction between bulk and thin-film BTO is particularly important for photonic devices: although bulk BTO has a relatively low Curie temperature, substrate-induced strain and clamping can substantially increase the ferroelectric transition temperature in thin films, thereby improving thermal stability during fabrication and operation.

Another key feature of BTO is the strong frequency dependence of its polarization response. In the direct current (DC) and radio frequency (RF) regimes, inverse-piezoelectric and ionic polarization mechanisms can contribute to the dielectric response \cite{chelladurai2025barium}. At optical frequencies, however, the response is dominated by electronic polarization \cite{chelladurai2025barium}. As a result, BTO can exhibit a large RF permittivity and a correspondingly large effective RF refractive index in microwave electrode structures, while its optical refractive index remains moderate. This large contrast between the RF and optical refractive indices is a central consideration for microwave-optical velocity matching in high-speed electro-optic modulators.

The same frequency-dependent polarization physics also explains why BTO exhibits exceptionally large EO coefficients but more moderate optical second-order nonlinear ($\chi^{(2)}$) coefficients. The Pockels effect involves one DC or RF electric field and can therefore benefit from slower inverse-piezoelectric and ionic polarization contributions, while all-optical $\chi^{(2)}$ processes such as second-harmonic generation are governed primarily by the electronic response at optical frequencies. Beyond these intrinsic optical and electro-optic properties, defect- and charge-related photorefractive effects and the temperature-dependent thermo-optic response can further influence device stability, drift, and tuning behavior. These considerations motivate the material-property discussion below.

\subsection{Structural properties}
\begin{figure}
    \centering
    \includegraphics[width=1\linewidth]{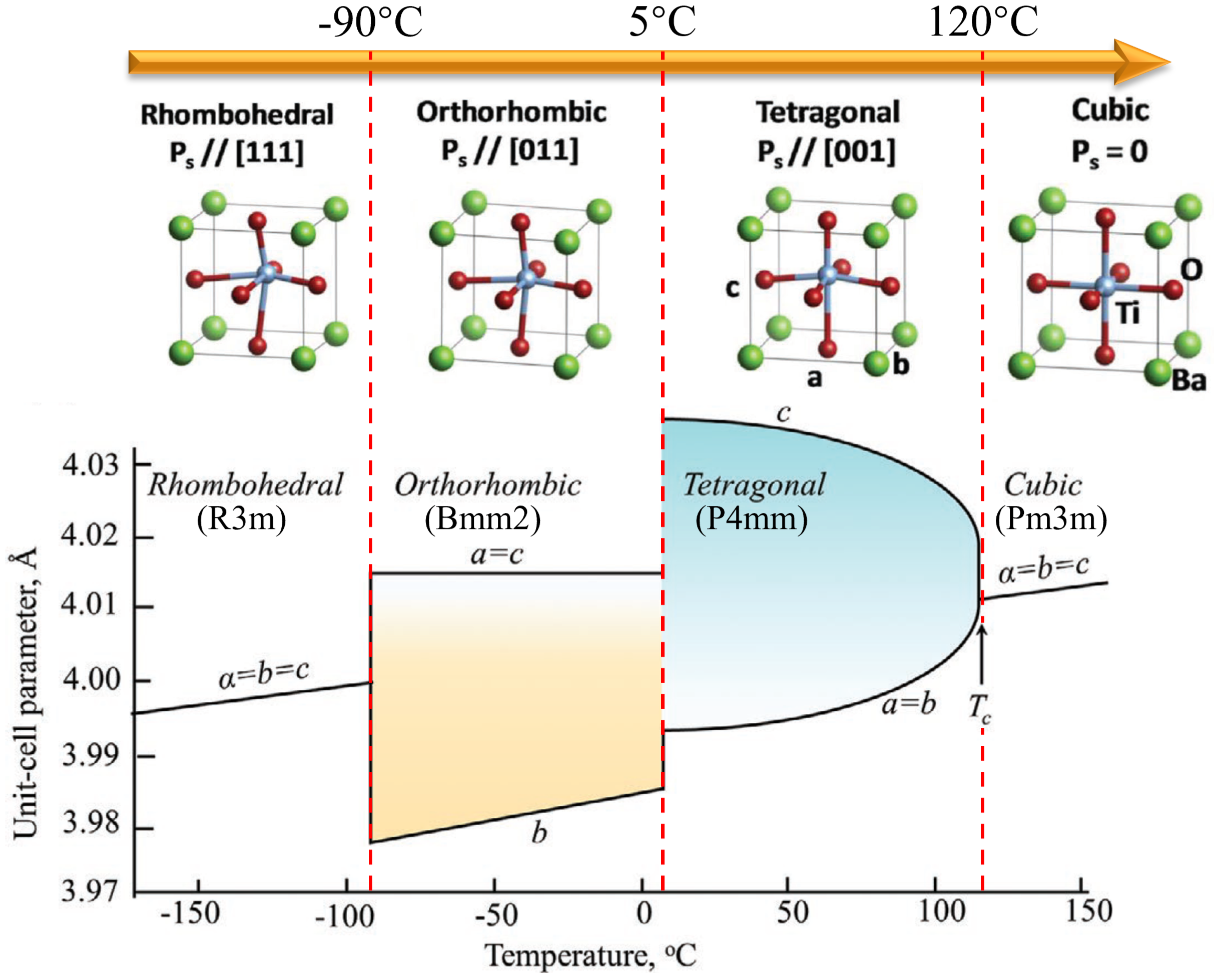}
    \caption{The temperature-phase diagram of bulk BTO, illustrating the crystal structures, the directions of spontaneous polarization \textbf{P}$_{\text{s}}$, and lattice constants of each phase, along with the corresponding phase transition temperatures. Reproduced with permission from Karvounis \textit{et al.}, Advanced Optical Materials 8, 2001249, \copyright 2020. Published by Wiley-VCH GmbH, licensed under the Creative Commons Attribution-NonCommercial \cite{karvounis2020barium}. Reproduced with permission from Deng \textit{et al.}, Advanced Sensor Research 3, 2300168 \copyright 2024. Published by Wiley-VCH GmbH, licensed under the Creative Commons Attribution \cite{deng2024recent}.}
    \label{fig:chap2-1}
\end{figure}

BTO is a representative perovskite-structured oxide with the general formula ABO$_3$, where the A-site is occupied by the relatively large Ba$^{2+}$ cation, the B-site by the high-valence Ti$^{4+}$ cation; the O$^{2-}$ anions form an octahedral coordination \cite{karvounis2020barium,li2019strong,deng2024recent}. A defining feature of BTO is its sequence of temperature-dependent phase transitions, which are crucial for its functional properties \cite{karvounis2020barium,deng2024recent,acosta2017batio3,merz1949electric,wemple1968dielectric,shu2001domain}. As illustrated in Fig. \ref{fig:chap2-1}, bulk BTO exhibits a centrosymmetric cubic phase (space group \textit{Pm3m}) above its Curie temperature $T_{\text{C}}$ of $\sim$120$^\circ$C. In this phase, the positive and negative charge centers coincide, resulting in the absence of spontaneous polarization and ferroelectric behavior. Upon cooling below $T_{\text{C}}$, BTO undergoes a phase transition to a tetragonal ferroelectric phase (space group \textit{P4mm}) in the temperature range of 5–120$^\circ$C. This phase is accompanied by a displacement of the Ti$^{4+}$ cation relative to the center of the oxygen octahedron along the [001] crystallographic direction (the $c$-axis). This ionic shift breaks the central symmetry, leading to an elongation of the $c$-axis, with the $c/a$ ratio exceeding unity (e.g., $\sim$1.01 at room temperature), and the emergence of a spontaneous polarization \textbf{P}$_{\text{s}}$ along the $c$-axis. The polarization direction in this ferroelectric phase can be reversed by an applied electric field.
With further cooling, BTO undergoes additional phase transitions. In a temperature range between -90$^\circ$C and 5$^\circ$C, bulk BTO transforms into an orthorhombic phase (space group \textit{Bmm2}), in which the spontaneous polarization \textbf{P}$_{\text{s}}$ is oriented along the [011] direction, corresponding to the face diagonal of the cubic unit cell. Finally, below -90$^\circ$C, BTO enters a rhombohedral phase (space group \textit{R3m}), where \textbf{P}$_{\text{s}}$ reorients along the [111] direction, aligned with the body diagonal of the cubic unit cell. 

\begin{figure}
    \centering
    \includegraphics[width=1\linewidth]{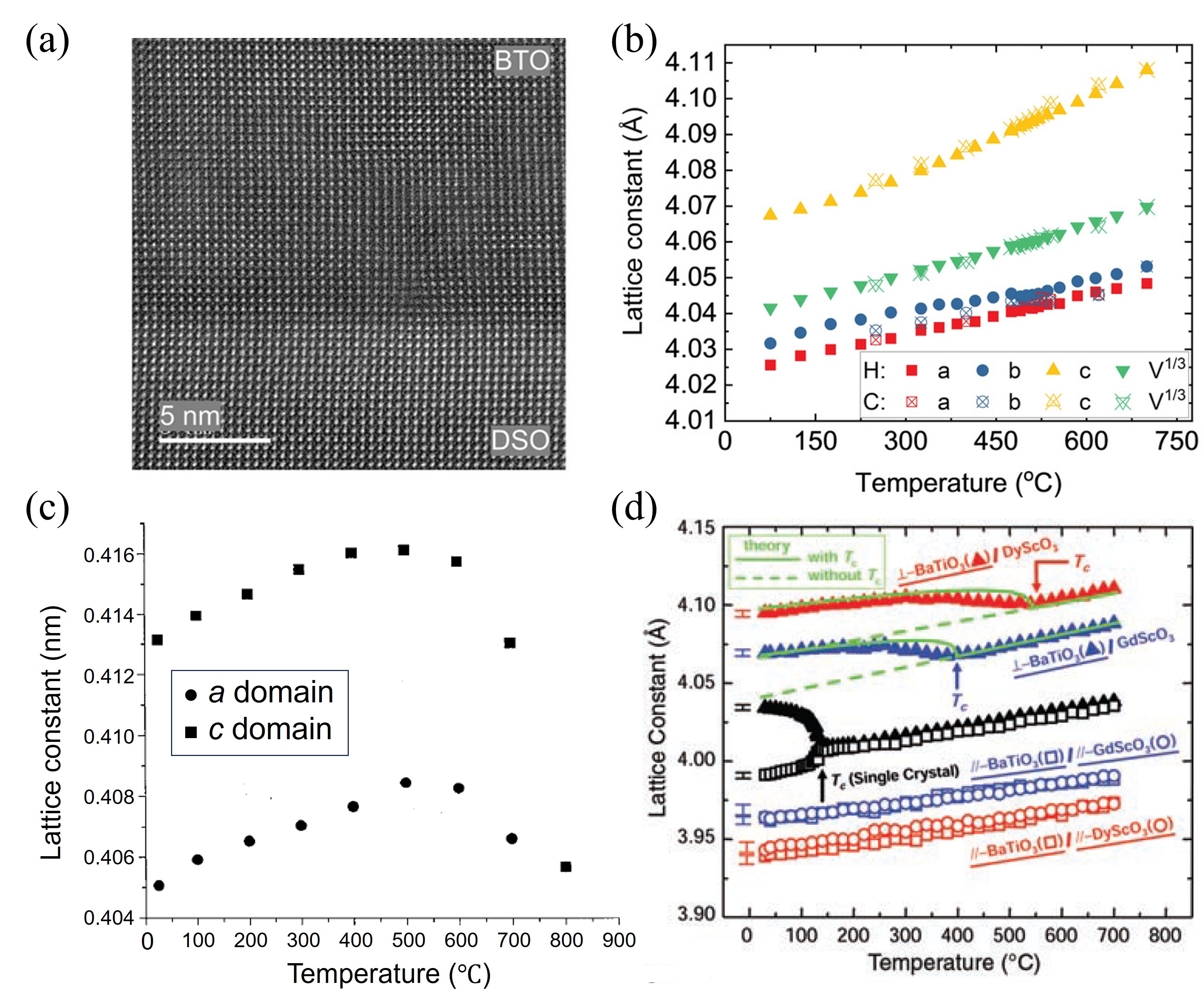}
    \caption{ Temperature dependence of lattice constants in epitaxial BTO thin films, demonstrating that Curie temperatures $T_{\text{C}}$ of several hundred degrees Celsius can be achieved through strain engineering. (a) A high-resolution cross-sectional STEM image of a 500-nm-thick BTO film grown on a DyScO$_3$ (DSO) substrate, indicating high crystalline quality at the BTO/DSO interface. (b) The temperature dependence of the BTO lattice constants $a$, $b$, and $c$, as well as the normalized unit-cell volume $V^{1/3}$, measured by X-ray diffractometry (XRD) reciprocal space mapping (RSM) during heating (solid symbols) and cooling (hollow symbols). The smooth change of lattice constants up to 700$^\circ$C indicates the absence of phase transitions or film cracking. Reproduced with permission from Cao \textit{et al.}, Advanced Materials 33, 2101128, \copyright 2021 Wiley-VCH GmbH\cite{cao2021barium}. 
    (c) Temperature-dependent lattice constants of a 600-nm-thick BTO film grown on a SrTiO$_3$ (STO) substrate, showing a phase transition at 800$^\circ$C where the lattice constants $a=c$. Reproduced from Li \textit{et al.}, Journal of applied physics 86, 4555–4558 (1999), with the permission of AIP Publishing\cite{li1999phase}.
    (d) Comparison of lattice-constant evolution for a 50-nm-thick BTO film on GdScO$_3$ (GSO) and DSO substrate, together with bulk BTO. The Curie temperatures of BTO on GSO and DSO substrates are $\sim$400$^\circ$C and $\sim$540$^\circ$C, respectively\cite{choi2004enhancement}. Reproduced from Choi \textit{et al.}, Science, DOI: 10.1126/science.1103218 (2004), AAAS. }
    \label{fig:chap2-2}
\end{figure}

For BTO-based optical devices, the tetragonal phase, which is stable at room temperature, is typically employed. Its non-centrosymmetric symmetry enables EO modulation and nonlinear optical interactions. However, the $T_{\text{C}}$ of bulk BTO is relatively low, at approximately 120$^\circ$C, which is significantly lower than that of LN ($\sim$1195$^\circ$C) \cite{smolenskii1966curie,miller1966temperature}. This substantial difference poses considerable challenges for thermal management during both the fabrication and operation of bulk BTO-based optical devices. Exceeding $T_{\text{C}}$ may induce structural deformation or cracking due to BTO's phase transition, accompanied by the complete loss of its EO and nonlinear optical properties.
Fortunately, in thin-film form, BTO exhibits significantly enhanced structural stability, and its $T_{\text{C}}$ is markedly increased \cite{li1999phase,choi2004enhancement,cao2021barium}. This improvement is primarily attributed to substrate-induced strain, where the clamping effect of the substrate stabilizes the lattice and constrains atomic displacements within the thin film. As reported in numerous studies, the $T_{\text{C}}$ of BTO thin films can be increased to several hundred degrees Celsius, as illustrated by representative examples in Fig. \ref{fig:chap2-2}. This improved thermal stability is a critical advantage for both microfabrication processes and practical applications of BTO thin-film photonic devices.

\subsection{\label{sec2-2}Permittivity and its associated properties}
The permittivity, $\epsilon=\epsilon_0\epsilon_r$, characterizes the response of a dielectric material to an applied electric field, where $\epsilon_0$ is the vacuum permittivity and $\epsilon_r$ donates the relative dielectric constant. It quantifies the extent to which charges - such as nuclei and electrons - within the material are displaced, or polarized, in response to the electric field \cite{grant2013electromagnetism}. Dielectric materials can exhibit several polarization mechanisms, each with a distinct frequency dependence. For BTO, three primary polarization mechanisms contribute to its dielectric response\cite{chelladurai2025barium}. The first mechanism arises from the inverse piezoelectric effect \cite{deng2024recent}, where an applied electric field induces mechanical deformation, along with lattice vibrations associated with acoustic phonons. This mechanism exhibits a low response frequency to the electric field, typically from DC to $\sim$100 MHz.
The second mechanism involves ionic polarization \cite{slater1950lorentz} present in ionic crystals, with an increased response frequency ranging from $\sim$100 MHz to several terahertz. The third mechanism is electronic polarization \cite{slater1950lorentz}, which is inherent to all atoms and responds on extremely short timescales, corresponding to optical (petahertz) frequencies. At sufficiently low electric field frequencies (DC to $\sim$1 MHz), all polarization mechanisms are able to fully respond to the electric field, resulting in the maximum dielectric constant, commonly referred to as the "static dielectric constant \cite{clemens1981dielectric}". When the electric-field frequency exceeds $\sim$100 MHz, the crystal is considered "clamped \cite{zgonik1994dielectric}", meaning that the inverse piezoelectric contribution becomes inactive, leading to a reduction in the dielectric constant. As the electric-field frequency further increases beyond the radio frequency (RF) range and into the optical regime, both piezoelectric and ionic polarization mechanisms become ineffective, leaving only electronic polarization as the active mechanism. 

In this section, we focus exclusively on the permittivity of BTO in the electric-field frequency range from DC to RF, along with related concepts of the loss tangent and RF effective refractive index. The permittivity in the optical regime is instead discussed in terms of the refractive index in Section \ref{Sec2.3}, which addresses the linear optical properties.

\subsubsection{Permittivity and loss tangent from DC to RF regime}
With an alternating electric field, the response of actual dielectric materials is not instantaneous. Internal processes within the material, such as mechanical deformation and ionic displacement, lag behind the changes in the applied electric field, resulting in energy dissipation. This loss behavior is described by defining the dielectric constant as a complex quantity \cite{bolton1948variation,roberts1947dielectric}: 
\begin{equation} \label{eq:2-1} 
\epsilon_{r} \equiv \epsilon_{r}^{\prime} + j \epsilon_{r}^{\prime\prime},
\end{equation}
where $j$ is the imaginary unit. The real part, $\epsilon_{r}^{\prime}$, characterizes the material's ability to polarize in response to the applied electric field, while the imaginary part, $\epsilon_{r}^{\prime\prime}$, serves as the dielectric loss term. This loss effect induces a phase lag $\delta$ between the polarization response \textbf{P} and the driving electric field \textbf{E}, known as the loss angle. The tangent of this angle, tan $\delta$, referred to as the loss tangent \cite{berge2007effect,ummethala2021hybrid}, is an essential parameter for quantifying dielectric losses. It is defined as the ratio of the imaginary part to the real part of the complex permittivity \cite{chelladurai2025barium}: 
\begin{equation} \label{eq:2-2} 
\tan \delta = \frac{\epsilon_{r}^{\prime \prime}}{\epsilon_{r}^{\prime}}.
\end{equation}

\begin{figure}
    \centering
    \includegraphics[width=1\linewidth]{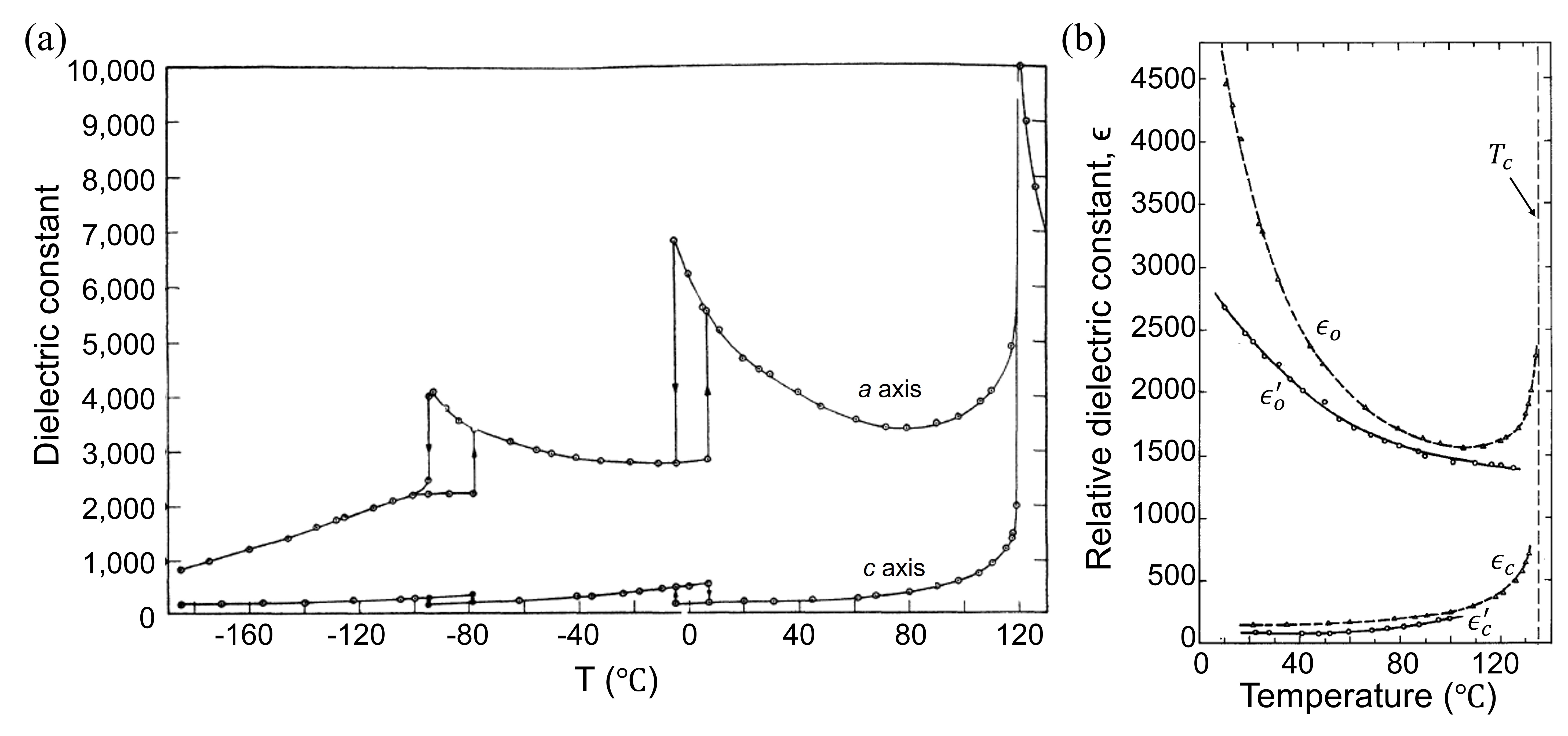}
    \caption{The temperature-dependent variation of the dielectric constant in BTO bulk materials. (a) Reproduced with permission from Merz \textit{et al.}, Physical Review 76, 1221 (1949)\cite{merz1949electric}. Copyright 1949 by the American Physical Society. (b) Reprinted from Journal of Physics and Chemistry of Solids, 29, Wemple \textit{et al.}, Dielectric and optical properties of melt-grown BaTiO$_3$,  1797–1803 (1968), with permission from Elsevier\cite{wemple1968dielectric}.}
    \label{fig:chap2-3}
\end{figure}

Since $\epsilon_r$ and tan $\delta$ are critical parameters in the design and performance of target devices, extensive experimental data on bulk BTO and BTO thin films fabricated using various methodologies have been reported in the literature \cite{merz1949electric,chelladurai2025barium,zgonik1994dielectric,powles1949measurement,von1950ferroelectricity,kazaoui1992high,laabidi1994indications,mcneal1998effect,hamano2003relative,petzelt2003fir,tsurumi2007ultrawide,rosa2017barium,bovtun2021ferroelectric}. Consensus results indicate that, at room temperature (23$^\circ$C) under a low-frequency electric field, the unclamped dielectric constant (static dielectric constant) of bulk BTO is  $\epsilon_{a}^{\prime}$$\sim$4400 along the $a$-axis and $\epsilon_{c}^{\prime}$$\sim$129 along the $c$-axis, while the clamped dielectric constant are reduced to  $\epsilon_{a}^{\prime}$$\sim$2200 and $\epsilon_{c}^{\prime}$$\sim$56, respectively. Table \ref{tab:2-1} presents the dielectric constants of several typical optical materials, with BTO exhibiting an exceptionally high value.
The dielectric constants also exhibit temperature dependence, undergoing significant discontinuity in the vicinity of the phase transition temperatures, as shown in Fig. \ref{fig:chap2-3}(a). Furthermore, the domain orientation, domain structure, and defect density in BTO crystals and thin films vary significantly depending on the fabrication processes employed, resulting in distinct frequency-dependent behaviors of both the dielectric constant and loss tangent under applied electric fields\cite{chelladurai2025barium,powles1949measurement,von1950ferroelectricity,kazaoui1992high,laabidi1994indications,mcneal1998effect,hamano2003relative,petzelt2003fir,tsurumi2007ultrawide,rosa2017barium,bovtun2021ferroelectric}.
Among reported studies, recent results by Chelladurai \textit{et al.}\cite{chelladurai2025barium} are notable, as they fabricated high-quality epitaxial $a$-axis-oriented single-crystal, multi-domain BTO thin films. In this case, the measured dielectric constant represents an effective average of the $a$- and $c$-axis responses, i.e. $\varepsilon_{r}^{\prime} \sim \left( \varepsilon_{a}^{\prime} + \varepsilon_{c}^{\prime} \right) / 2$. To date, this is the only work that has successfully achieved continuous permittivity measurements over an exceptionally wide frequency range from 100 MHz to 300 GHz, providing a reliable reference for thin-film BTO parameters, with $\epsilon_{r}^{\prime}$ of $\sim$1134 and $\sim$441 measured at 100 MHz and 300 GHz, respectively. In this study, a $\tan\delta \sim 0.43$ at an RF frequency of 300 GHz was also observed. This can be attributed to a general trend in dielectric materials: systems with larger real parts of the dielectric constant typically also exhibit larger imaginary components. This relationship can be described using the Debye model\cite{chelladurai2025barium}: 
\begin{equation} \label{eq:2-3} 
\varepsilon_{r}^{\prime}(f)+j\varepsilon_{r}^{\prime\prime}(f)=\varepsilon_{\infty}+\sum_{i}\Delta\varepsilon_{i}\int\mathcal{N}\left(\gamma,\gamma_{0i},\sigma_{i}\right)\frac{\gamma}{\gamma-jf}\mathrm{~d}\gamma,
\end{equation}
\begin{equation} \label{eq:2-4} 
\mathcal{N}\left(\gamma,\gamma_{0i},\sigma_{i}\right)=\frac{1}{\sigma_{i}\sqrt{2\pi}} e^{-\frac{1}{2}\left(\frac{\log\gamma-\log\gamma_{0i}}{\sigma_{i}}\right)^{2}},
\end{equation}
where $f$ is RF frequency, $\varepsilon_{\infty}$ is the constant permittivity that the model approaches at the high-frequency limit, $\Delta\varepsilon_{i}$ is the relaxation strength with $i$ represents different types of relaxation (for example, $i = 1$ for defect-related relaxation, $i = 2$ for Ti-ion relaxation, $i = 3$ for domain poling and so on). $\gamma$ denotes the relaxation frequency, and the normal distribution of logarithmic relaxation frequencies $\mathcal{N}\left(\gamma,\gamma_{0i},\sigma_{i}\right)$ is described by its center relaxation frequency $\gamma_{0i}$ and its width in log space $\sigma_{i}$. 

According to Eq. \ref{eq:2-3}, the real and imaginary parts of the dielectric constant share the same relaxation strength $\Delta\varepsilon_{i}$. In other words, for a given relaxation mechanism, a large $\Delta\varepsilon_{i}$ simultaneously leads to increased magnitudes of both the real part $\epsilon_{r}^{\prime}$ and the imaginary part $\epsilon_{r}^{\prime \prime}$. Consequently, achieving low dielectric loss requires minimizing relaxation processes with large relaxation strengths. This indicates the importance of fabricating defect-free films to suppress defect-related relaxation and of optimizing domain structure design for modulation of the normal distribution function \cite{mcneal1998effect,arlt1994dielectric}. 
From a structural perspective, an ideal signal-crystal, single-domain film is expected to exhibit the sharpest normal distribution peak, suggesting its potential to achieve exceptionally low losses over a wider frequency range except the peak. Alternatively, the multi-domain structure can be engineered through domain miniaturization, resulting in a broadened peak width that distributes the loss over a wide frequency band, thereby achieving relatively low loss across the entire frequency spectrum. 
Experimentally, low-loss behavior has been reported in several thin-film BTO systems. For example, a 500 nm-thick amorphous BTO film exhibits a tan $\delta$ of $\sim$0.03 at an RF frequency of 76 GHz\cite{ummethala2021hybrid}, while a $\sim$500 nm-thick polycrystalline film shows tan $\delta$ $<0.05$ at 10 GHz\cite{berge2007effect}. Although these examples are not epitaxial BTO films, they show that low dielectric loss can be achieved in BTO thin-film systems and motivate further efforts to reduce RF loss in epitaxial BTO.

\begin{table*}[t!]
\caption{Optoelectronic characteristics of some commonly used materials. The values are measured at room temperature.}
\label{tab:2-1}
\begin{ruledtabular}
\begin{tabular}{l l l l l l}

\textbf{Materials} &
\makecell[l]{\textbf{RF dielectric}\\ \textbf{constant}} &
\makecell[l]{\textbf{Second-order nonlinear}\\ \textbf{coefficient $\chi^{(2)}$ (pm/V)}} &
\makecell[l]{\textbf{Third-order nonlinear}\\ \textbf{refractive index $n_2$ (m$^2$/W)}} &
\makecell[l]{\textbf{Electro-optic}\\ \textbf{coefficient (pm/V)}} &
\makecell[l]{\textbf{Thermo-optic coefficient}\\ \textbf{TOC ($^\circ$C$^{-1}$)}} \\
\hline

\textbf{BaTiO$_3$} & $\varepsilon_a \sim 4400$ (unclamped) & $d_{15} = 17$ & $1.80 \times 10^{-18}$ & $r_{51} = r_{42} = 1300$ & $dn_e/dT = 1.8 \times 10^{-4}$ \\
& $\varepsilon_a \sim 2200$ (clamped) & $d_{31} = 15.7$ & @\SI{1550}{\nano\meter} \cite{riedhauser2025absorption} & $r_{33} = 105$ & (bulk, @\SI{633}{\nano\meter} \cite{buse1993refractive}) \\
& $\varepsilon_c \sim 129$ (unclamped) & $d_{33} = 6.8$ & & $r_{13} = r_{23} = 10.2$ & $dn_e/dT = 1.095 \times 10^{-5}$ \\
& $\varepsilon_c \sim 56$ (clamped) \cite{zgonik1994dielectric} & @\SI{1064}{\nano\meter} \cite{fleming2018handbook} & & (unclamped \cite{zgonik1994dielectric}) & (thin film, @\SI{1550}{\nano\meter} \cite{lin2025thermo}) \\
& & & & $r_{51} = r_{42} = 730$ & \\
& & & & $r_{33} = 40.6$ & \\
& & & & $r_{13} = r_{23} = 8$ & \\
& & & & (clamped \cite{zgonik1994dielectric}) & \\

\textbf{LiNbO$_3$} & $\varepsilon_a \sim 44$ & $d_{33} = 41.7$ \cite{hao2020second,boyd1964linbo3} & $1.80 \times 10^{-19}$ & $r_{33} = 30.9$ & $dn_e/dT = 3.34 \times 10^{-5}$ \\
& $\varepsilon_c \sim 27.9$ (clamped \cite{zhu2021integrated}) & $d_{31} = 4.3$ & @\SI{1550}{\nano\meter} \cite{zhu2021integrated} & $r_{51} = 32.6$ & (bulk, @\SI{1523}{\nano\meter} \cite{zhu2021integrated}) \\
& & $d_{22} = 2.1$ \cite{zhu2021integrated} & & $r_{13} = 9.6$ & \\
& & @\SI{1064}{\nano\meter} & & $r_{22} = 6.8$ \cite{zhu2021integrated} & \\

\textbf{LiTaO$_3$} & $\varepsilon_a \sim 38.3$ & $d_{33} = 13.8$ & $1.46 \times 10^{-19}$ & $r_{33} = 30.5$ & $d\Delta n/dT = 4.93 \times 10^{-5}$ \\
& $\varepsilon_c \sim 46.2$ \cite{zhu2021integrated} & $d_{31} = 0.85$ & @\SI{800}{\nano\meter} \cite{zhu2021integrated} & $r_{51} = 20$ & (bulk, @\SI{633}{\nano\meter} \cite{saadon2006thermo}) \\
& & @\SI{1064}{\nano\meter} \cite{zhu2021integrated} & & $r_{13} = 8.4$ & \\
& & & & $r_{22} = -0.2$ \cite{zhu2021integrated} & \\

\textbf{KTiOPO$_4$} & $\varepsilon_a \sim 11.6$ & $d_{31} = 6.5$ & $1.77 \times 10^{-19}$ & $r_{13} = 9.5$ & $dn_e/dT = 1.1 \times 10^{-5}$ \\
& $\varepsilon_b \sim 11.0$ & $d_{32} = 5.0$ & @\SI{1064}{\nano\meter} \cite{kulagin2003nonlinear} & $r_{23} = 15.7$ & $dn_e/dT = 1.3 \times 10^{-5}$ \\
& $\varepsilon_c \sim 15.4$ (clamped) \cite{bierlein1989potassium} & $d_{33} = 13.7$ & & $r_{33} = 36.3$ & $dn_e/dT = 1.6 \times 10^{-5}$ \\
& & $d_{24} = 7.6$ & & $r_{42} = 7.3$ & (bulk, @\SI{1064}{\nano\meter} \cite{bierlein1989potassium}) \\
& & $d_{15} = 6.1$ & & $r_{51} = 9.3$ \cite{bierlein1989potassium} & \\
& & @\SI{1064}{\nano\meter} \cite{bierlein1989potassium} & & & \\

\textbf{KNbO$_3$} & $\varepsilon_a \sim 37$ & $d_{33} = 19.6$ & $\sim 8\times 10^{-19}$ & $r_{13} = 10$ & $dn_e/dT = -4.5 \times 10^{-5}$ \\
& $\varepsilon_b \sim 780$ & $d_{31} = 10.8$ & @\SI{1064}{\nano\meter} \cite{bosshard1995kerr} & $r_{23} = -8$ & $dn_e/dT = 6.7 \times 10^{-5}$ \\
& $\varepsilon_c \sim 24$ (clamped) \cite{wiesendanger1973dielectric} & $d_{15} = 12.5$ & & $r_{33} = 50$ & (bulk, @\SI{852}{\nano\meter} \cite{chang1998heterodyne}) \\
& & @\SI{1064}{\nano\meter} \cite{shoji1997absolute} & & $r_{42} = 400$ & \\
& & & & $r_{51} = 120$ \cite{wiesendanger1973dielectric} & \\

\textbf{Si} & 11.7 \cite{zhu2021integrated} & 0 \cite{zhu2021integrated} & $5 \times 10^{-18}$ & 0 \cite{zhu2021integrated} & $dn/dT = 1.87 \times 10^{-4}$ \\
& & & @\SI{1550}{\nano\meter} \cite{zhu2021integrated} & & (bulk, @\SI{1500}{\nano\meter} \cite{frey2006temperature}) \\

\textbf{SiO$_2$} & 3.9 \cite{zhu2021integrated} & 0 \cite{zhu2021integrated} & $3 \times 10^{-20}$ & 0 \cite{zhu2021integrated} & $dn/dT = 8.16 \times 10^{-6}$ \\
& & & @\SI{1550}{\nano\meter} \cite{zhu2021integrated} & & (glass, @\SI{1550}{\nano\meter} \cite{rego2023temperature}) \\

\textbf{SiN} & 7.5 \cite{zhu2021integrated} & 0 \cite{zhu2021integrated} & $2.5 \times 10^{-19}$ & 0 \cite{zhu2021integrated} & $dn/dT = 4.7 \times 10^{-5}$ \\
& & & @\SI{1550}{\nano\meter} \cite{zhu2021integrated} & & (Amorphous film, \\
& & & & & @\SI{1510}{\nano\meter} \cite{zanatta2013thermo}) \\

\textbf{AlN} & 8.6 \cite{zhu2021integrated} & $d_{33} = 4.7$ & $2.3 \times 10^{-19}$ & $r_{13} = 0.67$ & $dn/dT \sim3 \times 10^{-5}$ \\
& & $d_{31} = 1.6$ & @\SI{1550}{\nano\meter} \cite{zhu2021integrated} & $r_{33} = -0.59$ \cite{zhu2021integrated} & (bulk, @\SI{1550}{\nano\meter} \cite{bowman2018optical}) \\
& & @\SI{1064}{\nano\meter} \cite{zhu2021integrated} & & & \\

\textbf{GaAs} & 16.9 \cite{zhu2021integrated} & $d_{36} = 170$ & $2.6 \times 10^{-17}$ \cite{zhu2021integrated} & $r_{41} = 1.43$ \cite{zhu2021integrated} & $dn/dT \sim2.19 \times 10^{-4}$ \\
& & @\SI{1064}{\nano\meter} \cite{shoji1997absolute} & & & (bulk, @\SI{1523}{\nano\meter} \cite{della2000temperature}) \\

\end{tabular}
\end{ruledtabular}
\end{table*}

\subsubsection{RF effective refractive index}
The refractive index $n$ of a dielectric material is related to its relative permittivity $\varepsilon_r$ by \cite{grant2013electromagnetism},
\begin{equation}
n = \sqrt{\varepsilon_{r}}.
\end{equation}
BTO material with an exceptionally high dielectric constant therefore tends to exhibit large refractive indices in the RF frequency regime. And it should be noted that the  effective RF refractive index, $n_{\text{RF}}$, is not an intrinsic refractive index of BTO itself, but an effective microwave refractive index determined by the full electrode--BTO--substrate structure and the corresponding microwave-mode distribution. Therefore, $n_{\text{RF}}$ depends not only on the BTO permittivity, but also on the BTO thickness, substrate and cladding materials, and electrode geometry. A thicker BTO layer or stronger RF-field overlap with the high-permittivity BTO region generally increases the effective microwave capacitance and thus increases $n_{\mathrm{RF}}$, whereas thinner BTO slabs or geometries with reduced RF-field confinement in BTO can lower it. This explains why, for devices with BTO thin-film slab thicknesses from 35 to 570 nm, the reported effective RF refractive index $n_{\text{RF}}$ lies in the range of $\sim$3-5 \cite{tang2004electrooptic,sun2005performance,sun2006low,hu2015modeling,sun2015theoretical,castera2016towards,girouard2017modulator,xing2022membrane}. Moreover, a recent theoretical study \cite{shang2026design} further identified thinning the residual BTO slab as a key design principle for reducing the RF effective refractive index and achieving microwave-optical velocity matching in high-speed BTO modulators. In addition, because the RF permittivity of BTO is dispersive, $n_{\text{RF}}$ is also frequency dependent and generally decreases as the slow polarization contributions relax at higher RF frequencies.

\subsection{\label{Sec2.3}Linear optical properties}
The investigation of the linear optical properties of materials fundamentally involves analysis of the dielectric constant in the optical frequency range, where, for BTO materials, the dielectric constant is determined exclusively by electronic polarization \cite{chelladurai2025barium,slater1950lorentz}. In the optical regime, the complex refractive index $\tilde{n} = n + j\kappa$ is commonly used to represent the dielectric response, where the real part $n$ determines the refractive index, while the imaginary part $\kappa$ governs the optical loss. 

\begin{figure}
    \centering
    \includegraphics[width=1\linewidth]{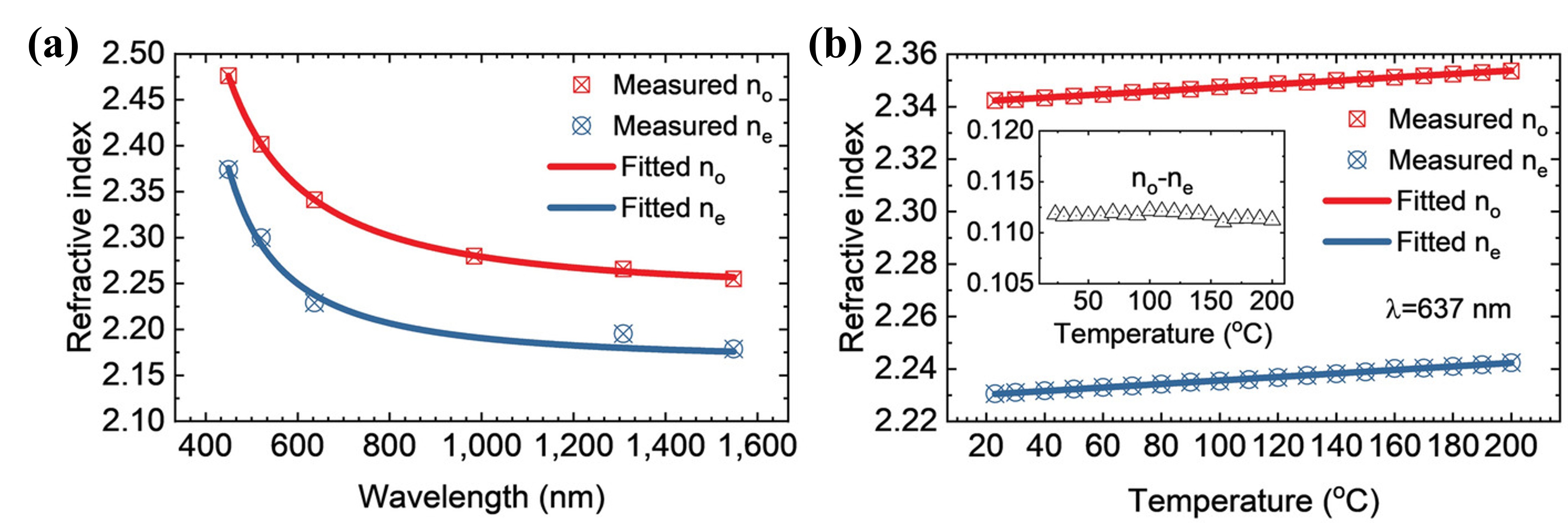}
    \caption{(a) Refractive indices of BTO growth of DSO substrate as a function of wavelength. $n_\text{o}$, $n_\text{e}$ are measured at 450, 521, 637, 984, 1308, 1548 nm and fitted with the Sellmeier equation at room temperature. (b) Measured refractive indices of BTO as a function of temperature are linear fitted indicating phase stability of BTO films up to 200$^\circ$C. The difference between $n_\text{o}$ and $n_\text{e}$ are illustrated in the inset indicating a stable birefringence.Reproduced with permission from Cao \textit{et al.}, Advanced Materials 33, 2101128 \copyright 2021 Wiley-VCH GmbH\cite{cao2021barium}.}
    \label{fig:chap2-4}
\end{figure}

For the real part $n$, its dependence on wavelength $\lambda$ defines the optical dispersion relation. Numerous experimental studies on bulk and thin-film BTO have measured this dispersion relation within the visible and near-infrared spectral ranges, as illustrated in Fig. \ref{fig:chap2-4}. At room temperature in the tetragonal phase of BTO, the ordinary axes are the $a$- and $b$-axes, while the extraordinary axis is the $c$-axis. The corresponding refractive indices are denoted as $n_\text{o}$ and $n_\text{e}$, respectively. The wavelength dependence of $n$ is typically described using the empirical Sellmeier equation \cite{buse1993refractive,drdomenico1969oxygen}: 
\begin{equation}
n^{2}(\lambda)=A+\sum_{i}\frac{B_{i}\lambda^{2}}{\lambda^{2}-C_{i}},
\label{eq:2-6}
\end{equation}
where the coefficients $A$, $B_i$ and $C_i$ are experimentally determined Sellmeier parameters. Variants of the Sellmeier equation have also been developed to account for refractive-index changes induced by temperature, pressure, or other external factors. Figure \ref{fig:chap2-4} shows refractive index measurements as a function of wavelength and temperature for a BTO thin film grown on a DSO substrate, as reported by Cao \textit{et al.}\cite{cao2021barium}. As shown in Fig. \ref{fig:chap2-4}(a), over an extended wavelength measurement range of 450-1550 nm at 20$^\circ$C, the extracted Sellmeier coefficients are as follows: for the ordinary index $n_\text{o}$, $A = 1$, $B_1 = 0.69661$, $C_1 = 294.82$ nm$^2$, $B_2 = 1.33348$, $C_2 = 174.92$ nm$^2$, $B_3 = 1.99513$, and $C_3 = 172.02$ nm$^2$; for the extraordinary $n_\text{e}$, $A = 3$, $B_1 = 0.93217$, $C_1 = 319.52$ nm$^2$, $B_2 = 0.76066$, and $C_2 \sim0$ nm$^2$.
Regarding the temperature dependence of birefringence, thin-film BTO exhibits behavior that differs markedly from that of bulk BTO. Owing to the substantial enhancement of the Curie temperature $T_\text{C}$ in thin films, the birefringence remains largely unchanged even at temperatures exceeding 200$^\circ$C, as shown in Fig. \ref{fig:chap2-4}(b).

\begin{figure}
    \centering
    \includegraphics[width=0.5\linewidth]{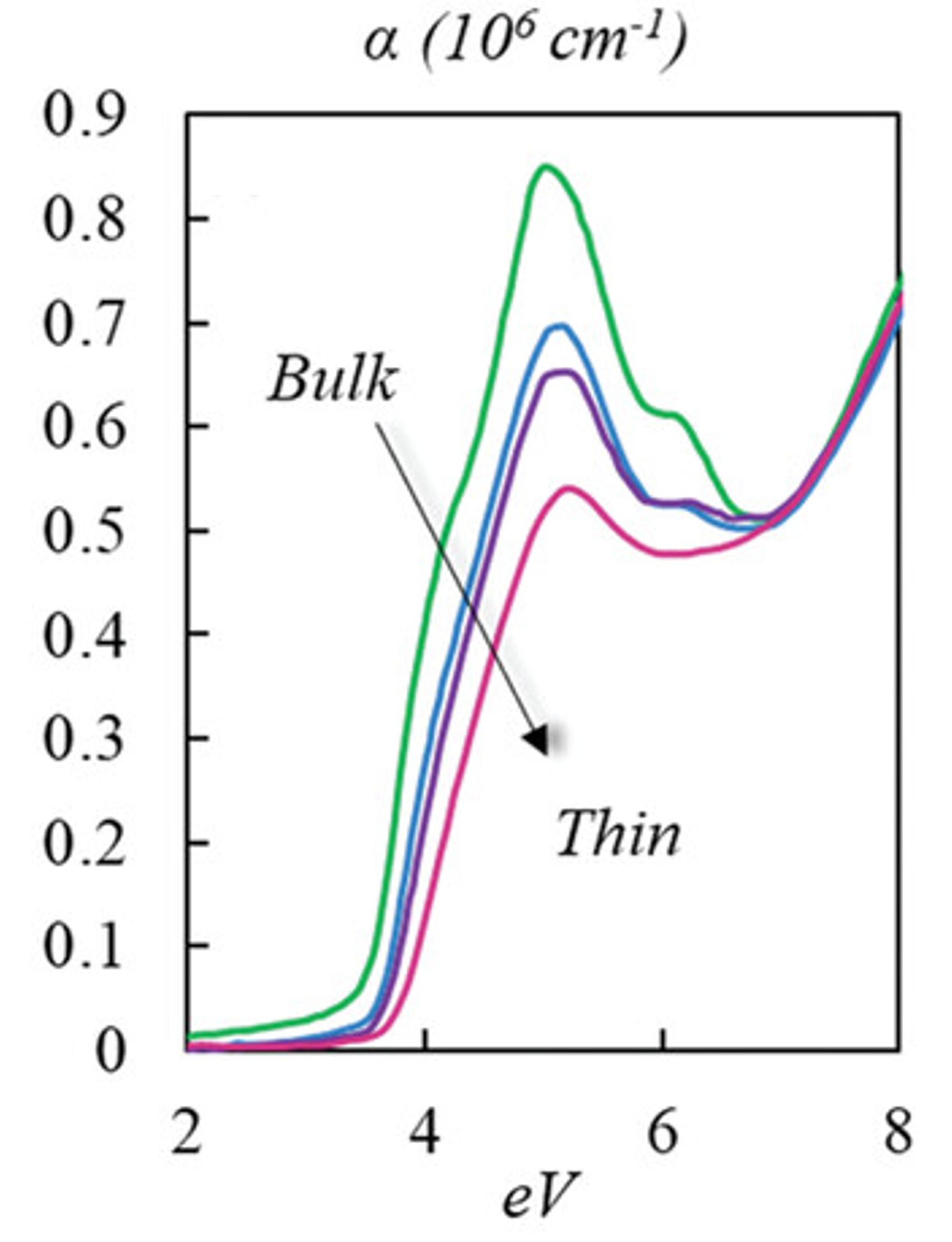}
    \caption{Absorption coefficient $\alpha$ spectra are modified as the crystal thickness; bulk (green), 20 (blue), 11 (purple), and 8 nm (magenta) film. Reproduced from Chernova \textit{et al.}, Applied Physics Letters 106, 192903(2015), with the permission of AIP Publishing\cite{chernova2015strain-controlled}.}
    \label{fig:chap2-5}
\end{figure}

For the imaginary part $\kappa$, its dependence on wavelength $\lambda$ determines the optical transparency window of the material. As illustrated in Fig. \ref{fig:chap2-5}, the relationship between optical loss and photon energy reveals that, for bulk BTO, optical losses remain minimal for photon energies below 3.2 eV, corresponding to wavelengths above 387 nm. This photon energy, commonly referred to as the bandgap $E_g$, represents the threshold below which photons do not possess sufficient energy to be absorbed by BTO, thereby preventing electronic transitions from the valence band to the conduction band. Consequently, the material remains transparent to photons with energies below $E_g$.
For thin films, the bandgap $E_g$ is often further increased due to strain effects induced by the substrate. This behavior indicates that the transparency window of BTO can safely spans both the visible and infrared spectral ranges, supporting its suitability for optical devices operating in these wavelength bands.
In addition, as evident from the Sellmeier equation in Eq. \ref{eq:2-6}, the coefficient $A$ provides an approximation of short-wavelength (e.g., ultraviolet) absorption contributions to the refractive index at longer wavelengths, while each term in the summation represents an absorption resonance with strength $B_i$ at a characteristic wavelength $\sqrt{C_i}$. Notably, the fitted resonance wavelengths are all shorter than the wavelength corresponding to the bandgap energy $E_g$.

\subsection{\label{optical nonlinear}Nonlinear optical properties}
In addition to the dielectric response determined by linear polarizability and characterized by the dielectric constant $\epsilon_r$, higher-order polarization responses of the medium give rise to nonlinear optical properties \cite{shenprinciples}. The relationship between the total polarization vector \textbf{P} and the applied electric field vector \textbf{E} in the medium is expressed as \cite{drummond2014quantum}: 
\begin{equation}
\mathbf{P}(\mathbf{E}) = \mathbf{P}^{\mathrm{L}} + \mathbf{P}^{\mathrm{NL}} = \varepsilon_0 \sum_{n>0} \boldsymbol{\chi}^{(n)} : \mathbf{E}^{\otimes n},
\label{eq:2-7}
\end{equation}
where
\begin{equation}
\mathbf{P}^{\mathrm{L}} = \varepsilon_0 \boldsymbol{\chi}^{(1)} \cdot \mathbf{E} = \varepsilon_0 (\varepsilon_r - 1) \cdot \mathbf{E}
\label{eq:2-8}
\end{equation}

\noindent represents the linear response, which determines the optical refractive index through the linear susceptibility tensor $\boldsymbol{\chi}^{(1)}$ or equivalently the dielectric constant tensor $\boldsymbol{\varepsilon}_{r}$. The term \textbf{P}$_{\text{NL}}$ denotes the nonlinear polarization, where the notation $\mathbf{E}^{\otimes n}$ indicates the vector Kronecker product. When $n = 2$ or $n = 3$, the polarization response corresponds to second-order or third-order optical nonlinearities, respectively.

\subsubsection{$\chi^{(2)}$ Nonlinearity}
The second-order nonlinear polarization vector \textbf{P}$_{\text{NL}}$ can be expressed as
\begin{equation}
\mathbf{P}^{\mathrm{NL}} = \varepsilon_0 \boldsymbol{\chi}^{(2)} : \mathbf{E} \otimes \mathbf{E},
\label{eq:2-9}
\end{equation}
or, in the more commonly used component form,
\begin{equation}
P_{j}^{\mathrm{NL}}(\omega_{p}) = \varepsilon_{0} \sum_{k,l} \chi_{jkl}^{(2)} (-\omega_{p}, \omega_{s}, \omega_{i}) E_{k}(\omega_{s}) E_{l}(\omega_{i}).
\label{eq:2-10}
\end{equation}
Here, $E_k$ and $E_l$ are the driving electric fields with polarization directions of $k$ and $l$, with angular frequencies $\omega_s$ and $\omega_i$, respectively, generating a nonlinear polarization component $P_j^{\text{NL}}$ polarized along direction $j$ at frequency $\omega_p = \omega_s + \omega_i$. The subscripts $p$, $s$, and $i$ correspond to the pump, signal, and idler light, respectively.  Here we implicitly assume alternating fields, i.e., propagating light of three distinct frequencies. When a static field is applied, the physical effect is similar, but the material's response is typically referred to as the electro-optic (EO) effect and the notation is different; see Section E below.

As a third-rank tensor, $\boldsymbol{\chi}^{(2)}$ contains 27 components. However, due to the intrinsic permutation symmetry of the electric fields (i.e., the equivalence of $\chi_{jkl}^{(2)}(-\omega_p,\omega_s,\omega_i)$ and $\chi_{jkl}^{(2)}(-\omega_p,\omega_i,\omega_s)$), as well as frequency permutation symmetry such as $\chi_{jkl}^{(2)}(-\omega_p,\omega_s,\omega_i) = \chi_{kjl}^{(2)}(\omega_s,-\omega_p,\omega_i)$, the number of independent components can be significantly reduced. Under the additional assumptions that the medium is transparent over the relevant optical frequency range and that dispersion can be neglected, the Kleinman symmetry applies. By using the nonlinear coefficients $d_{jkl}$ to express the $\chi_{jkl}^{(2)}$ components, where $d_{jkl}=\chi_{jkl}^{(2)}/2$ \cite{boyd2008nonlinear}, the $\chi^{(2)}$ tensor can be conveniently represented in the contracted 3×6 \textbf{d}-matrix form,
\begin{equation}
\mathbf{d} = \left(\begin{array}{cccccc}
d_{11} & d_{12} & d_{13} & d_{14} & d_{15} & d_{16} \\
d_{21} & d_{22} & d_{23} & d_{24} & d_{25} & d_{26} \\
d_{31} & d_{32} & d_{33} & d_{34} & d_{35} & d_{36}
\end{array}\right),
\label{eq:2-11}
\end{equation}
and the polarization vector \textbf{P}$^{\text{NL}}$ can be expressed as \cite{yariv1989quantum,bihari1994investigation}
\begin{equation}
\mathbf{P}^{\mathrm{NL}} = 2\varepsilon_{0}
\left(\begin{array}{cccccc}
d_{11} & d_{12} & d_{13} & d_{14} & d_{15} & d_{16} \\
d_{21} & d_{22} & d_{23} & d_{24} & d_{25} & d_{26} \\
d_{31} & d_{32} & d_{33} & d_{34} & d_{35} & d_{36}
\end{array}\right)
\left(\begin{array}{c}
E_{x}^{2} \\
E_{y}^{2} \\
E_{z}^{2} \\
2E_{y}E_{z} \\
2E_{x}E_{z} \\
2E_{x}E_{y}
\end{array}\right).
\label{eq:2-12}
\end{equation}
In this notation, the first subscript of each \textbf{d}-matrix elements denotes the polarization direction of the pump light, where indices 1, 2, and 3 correspond to the \textit{x}, \textit{y}, and \textit{z} directions, respectively. The second subscript specifies the polarization configuration of the interacting optical fields, with indices 1–6 representing, respectively: $xx$, $yy$, $zz$, $yz$, $xz$, and $xy$ field combinations.

For BTO in the tetragonal phase, the corresponding \textbf{d}-matrix takes the form \cite{karvounis2020barium,miller1964optical,bihari1994investigation}
\begin{equation}
\mathbf{d} = \left(\begin{array}{cccccc}
0 & 0 & 0 & 0 & d_{15} & 0 \\
0 & 0 & 0 & d_{24} & 0 & 0 \\
d_{31} & d_{32} & d_{33} & 0 & 0 & 0
\end{array}\right),
\label{eq:2-13}
\end{equation}
where symmetry between the \textit{x}- and \textit{y}-axes leads to the relations $d_{15} = d_{24}$ and $d_{31} = d_{32}$. Both theoretical calculations of higher-order polarization and experimental measurements based on second-harmonic generation (SHG) have reported values for these coefficients. Typical values at an optical wavelength of 1064 nm are $d_{15} = 17$ pm/V, $d_{31} = 15.7$ pm/V, and $d_{33} = 6.8$ pm/V\cite{fleming2018handbook}. 

A comparison of second-order nonlinear coefficients for several representative materials is provided in Table \ref{tab:2-1}. Although the $\chi^{(2)}$ response of BTO is not as large as that of LN, which exhibits a maximum $d_{33}$ of approximately 41.7 pm/V, BTO nonetheless remains a promising second-order nonlinear optical material. When considered together with its exceptionally large electro-optic coefficients, BTO holds significant promise as a multifunctional platform for fully integrated photonic and electro-optic applications. Moreover, it should be noted that the optical $\chi^{(2)}$ coefficients and the EO coefficients of BTO correspond to different frequency regimes of the second-order response. For optical $\chi^{(2)}$ processes, all interacting fields are at optical frequencies, so the response is mainly governed by the electronic polarization. Therefore, the optical nonlinear coefficients of BTO, such as $d_{15}$, $d_{31}$, and $d_{33}$, are moderate and do not scale directly with its low-frequency EO coefficients. In contrast, the Pockels effect involves one optical field and one DC or RF electric field, allowing slower polarization mechanisms, including inverse piezoelectric response and ionic displacement contribution, to enhance the EO response depending on the modulation frequency. This frequency-dependent distinction explains why BTO can exhibit exceptionally large EO coefficients while having optical nonlinear coefficients that are not correspondingly large.

\subsubsection{\label{chi3}$\chi^{(3)}$ Nonlinearity}
The third-order nonlinear polarization vector \textbf{P}$^{\text{NL}}$ is given by \cite{drummond2014quantum}
\begin{equation}
\mathbf{P}^{\mathrm{NL}} = \varepsilon_{0} \boldsymbol{\chi}^{(3)} : \mathbf{E}^{\otimes 3},
\label{eq:2-14}
\end{equation}
or, in the more commonly used component form,
\begin{equation}
P_{j}^{\mathrm{NL}} = \varepsilon_{0} \sum_{k,l,m} \chi_{jklm}^{(3)} E_{k} E_{l} E_{m}.
\label{eq:2-15}
\end{equation}
Here, $E_k$, $E_l$ and $E_m$ denote the driving electric fields with polarization directions of $k$, $l$ and $m$, with frequencies of $\omega_{p2}$, $\omega_s$ and $\omega_i$, respectively, generating a $P_j^{NL}$ with polarization direction of $j$ and frequency of $\omega_{p1} + \omega_{p2} = \omega_s + \omega_i$, where the subscripts $p1$, $p2$, $s$, and $i$ denote the pump 1, pump 2, signal, and idler light, respectively. This nonlinear interaction is commonly referred to as four-wave mixing, and the associated $\chi^{(3)}$ response is known as Kerr nonlinearity.

Due to its higher-order nature, the magnitude of the $\chi^{(3)}$ coefficient is typically several orders of magnitude smaller than that of $\chi^{(2)}$, with units of (pm/V)$^2$. A more commonly used parameter for characterizing third-order nonlinearity in optical materials is the nonlinear refractive index $n_2$, which describes the intensity-dependent refractive index change induced by nonlinear optical effects\cite{karvounis2020barium,drummond2014quantum},
\begin{equation}
n(I) \sim n_{1} + n_{2} I,
\label{eq:2-16}
\end{equation}
where $n_1$ is the linear optical refractive index and $I$ is the optical intensity. The unit of $n_2$ is m$^2$/W, and its relationship with the $\chi^{(3)}$ coefficient is given by \cite{drummond2014quantum}
\begin{equation}
n_{2} = \frac{3\chi^{(3)}}{4\varepsilon_{0} n_{1}^{2} c}.
\label{eq:2-17}
\end{equation}

Experimentally, the nonlinear refractive index of BTO thin films has been measured to be $1.80 \times 10^{-18}$ m$^2$/W \cite{riedhauser2025absorption}, corresponding to a $\chi^{(3)}$ coefficient of approximately $2.55 \times 10^{4}$ (pm/V)$^2$. Compared with commonly used $\chi^{(3)}$ materials such as silicon nitride (SiN) and LN \cite{riedhauser2025absorption}, the $\chi^{(3)}$ coefficient of BTO thin films is enhanced by approximately one order of magnitude, as shown in Table \ref{tab:2-1}. Furthermore, metal-ion doping of BTO thin films has been shown to further increase the $\chi^{(3)}$ coefficient by several orders of magnitude, in some cases surpassing that of III-V materials\cite{yang2002z,yang2002rh,shi1999nonlinear,wang2002iron}, as detailed in Table \ref{tab:2-2}. These results highlight the strong potential of BTO thin films for $\chi^{(3)}$-based nonlinear optical applications.

\begin{table*}[t!]
\caption{Nonlinear refractive index $n_2$ of undoped and doped BTO materials, comparing with III-V materials.}
\label{tab:2-2}
\begin{ruledtabular}
\begin{tabular}{l l l}

\textbf{Materials} & 
\makecell[c]{\textbf{Nonlinear refractive index} $\bm{n_{2}}$ (\si{\meter\squared\per\watt})} & 
\textbf{Reference} \\
\hline

BTO thin film & $1.80 \times 10^{-18}$ & \cite{riedhauser2025absorption} \\
Rh:BTO thin film & $2.94 \times 10^{-12}$ ($\chi^{(3)}=3.59 \times 10^{-7}$ esu) & \cite{yang2002z} \\
Rh:BTO thin film & $4.66 \times 10^{-12}$ ($\chi^{(3)}=5.71 \times 10^{-7}$ esu) & \cite{yang2002rh} \\
Ce:BTO quantum dot & $1.80 \times 10^{-13}$ ($\chi^{(3)}=2.21 \times 10^{-8}$ esu) & \cite{shi1999nonlinear} \\
Fe:BTO thin film & $5.87 \times 10^{-12}$ ($\chi^{(3)}=7.18 \times 10^{-7}$ esu) & \cite{wang2002iron} \\
GaAs & $2.6 \times 10^{-17}$ & \cite{zhu2021integrated} \\
GaP & $1.13 \times 10^{-17}$ & \cite{wilson2020integrated} \\
Al$_{0.2}$Ga$_{0.8}$As & $2.60 \times 10^{-17}$ & \cite{gao2022probing} \\

\end{tabular}
\end{ruledtabular}
\end{table*}

\subsection{\label{EOC}Electro-optic coefficients}
The electro-optic (EO) effect originates from second-order nonlinear ($\chi^{(2)}$) processes in which one of the interacting frequencies lies in the constant-field or RF regime $\Omega$ \cite{sauter1996nonlinear},
\begin{equation}
\mathbf{P}^{\mathrm{EO}}(\omega+\Omega)=\varepsilon_{0}\boldsymbol{\chi}^{(2)}(-(\omega+\Omega),\omega,\Omega):\mathbf{E}(\omega)\otimes\mathbf{E}(\Omega),
\label{eq:2-18}
\end{equation}
where $\Omega$ is significantly lower than the optical frequency $\omega$, reflecting the modulation effect of $\Omega$ on $\omega$:
\begin{equation}
\mathbf{P}^{\mathrm{EO}}(\omega)=\varepsilon_{0}\boldsymbol{\chi}^{(2)}(-\omega,\omega,\Omega):\mathbf{E}(\omega)\otimes\mathbf{E}(\Omega),
\label{eq:2-19}
\end{equation}
that is to say, under the modulation of electric field \textbf{E}($\Omega$), the optical frequency remains unchanged, while an additional polarization is induced, thereby modifying the total polarization vector. Consequently, this leads to a variation of the refractive index of the optical field.
It is important to emphasize that the magnitude of the EO $\chi^{(2)}$ response differs significantly from that of the nonlinear optical-frequency $\chi^{(2)}$.

This difference arises from the distinct polarization mechanisms active in different frequency regimes, as discussed in Section \ref{sec2-2}. 
Specifically, polarization in the optical frequency band is governed primarily by electronic polarization, whereas in the RF frequency band additional contributions from ionic polarization and mechanical effects become significant. Consequently, for BTO, which exhibits a much stronger polarization response at radio frequencies than at optical frequencies, the EO $\chi^{(2)}$ is substantially larger than the corresponding nonlinear optical $\chi^{(2)}$ coefficient. 

In the study of the EO effect, the EO coefficient $r$ is commonly used as an alternative representation of the EO $\chi^{(2)}$ response. The modification of the refractive-index ellipsoid under an applied electric field can be expressed as \cite{wemple1969oxygen,vasudevan2023domain,kim2023nature}:
\begin{equation}
  \begin{split}
    &\left(\frac{1}{n_{x}^{2}}+\Delta\left(\frac{1}{n^{2}}\right)_{x}\right)x^{2}
     +\left(\frac{1}{n_{y}^{2}}+\Delta\left(\frac{1}{n^{2}}\right)_{y}\right)y^{2} \\ &\qquad +\left(\frac{1}{n_{z}^{2}}+\Delta\left(\frac{1}{n^{2}}\right)_{z}\right)z^{2} \\
    &\qquad +2\Delta\left(\frac{1}{n^{2}}\right)_{yz}yz
     +2\Delta\left(\frac{1}{n^{2}}\right)_{xz}xz
     +2\Delta\left(\frac{1}{n^{2}}\right)_{xy}xy
     =1,
  \end{split}
  \label{eq:2-20}
\end{equation}
where
\begin{equation}
\begin{pmatrix}
\Delta\left(\frac{1}{n^{2}}\right)_{x} \\
\Delta\left(\frac{1}{n^{2}}\right)_{y} \\
\Delta\left(\frac{1}{n^{2}}\right)_{z} \\
\Delta\left(\frac{1}{n^{2}}\right)_{yz} \\
\Delta\left(\frac{1}{n^{2}}\right)_{xz} \\
\Delta\left(\frac{1}{n^{2}}\right)_{xy}
\end{pmatrix}
=
\begin{pmatrix}
r_{11} & r_{12} & r_{13} \\
r_{21} & r_{22} & r_{23} \\
r_{31} & r_{32} & r_{33} \\
r_{41} & r_{42} & r_{43} \\
r_{51} & r_{52} & r_{53} \\
r_{61} & r_{62} & r_{63}
\end{pmatrix}
\begin{pmatrix}
E_{x} \\
E_{y} \\
E_{z}
\end{pmatrix},
\label{eq:2-21}
\end{equation}
the values of \textbf{r} matrix elements reflect the modulation capabilities of different components of the refractive index ellipsoid. For BTO material, the \textbf{r} matrix is \cite{karvounis2020barium,chen2006analysis,castera2015influence}
\begin{equation}
\begin{pmatrix}
0 & 0 & r_{13} \\
0 & 0 & r_{23} \\
0 & 0 & r_{33} \\
0 & r_{42} & 0 \\
r_{51} & 0 & 0 \\
0 & 0 & 0
\end{pmatrix},
\label{eq:2-22}
\end{equation}
where $r_{13} = r_{23}$ and $r_{42} = r_{51}$ owning to the symmetry between $x$- and $y$-axes. Because both second-order optical nonlinearity and the EO effect originate from the same $\chi^{(2)}$ tensor, the nonzero tensor elements of the \textbf{d}-matrix and the \textbf{r}-matrix exhibit a similar structure.

For bulk BTO, experimentally measured EO coefficients are summarized in Table \ref{tab:2-1}. Here, the unclamped coefficients correspond to sufficiently low electric field frequencies (DC to $\sim$1 MHz), while the clamped coefficients correspond to electric field frequency exceeding $\sim$100 MHz \cite{chelladurai2025barium}. As discussed earlier, the EO coefficients are larger under unclamped conditions due to the additional contribution from mechanical effects. Among these coefficients, the off-diagonal $r_{42}$ (or $r_{51}$) exhibits the largest magnitude, with an unclamped value of 1300 pm/V and clamped value of 730 pm/V \cite{zgonik1994dielectric}. These values are at least an order of magnitude higher than those of conventual EO materials, such as LN, whose largest EO coefficient is $r_{33}$ =31.8 pm/V \cite{zhu2021integrated,wemple1968relationship}. This exceptional EO response is a primary reason why BTO has attracted significant interest as an EO material.

In addition, the relationship between the EO coefficient $r$ and the RF permittivity $\epsilon_r$ can be described by an empirical formula known as Miller's Rule \cite{chelladurai2025barium}:
\begin{equation}
r \sim 2\delta\frac{\left(n_{0}^{2}-1\right)^{2}}{n_{0}^{4}}(\varepsilon_{r}-1),
\label{eq:2-23}
\end{equation}
where $n_0$ is the optical refractive index, and $\delta$ is the Miller’s coefficient. This relation highlights the general trend that materials with larger dielectric constants tend to exhibit larger EO coefficients, reflecting the strong polarization capability intrinsic to BTO.

\begin{figure}
    \centering
    \includegraphics[width=1\linewidth]{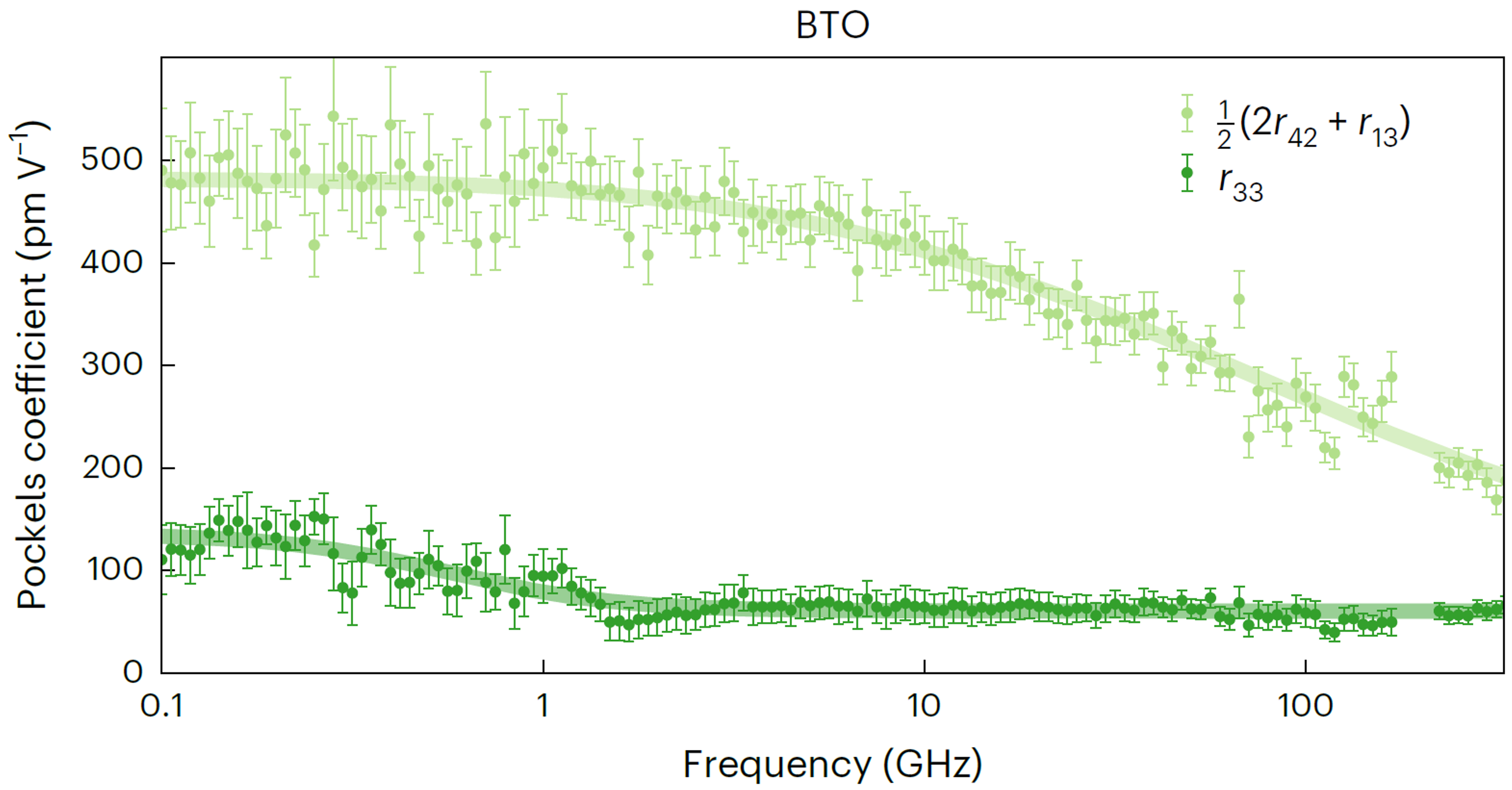}
    \caption{Pockels coefficient measurements for BTO, with $r_{33}$ in dark green, $r_{42}$ in light green and model fits to the data in solid lines. Note that typically $2r_{42} \gg r_{13}$ so that $(r_{13} + 2r_{42}) /2 \approx r_{42}$. Reproduced with permission from Chelladurai \textit{et al.}, Nature Materials 24, 868-875 (2025). Copyright 2025 authors, licensed under a Creative Common Attribution (CC BY) license\cite{chelladurai2025barium}.}
    \label{fig:chap2-6}
\end{figure}

For BTO thin films, the quality of film growth determines the value of the effective EO coefficient. In the case of high-quality single-crystal, single-domain BTO thin films, the EO coefficient can be comparable to that of bulk materials, with the highest reported unclamped result currently reaching 1268 pm/V \cite{lin2025giant}. As the frequency of the RF field increases, the mechanical effects and ionic polarization undergo relaxation, resulting in a reduction of the EO coefficient. Regarding the variation relation of the EO coefficient with the RF frequency, the few available reference results come from recent work published by Chelladurai \textit{et al.} \cite{chelladurai2025barium}, where RF frequency was varied from 100 MHz to 300 GHz, as shown in Fig. \ref{fig:chap2-6}. Due to the non-ideal quality of the $a$-axis-oriented poly-domain BTO thin films grown for this work, the measured EO coefficient is lower than that of bulk BTO. At an RF frequency of 100 MHz, the $r_{42}$ was $\sim$481 pm/V, and it dropped to $\sim$268 pm/V at 300 GHz. These results indicate that BTO thin films with reduced defect density and well-controlled domain orientation are essential for maximizing the EO coefficient across the RF frequency spectrum. Further experimental and theoretical investigations are therefore needed to fully understand the frequency dependence of the EO response in high-quality BTO thin films and to guide the design of future EO devices.

\subsection{Photorefractive effect}
\begin{figure}
    \centering
    \includegraphics[width=1\linewidth]{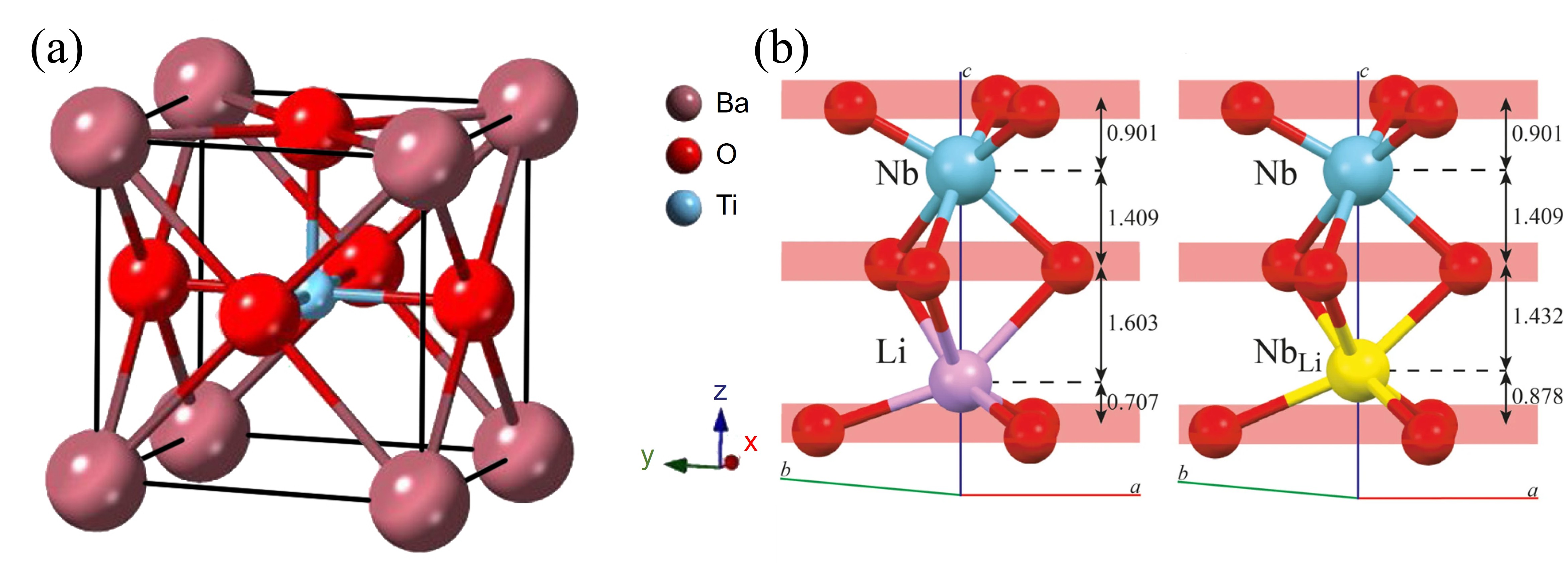}
    \caption{(a) The oxygen vacancies of BTO material. Reprinted from Applied Catalysis B: Environmental 279, 119340, Wang \textit{et al.}, “Impact of oxygen vacancy occupancy on piezo-catalytic activity of BaTiO$_3$ nanobelt,” 2020 with permission from Elsevier \cite{wang2020impact}. (b) The conventional configuration of the LN crystal structure (left) and the anti-site niobium defect structure (right). Reproduced with permission from Palantnikov \textit{et al.}, Materials 16, 732. Copyright 2023, Authors, licensed under a Creative Common Attribution (CC BY) license \cite{palatnikov2023growing}.}
    \label{fig:chap2-7}
\end{figure}

The photorefractive effect originates from defect-related energy levels within a material \cite{klein2005photorefractive,klein1986photorefractive}. When photons with frequency matching these defect energy levels are absorbed, spatially separated charge carriers are excited and subsequently generate a space-charge electric field. This internal electric field, through EO effect, modulates the refractive index of the material, leading to the observed photorefractive phenomenon. 
For intrinsic BTO materials, the dominant intrinsic defects are oxygen vacancies, as shown in Fig. \ref{fig:chap2-7}(a). These defects are reported to introduce relatively shallow donor energy levels, which are likely to be thermally ionized at room temperature and therefore do not effectively participate in the photorefractive process\cite{klein1986photorefractive}. As a result, intrinsic BTO exhibits a weak photorefractive response within its optical transparency window (photon energy $<$ 3.2 eV \cite{karvounis2020barium}). This is markedly different from the pronounced photorefractive effect observed in materials such as LN under visible light. In LN, a dominant intrinsic defect known as the anti-site niobium defect, where Nb$^{5+}$ ions occupy Li$^+$ ion sites, acts as an efficient photorefractive center \cite{palatnikov2023growing,qin2025electronic}, as illustrated in Fig. \ref{fig:chap2-7}(b). Consequently, LN exhibits pronounced photorefractive effects in the visible wavelength range \cite{Zheng2022LNphotorefractive}. For applications where suppression of photorefractive effects is desirable, intrinsic BTO therefore represents a promising material platform.

On the other hand, when intrinsic BTO is intentionally doped with metal ions that act as photorefractive centers, the photorefractive effect can be significantly enhanced owing to BTO's giant EO coefficient. This phenomenon was initially reported in early-stage BTO crystals containing Fe$^{3+}$ and Fe$^{2+}$ ion impurities \cite{klein1986photorefractive}. Subsequent investigations have explored doping with elements such as Co \cite{rytz1990photorefractive}, Ce \cite{yang1995photorefractive}, Ru \cite{lin2003increasing}, and Nb \cite{chang1999optical} to induce the photorefractive effect, enabling applications including photorefractive information storage and holographic display by converting optical information into crystal refractive index information. With continued advances in BTO thin-film growth technologies, precise control over dopant species and concentrations is becoming increasingly feasible, opening opportunities to leverage photorefractive properties for diverse device applications.

\subsection{Thermo-optic properties}
The thermo-optic (TO) properties of materials characterize the variation of optical refractive index with temperature. The thermal optical coefficient (TOC) is defined as \cite{lin2025thermo}
\begin{equation}
\mathrm{TOC} = \frac{\mathrm{d}n}{\mathrm{d}T},
\label{eq:2-24}
\end{equation}
where the temperature dependence of the refractive index arises from two competing contributions. A negative contribution originates from thermal expansion, whereby an increase in temperature reduces the material density and thus decreases the refractive index. In contrast, the positive contribution stems from the temperature-dependent excitonic band gap. Reported TOC values for BTO indicate that bulk crystals exhibit $\text{d}n_e/\text{d}T = 1.8 \times 10^{-4}$ $^\circ$C$^{-1}$ at a wavelength of 633 nm near $T = 20$$^\circ$C \cite{buse1993refractive}. For thin film BTO, a smaller value of $\text{d}n_o/\text{d}T = 1.095 \times 10^{-5}$ $^\circ$C$^{-1}$ has been reported at a wavelength of 1550 nm near $T = 20$$^\circ$C \cite{lin2025thermo}. In both cases, BTO exhibits a positive TOC, indicating that the excitonic bandgap contribution dominates over thermal expansion. TO devices have been demonstrated on various materials\cite{lin2023integrated,chen2021highly,parra2024silicon}, and Table \ref{tab:2-1} summarizes the corresponding TOC values. The relatively favorable TO properties of BTO further support its potential for high-performance thermo-optic device applications, either as a primary tuning mechanism or as a complementary effect in EO-based photonic systems.

\section{Epitaxial Growth of Thin Film BTO}

\begin{table*}[t!]
\caption{\label{tab:BTO_growth_summary}
Summary of representative growth methods for BTO thin films and the key material properties reported in the selected literature. For studies evaluating several substrates, only the one exhibiting the best performance is summarized. Thickness values represent the maximum film thickness reported.}
\begin{ruledtabular}
\begin{tabular}{c c c c c c c}

\textbf{Methods} &
\textbf{Year} &
\makecell{\textbf{Substrate and}\\ \textbf{buffer layer}} &
\textbf{Orientation} &
\makecell{\textbf{Thickness}\\ \textbf{(nm)}} &
\makecell{\textbf{XRD rocking}\\ \textbf{curve FWHM ($^\circ$)}} &
\makecell{\textbf{RMS}\\ \textbf{roughness (nm)}}
\\ \hline

PLD \cite{nashimoto1992epitaxial} & 1992 & MgO on GaAs & c & 44 & 1.8 & -- \\
PLD \cite{kim1995pulsed} & 1995 & MgO & a & 200 & 0.6 & -- \\
PLD \cite{beckers1998structural} & 1998 & MgO & c & 950 & 0.36 & 1.1 \\
PLD \cite{lisoni2001growth} & 2001 & MgO on Sapphire & a and c & 440 & 0.7 & 1 \\
PLD \cite{petraru2002ferroelectric} & 2002 & MgO & a and c & 1000 & -- & 1.1 \\
PLD \cite{dicken2008electrooptic} & 2008 & SRO on MgO & c & 400 & -- & -- \\
PLD \cite{niu2011epitaxy} & 2011 & STO on Si & c & 90 & 0.7 & 0.32 \\
PLD \cite{cao2021barium} & 2021 & DSO & c & 850 & 0.075 & 0.754 \\
PLD \cite{kim2022controlling} & 2022 & MgO & c & 500 & -- & 0.23 \\
PLD \cite{cao2022characterization} & 2023 & SRO on DSO & c & 860 & 0.41 & 0.279 \\
PLD \cite{winiger2024pld} & 2024 & MgO & -- & 96 & -- & -- \\
PLD \cite{wen2024enhanced} & 2024 & STO & a & 270 & 0.046 & 1.2 \\
Laser MBE \cite{cui1997crystallographic} & 1997 & STO & c & 44 & 0.39 & 0.29 \\
Laser MBE \cite{zhao2000thickness} & 2000 & STO & a and c & 400 & 0.4 & 0.14 \\
Laser MBE \cite{wei2005effect} & 2005 & MgO on Si & c & 30 & -- & 0.34 \\
MBE \cite{niu2007epitaxial} & 2007 & STO on Si & Mixed a/c & 30 & 0.9 & 2 \\
MBE \cite{huang2009electrical} & 2009 & STO on GaAs & c & 150 & -- & -- \\
MBE \cite{mazet2014structural} & 2014 & STO on Si & c & 20 & -- & 0.35 \\
MBE \cite{xiong2014active} & 2014 & STO on SOI & a & 80 & 0.5 & 0.4 \\
MBE \cite{merckling2019epitaxial} & 2019 & STO on Si & c & 37 & 0.59 & 0.42 \\
MBE \cite{abel2019large} & 2019 & STO on Si & a & 225 & 0.3 & 0.4 \\
MBE \cite{reynaud2022microstructural} & 2022 & STO on Si & Mixed a/c & 110 & 0.28 & -- \\
MBE \cite{abbasi2022ferroelectric} & 2022 & Nb-doped STO & c & 15 & -- & 0.428 \\
MBE \cite{reynaud2023si} & 2023 & MgO/STO on Si & a & 100 & -- & 1.06 \\
MBE \cite{manjeshwar2025ferroelectric} & 2025 & SRO on Nb-doped STO & c & 40 & 0.71 & -- \\
MBE \cite{haque2025heterogeneous} & 2025 & SAO on STO & c & 30 & 0.61 & 0.27 \\
RF sputtering \cite{kim1995structural} & 1995 & MgO & c & 200 & 0.9 & -- \\
RF sputtering \cite{kim2014ridge} & 2014 & MgO & c & 1000 & 0.86 & 0.81 \\
RF sputtering \cite{posadas2021thick} & 2021 & STO on SOI & a & 1000 & 0.4 & -- \\
RF sputtering \cite{posadas2023rf} & 2023 & STO on SOI & c & 105 & 0.6 & 0.4 \\
RF sputtering \cite{raju2025high} & 2025 & STO on SOI & a & 300 & 0.52 & 1 \\
MOCVD \cite{gill1996thin} & 1996 & MgO & a & 300 & -- & 7 \\
MOCVD \cite{tang2004low} & 2004 & MgO & c & 570 & -- & 7 \\
MOCVD \cite{meier2006integration} & 2006 & MgO/STO on Si & c & 100 & 1.3 & 12 \\
HV-CVD \cite{reinke2017low} & 2017 & STO on Si & a & 200 & 1.24 & 0.2 \\
HV-CVD \cite{szmyt2025high} & 2025 & STO on Si & a & 80 & 0.77 & 0.16 \\
CSD \cite{edmondson2020epitaxial} & 2020 & STO on Si & a & 85 & 1.23 & 1.4 \\
CSD \cite{zhang2024hybrid} & 2024 & SOI & -- & 150 & -- & 5 \\
ALD \cite{ngo2014epitaxial} & 2014 & STO on Si & c & 20 & 0.74 & -- \\
ALD \cite{lin2019atomic} & 2019 & STO on Si & c & 66 & 0.8 & -- \\

\end{tabular}
\end{ruledtabular}
\end{table*}

The preparation of high-quality BTO thin films on suitable substrates is one of the most critical steps in realizing BTO-based devices. The exceptional properties of BTO that are exhibited in bulk material are the major reason for its significant attention. Thus, it is crucial to deposit high-quality BTO thin films that retain bulk-like properties. To obtain high-quality BTO films, both the choice of substrate and the deposition method are essential. The choice of substrates with less lattice mismatch compared to BTO is important to reduce strain within the film, which can otherwise degrade crystallinity and impair the properties of the BTO thin film.

In this section, we categorize the substrates used for BTO thin film growth into two main groups: perovskite-based substrates and Si platforms. In the early development of BTO thin films, perovskite materials with lattice constants close to that of BTO and lower refractive indices were commonly used to enable epitaxial growth of BTO-on-insulator (BTOI). Materials such as MgO\cite{norton1991epitaxy,petraru2002ferroelectric,tang2004low}, CeO$_2$\cite{lin1995growth}, DyScO$_3$ (DSO)\cite{cao2021barium,cao2022characterization,Lin2024single}, and SrTiO$_3$ (STO)\cite{cavanagh2025effect,cui1997crystallographic,zhao2000thickness,wang2020polarization} have been successfully employed to achieve epitaxial thin-film BTO growth. Besides the perovskite substrates, the integration of BTO with Si has also attracted significant interest in recent years, as it would hypothetically enable efficient and low-loss EO modulation on a Si platform\cite{eltes2019batio3,psiquantum_team2025manufacturable}. However, the large lattice mismatch between Si and BTO is a major challenge to achieve high-quality BTO thin film. To address this, buffer layers such as MgO\cite{buchal1998epitaxial,kim1997preparation} and STO\cite{merckling2019epitaxial,abel2019large} are commonly introduced to facilitate epitaxial BTO growth on Si. With the assistance of these buffer layers, high-performance EO devices have been reported and demonstrated on the Si platform. It is worth noting that various waveguide structures have been designed to realize BTO-based integrated photonics on both BTOI and Si platforms. The details of BTO-based waveguide structures in both platforms will be further described in Section \ref{section4}.

Table \ref{tab:BTO_growth_summary} summarizes the reported epitaxial growth of BTO thin films using various deposition techniques, including pulsed laser deposition (PLD), molecular-beam epitaxy (MBE), radio frequency (RF) sputtering, chemical vapor deposition (CVD), atomic layer deposition (ALD) and chemical solution deposition (CSD). The selected literature is arranged chronologically from the early development of thin film BTO to recent works with their corresponding characteristics. X-ray diffraction (XRD) is a common technique employed to evaluate the crystallinity of thin film materials\cite{khan2020experimental,zhu1987x-ray}. From XRD measurements, both in-plane and out-of-plane lattice constants can be extracted, while the full width at half maximum (FWHM) of the rocking curve indicates the degree of crystal alignment. Smaller FWHM values correspond to better crystal quality. The orientation of the BTO thin film is crucial because its EO and ferroelectric properties are anisotropic; exploitation of the EO effect necessitates careful consideration of the direction of the light's polarization and the applied electric field orientation relative to the crystal axes\cite{vasudevan2023domain}. The reported film thicknesses of various grown crystal orientations in the literature are also included. Due to the different growth mechanisms of each deposition method, the achievable growth rates can vary. Therefore, some methods are more suitable for applications that require thicker BTO films, while others are better suited for thinner films with higher crystalline quality. Finally, the surface roughness typically measured by atomic force microscope (AFM) reflects the smoothness of the film surface\cite{sundararajan2002development}. A smoother surface reduces optical scattering and propagation loss in waveguides and is also important for subsequent deposition or wafer-bonding processes.  It should nonetheless be mentioned that stoichiometry is also an important consideration, as Cavanagh \textit{et al}. found that excess titanium can form pervasive defects in BTO films despite the appearance of low surface roughness and high-quality diffraction that would otherwise indicate high quality\cite{cavanagh2025effect}.

\subsection{Pulsed laser deposition (PLD)}
Pulsed laser deposition (PLD) is one of the earliest techniques used to achieve the growth of thin-film BTO and remains a promising method for epitaxial deposition on various substrates, as demonstrated in recent years. In PLD, a pulsed laser ablates material from a BTO target, and the ablated species are deposited onto a heated substrate in an oxygen atmosphere. Compared to other deposition methods, PLD offers a relatively high growth rate, enabling the fabrication of thicker films within shorter deposition times. However, due to the limited size of the plasma plume generated during laser ablation, the deposition area is relatively small, making it challenging to scale up to large-area wafers.

\begin{figure}
    \centering
    \includegraphics[width=1\linewidth]{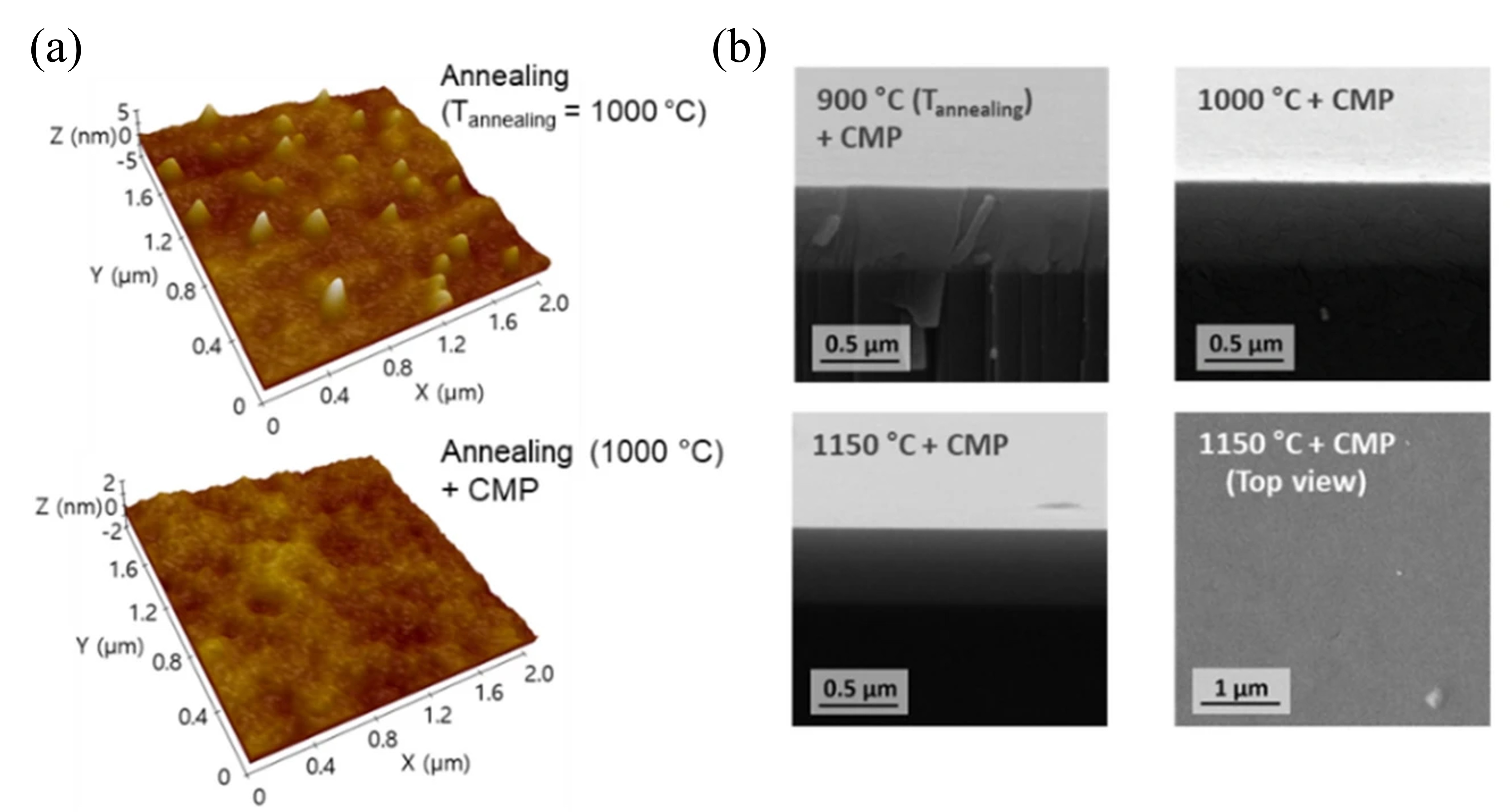}
    \caption{(a) AFM images of BTO thin films annealed at 1000 \degree C before and after polishing. (b) Cross-sectional SEM images of BTO films annealed at 900 \degree C, 1000 \degree C, and 1150 \degree C, and the top view of the 1150 \degree C-annealed film after CMP. Reproduced with permission from Kim \textit{et al.}, Scientific Reports 12, 5363 (2022) Copyright 2024 authors, licensed under a Creative Common Attribution (CC BY) \cite{kim2022controlling}.}
    \label{fig:3kim2022}
\end{figure}

One of the earliest demonstrations of PLD-grown BTO thin film was reported by Norton \textit{et al.}\cite{norton1991epitaxy} in 1991, where BTO was deposited onto (001)-oriented MgO substrates at 670 \degree C in a 400 mTorr oxygen ambient. This high deposition temperature was necessary to achieve epitaxial growth on MgO, and the resulting BTO films exhibited a well-aligned $c$-axis crystal orientation (the $c$ axis being perpendicular to the film's surface). The film quality was confirmed using ion-channeling and TEM, showing aligned grains despite the ~5\% lattice mismatch between BTO and MgO. Later, Kim \textit{et al.}\cite{kim1995pulsed} demonstrated the growth of $a$-axis oriented BTO thin films on MgO(001) under different PLD conditions. In their work, BTO was deposited at 720 \degree C in a 140 mTorr of oxygen ambient, and XRD measurements showed a narrow rocking curve FWHM of 0.6 \degree for the (200) peak, indicating a highly oriented $a$-axis film. They also investigated the effect of laser repetition rate on film morphology and found that lower repetition rates produced smoother surfaces, with the best surface smoothness obtained at 2 Hz. More recently, Kim \textit{et al.}\cite{kim2022controlling} reported ultra-smooth BTO films growth on MgO using PLD. They observed that recrystallization during post-annealing roughened the film surface, and to mitigate this effect, a CMP process was employed to restore an ultra-smooth surface, as shown in Fig. \ref{fig:3kim2022}(a) and \ref{fig:3kim2022}(b). Additionally, XRD and reciprocal space mapping (RSM) measurements revealed a transformation from $a$-axis–oriented domains to $c$-axis–oriented domains, attributed to the relaxation of out-of-plane lattice strain during post-annealing.

Besides being used directly as a substrate, MgO has also been employed as an intermediate buffer layer to assist growth on other substrates that have a large lattice constant mismatch with BTO. A notable example is the work of Nashimoto \textit{et al.}\cite{nashimoto1992epitaxial}, who demonstrated for the first time epitaxial BTO growth on GaAs(001) using a thin MgO(001) buffer layer deposited by PLD. They found that the best quality MgO was obtained at 350 \degree C in $5 \times 10^{-6}$ Torr O$_2$, with a (002) rocking curve FWHM of 1.4\degree. Although in this work, the buffer layer was typically about 440 \si{\angstrom} thick, they found that only $\sim$40 \si{\angstrom} of MgO is sufficient for epitaxial growth of BTO. The 600 \si{\angstrom} thick BTO film was deposited at 780 \degree C in 1.2 mTorr O$_2$, resulting in a $c$-axis orientated film with a (002) rocking curve FWHM of 1.84\degree. This work established the feasibility of integrating BTO onto III-V substrates through an MgO buffer layer. Another example of using MgO as a buffer layer to enable epitaxial BTO growth on lattice-mismatched substrates was reported by Lisoni \textit{et al.}\cite{lisoni2001growth}, who demonstrated epitaxial BTO film growth on sapphire, a silicon-compatible optical substrate, through an MgO buffer layer. In their work, MgO was first deposited on $r$-cut sapphire using PLD at 750 \degree C in $2 \times 10^{-3}$ mbar O$_2$, resulting in high quality film that has a (002) rocking curve FWHM less than 0.5\degree and an rms roughness of 1.7 nm. Then, 400–450 nm thick BTO films were grown on this MgO buffer at 1000 \degree C and 1050 \degree C. At 1000 \degree C, the BTO films could be either $a$- or $c$-oriented, whereas at 1050 \degree C the films were consistently $a$-axis oriented due to the thermal expansion mismatch between BTO and the sapphire substrate. In both works, the MgO layer serves to reduce the lattice mismatch and establish a template for epitaxial growth of BTO thin films.

\begin{figure}
    \centering
    \includegraphics[width=1\linewidth]{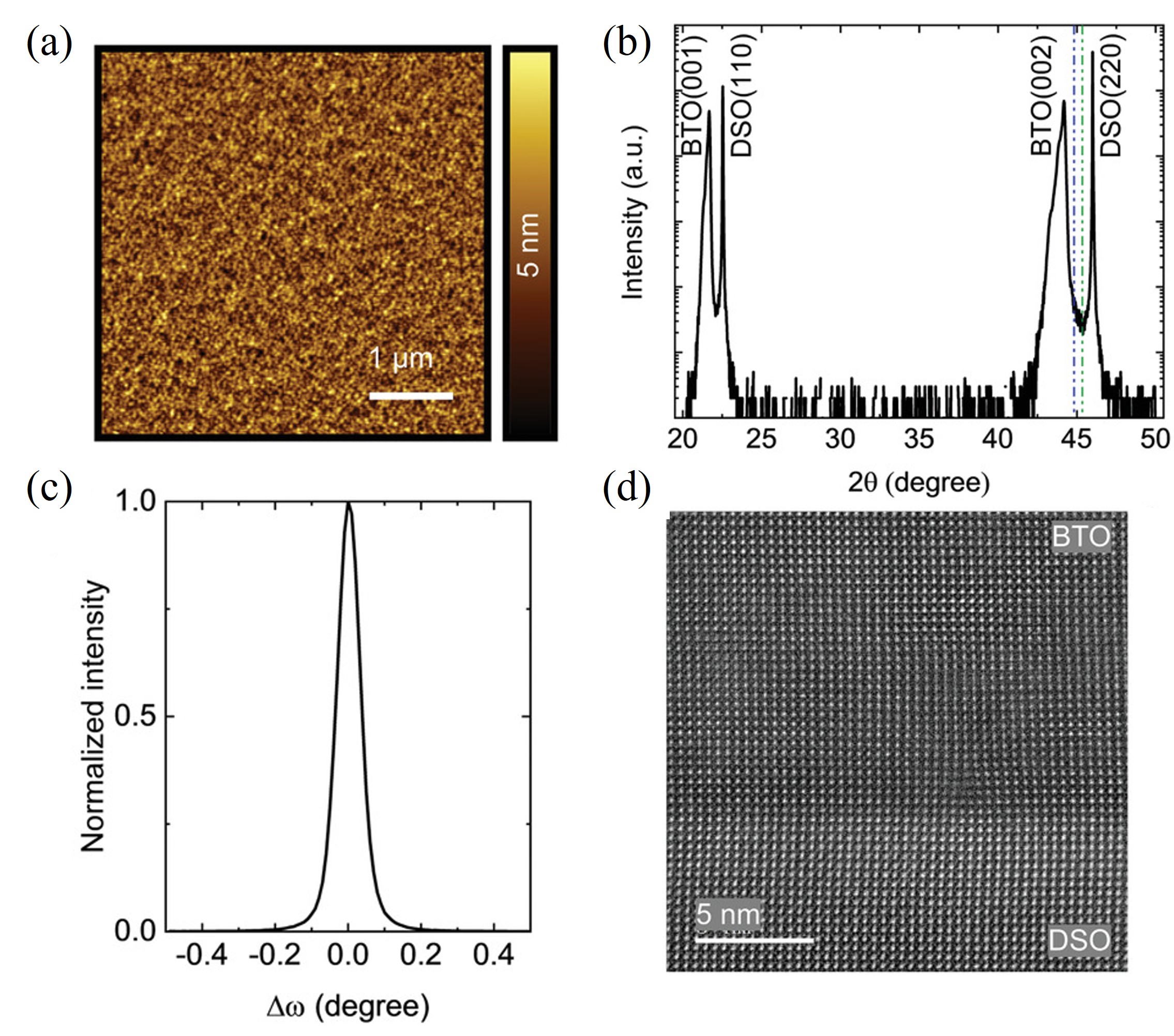}
    \caption{(a) AFM morphology image of BTO thin film growth on DSO substrate using PLD. (b) XRD scan of the peaks of BTO thin fim and DSO substrate. (c) The rocking curve of the BTO (002) peak indicating a small FWHM of 0.075\degree. (d) High-resolution STEM cross-sectional image of the BTO/DSO interface, showing the high crystalline quality. Reproduced with permission from Cao \textit{et al.}, Advanced Materials 33, 2101128 \copyright 2021 Wiley-VCH GmbH\cite{cao2021barium}.}
    \label{fig:3cao2021}
\end{figure}

Beyond MgO-based platforms, significant progress has also been made in achieving extremely high-quality PLD-grown BTO films on lattice-matched perovskite substrates. Notably, the lowest reported XRD rocking-curve FWHM for BTO thin films (below 0.1\degree) was achieved using PLD, highlighting the exceptional crystal quality attainable with this technique. Cao \textit{et al.}\cite{cao2021barium} demonstrated BTO growth on several scandate-based substrates with small lattice mismatch, among the tested materials, DSO provided the best film quality due to its minimal mismatch compared to BTO ($\sim$1\%). In their work, the BTO was deposited at 650 \degree C and under 10 mTorr oxygen pressure and subsequently post-annealed in an oxygen environment at 520 \degree C for 30 min. This process produced $c$-axis oriented high quality BTO film on DSO with a narrow rocking curve FWHM of 0.075\degree at the (002) peak and a rms surface roughness of 0.75 nm, as shown in Fig. \ref{fig:3cao2021}. Additionally, the thermal stability of the BTO thin film clamped by the DSO showed no phase transition observe up to 700 \degree C, indicating a significant increased Curie temperature compared to bulk BTO.

\begin{figure}
    \centering
    \includegraphics[width=1\linewidth]{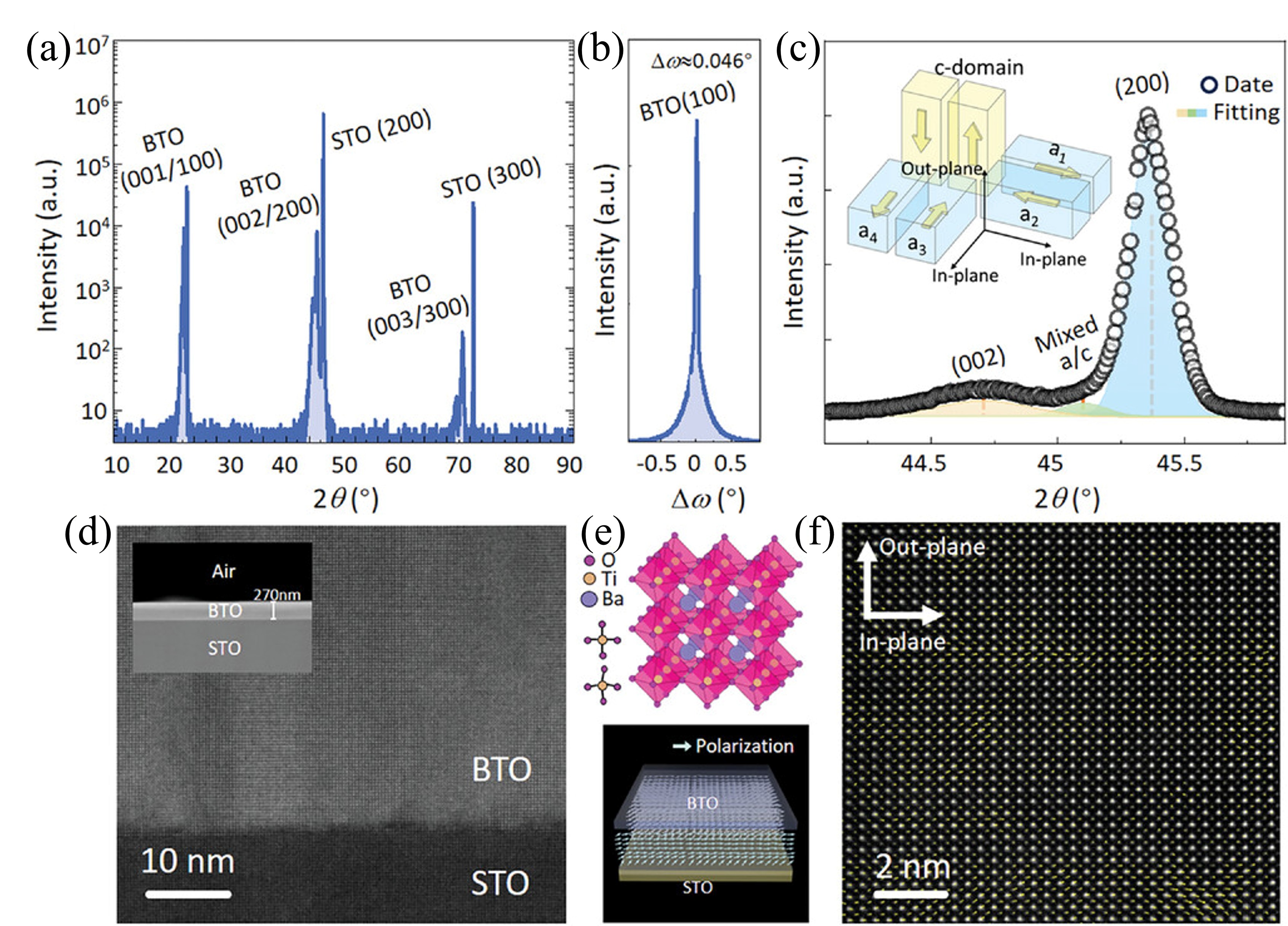}
    \caption{(a) XRD scan of BTO thin film growth on STO substrate using PLD. (b) Rocking curve of BTO (100) peak with a narrow FWHM of 0.046\degree. (c) XRD spectrum of BTO (002)/(200) peaks, indicating that the BTO thin film is 96\% $a$-axis-orientation. (d) Cross-sectional image of the BTO/STO interface. (e) Schematic of the polarization of BTO thin film. (f) A high-resolution cross-sectional image showing high crystalline quality. Reproduced with permission from Wen \textit{et al.}, Advanced Optical Materials 12, 2303058 (2024)\cite{wen2024enhanced}. \copyright 2024 Wiley-VCH GmbH.}
    \label{fig:3wen2024enhanced}
\end{figure}

Similar high-quality epitaxial films were recently reported by Wen \textit{et al.}\cite{wen2024enhanced}, who used PLD to grow epitaxial BTO on STO, as shown in Fig. \ref{fig:3wen2024enhanced}. In this study, they tuned the oxygen pressures during growth at 750 \degree C to engineered domain alignment. With an optimized oxygen pressure of 0.2 mbar during deposition and recrystallization, they achieved about 96\% $a$-axis oriented BTO films on STO(100) substrate. The resulting BTO films exhibited outstanding crystal quality, with an exceptionally narrow rocking-curve FWHM of 0.046\degree at the (100) peak and a rms surface roughness of about 1.2 nm. These works demonstrate that PLD-grown BTO films can achieve excellent crystalline and surface quality required for optical and EO applications.

\begin{figure}
    \centering
    \includegraphics[width=1\linewidth]{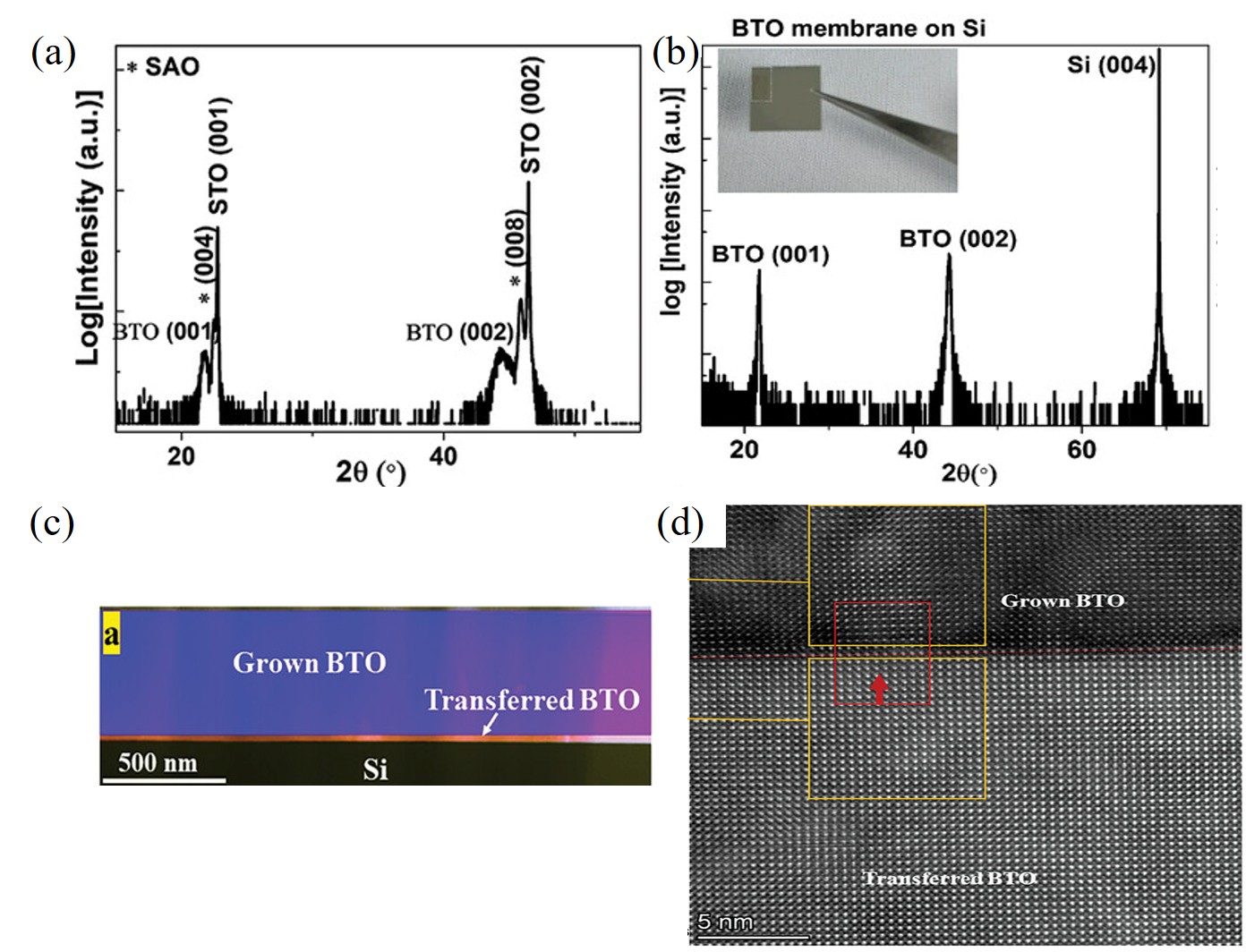}
    \caption{(a) XRD scan of BTO thin film growth on SAO and STO substrate using PLD. (b) XRD scan of the transferred membrane on Si. (c) HAADF-STEM image of the grown BTO thin film on transferred BTO and (d) zoomed in for atomic structure. Reproduced with permission from Haque \textit{et al.}, Advanced Functional Materials 35, 2413515 (2025)\cite{haque2025heterogeneous}. \copyright 2025 Wiley-VCH GmbH.}
    \label{fig:3haque2025heterogeneous}
\end{figure}

Furthermore, Haque \textit{et al.}\cite{haque2025heterogeneous} presented an approach for integrating high-quality BTO films onto Si platform using a water-soluble sacrificial layer with the help of PLD growth. They first grew a 30 nm-thick epitaxial BTO film on a water-soluble Sr$_3$Al$_2$O$_6$ (SAO) sacrificial layer deposited on an STO substrate, achieving a rocking-curve FWHM of 0.61\degree, as illustrated in Fig. \ref{fig:3haque2025heterogeneous}(a). The BTO membrane was subsequently transferred onto a Si substrate by dissolving the sacrificial layer as shown in Fig. \ref{fig:3haque2025heterogeneous}(b). The transferred BTO films retained high crystalline quality and can served as a template for the subsequent growth of thicker PLD-deposited BTO films. HAADF-STEM analysis revealed a sharp BTO/Si and grown BTO/transferred BTO interface with no detectable elemental interdiffusion, as shown in Fig. \ref{fig:3haque2025heterogeneous}(c) and \ref{fig:3haque2025heterogeneous}(d), respectively. This transfer-based approach broadens the range of substrates that can accommodate epitaxial BTO films, including technologically important yet lattice-mismatched materials such as silicon and sapphire. Consequently, it establishes a promising route for combining complex oxide heterostructures with advanced semiconductor technologies.

\subsection{Molecular beam epitaxy (MBE)}
Molecular beam epitaxy (MBE) is another major technique for the epitaxial growth of BTO thin films, especially for the recently popular hybrid integrated BTO/Si platform. MBE offers nearly atomic-layer control over composition, flux, and interface engineering. In this technique, elemental sources such as Ba, Ti, or Sr are evaporated under ultrahigh-vacuum conditions and react with either molecular or plasma-activated oxygen to form complex oxides with precise stoichiometry control\cite{mazet2015review}. In addition, in-situ monitoring methods such as reflection high-energy electron diffraction (RHEED) enable real-time assessment of crystallinity and layer-by-layer growth, which is particularly valuable for multicomponent perovskite systems. One disadvantage of MBE is its relatively slow deposition rate, which makes it challenging to grow thick films for certain applications, and the process control is more complex compared to other deposition techniques. Nevertheless, MBE has enabled several exceptional demonstrations of high-quality BTO thin films, particularly in recent works focused on hybrid integration with silicon.

In the early stages of developing BTO thin film growth using MBE, most studies focused on deposition onto bulk substrates that exhibit small lattice mismatches with BTO and have lower refractive indices. One of the earliest demonstrations of MBE-grown BTO was reported by Cui \textit{et al.}\cite{cui1997crystallographic}, who used laser MBE to deposit BTO thin films on STO(001) substrates. In their work, BTO was grown at 750 \degree C under a vacuum of ~10$^{-3}$ Torr, resulting in an epitaxial film approximately 44 nm thick. XRD measurements showed that the film was $c$-axis-oriented, and the (002) rocking curve exhibited a narrow FWHM of 0.39\degree, indicating good crystalline quality. Also, AFM measurement confirmed a smooth surface with a rms roughness of ~0.5 nm. Later, Zhao \textit{et al.}\cite{zhao2000thickness} further investigated the growth BTO thin film on STO(001) using laser MBE, focusing on how film thickness and oxygen pressure influence the film orientation. They first varied the BTO thickness from 10 to 400 nm under an oxygen pressure of 2 $\times$ 10$^{-4}$ Pa and used XRD to examine the tetragonality of the films. Their results showed that the c/a lattice-constant ratio decreased with increasing thickness and the films remained $c$-axis-oriented. Under this oxygen pressure, rms surface roughness values as low as 0.14 nm and a rocking-curve FWHM of 0.4\degree were measured for 90 nm and 40 nm thick films, respectively. They then explored the effect of oxygen pressure in the range of 2 $\times$ 10$^{-4}$ to 12 Pa using 400 nm thick films and observed a clear transition in domain orientation from $c$-axis to $a$-axis as the pressure increased. Together, these works demonstrated that laser-MBE can produce high-quality and well-oriented epitaxial BTO films, establishing an important foundation for subsequent MBE-based heterostructure growth and device applications.

Following the early demonstrations of MBE-grown BTO on perovskite substrates, several studies explored the integration of BTO thin films onto the Si photonic platform. Due to the large lattice mismatch between Si and BTO, using a high-quality buffer layer is essential to mitigate the mismatch and enable epitaxial BTO growth. Wei \textit{et al.}\cite{wei2005effect} demonstrated the deposition of BTO on Si(001) using a thin MgO buffer layer grown by laser MBE. By varying the laser fluence from 4 to 8 J/cm$^2$, they found that higher fluence produced biaxially textured MgO, while lower fluence resulted in poorer crystallinity. As a result, 30 nm thick BTO film grown on the biaxially textured MgO buffer achieved fully $c$-axis-orientation and exhibited a rms surface roughness of 0.34 nm. Although MgO can serve as a buffer for BTO/Si integration, the most widely used buffer material for MBE is STO, which provides better lattice matching with BTO and can be grown with high crystalline quality using MBE. Mazet \textit{et al.}\cite{mazet2014structural} demonstrated the growth of BTO thin films on Si using a 4 nm STO buffer layer and investigated the MBE growth conditions, specifically the oxygen pressure and growth temperature, required to obtain high-quality epitaxial BTO on Si. They found that under lower oxygen pressure of 1 $\times$ 10$^{-7}$ Torr, BTO film with a thickness of 16-20 nm exhibited a rms roughness of 0.35 nm and $c$-axis-orientation. Also, the optimal growth temperature window was found to be 440–525 \degree C. Temperatures below this range led to insufficient crystallization, while higher temperatures resulted in mixed $a$- and $c$-oriented domains. By employing such relatively low growth temperatures together with rapid cooling, the authors could suppress the interfacial oxidation and limit the SiO$_2$ layer between the STO buffer and the Si substrate to ~1 nm. Ferroelectric switching in the BTO films was also confirmed using PFM, demonstrating the functional quality of the epitaxial layers.

\begin{figure}
    \centering
    \includegraphics[width=1\linewidth]{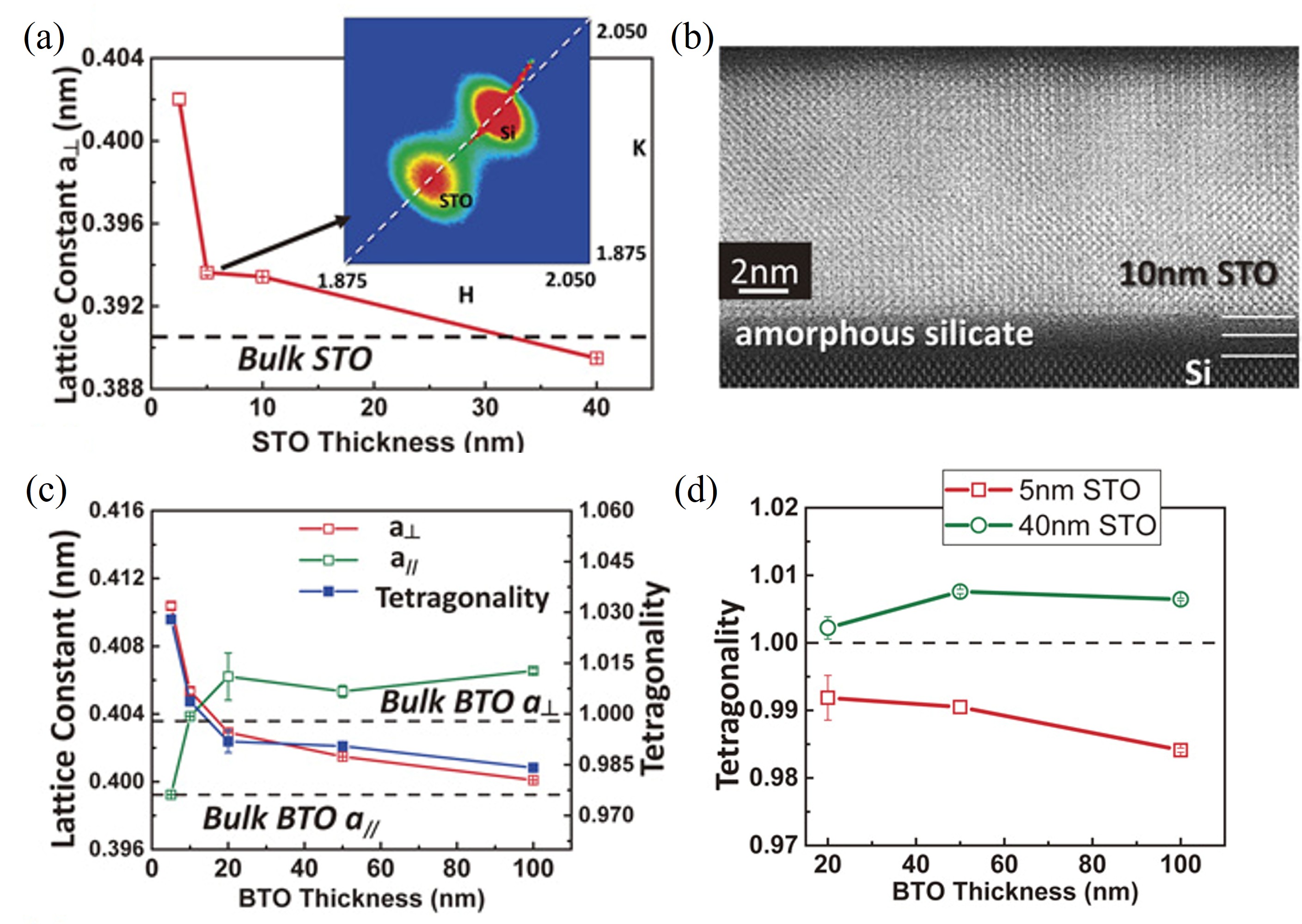}
    \caption{(a) Change in the BTO out-of-plane lattice constant (a$_\bot$) as a function of STO buffer layer thickness on Si substrate. (b) Cross-sectional image of an amorphous sillicate interfacial layer between the Si substrate and the STO buffer, due to the instable thermodynamic system. (c) Out-of-plan (a$_\bot$) and in-plan (a$_{/\!\!/}$) lattice constants of BTO films with different thicknesses grown on a 5 nm thick STO buffer layer. (d) Tetragonality of BTO films grown on 5 and 40 nm thick STO buffer layers as a function of BTO thickness. Reproduced with permission from Hsu \textit{et al.}, "Controlled orientation of molecular-beam-epitaxial BaTiO$_3$ on Si(001) using thickness engineering of BaTiO$_3$ and SrTiO$_3$ buffer layers," Applied Physics Express 10, 065501 (2017), published 16 May 2017; DOI 10.7567/APEX.10.065501 \copyright The Japan Society of Applied Physics. Reproduced by permission of IOP Publishing Ltd. All rights reserved. \cite{hsu2017controlled}.}
    \label{fig:3hsu2017}
\end{figure}

More recently, Hsu \textit{et al.}\cite{hsu2017controlled} investigated how BTO orientation on Si(001) can be controlled through thickness engineering of both the STO buffer layer and the BTO film. In their work, STO was first deposited on Si and recrystallized to form high-quality buffer layers with thicknesses of either 5 or 40 nm, followed by BTO growth at 630 \degree C. 
They found that BTO orientation is strongly governed by strain effects originating from lattice mismatch and differential thermal expansion between Si, STO, and BTO. 
First, STO buffer layers with different thicknesses were prepared for BTO growth. As a result, the authors observed that for thin STO layers, compressive strain from the Si substrate leads to a large out-of-plane lattice constant in the BTO thin film, whereas thicker STO buffer layers enable partial strain relaxation, resulting in a reduced out-of-plane lattice constant, as shown in Fig. \ref{fig:3hsu2017}(a). In addition, when BTO was grown on a 5 nm STO layer, increasing the BTO thickness induced a transition from $c$-axis- to $a$-axis-orientation, as illustrated in Fig. \ref{fig:3hsu2017}(c). Finally, Fig. \ref{fig:3hsu2017}(d) summarized the orientation of BTO films with various thickness grown on 5 nm and 40 nm STO layer. The results show that BTO grown on a 5 nm STO buffer is $a$-axis oriented, while BTO films grown on 40 nm thick STO remain $c$-axis orientation. These results indicate the importance of STO buffer layer thickness and strain engineering in controlling BTO orientation on the Si platform.

\begin{figure}
    \centering
    \includegraphics[width=0.8\linewidth]{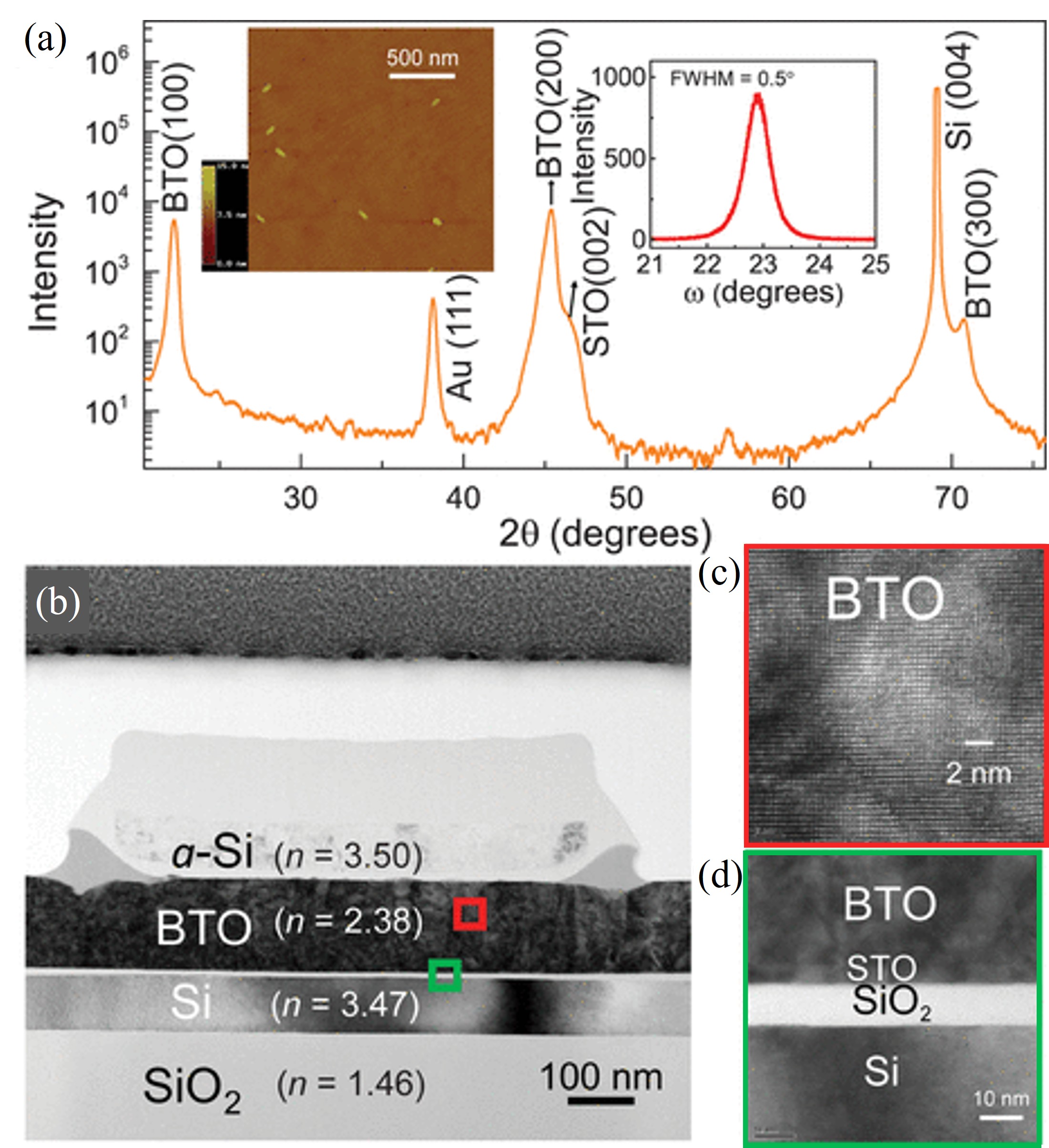}
    \caption{(a) XRD scan of the BTO thin film growth on a STO-buffered SOI platform. The insets show the AFM top-surface morphology of the BTO film and the rocking curve of the BTO (200) reflection. (b) Cross-sectional image of the hybrid Si Strip-loaded BTO waveguide on the SOI platform. High-resolution images of (c) the epitaxial BTO thin film and (d) the BTO/STO/Si interface, revealing an amorphous SiO$_2$ interlayer formed due to silicon oxidation\cite{xiong2014active}. Reprinted with permission from Xiong \textit{et al.}, Nano Letters 14, 1419–1425, Copyright 2014 American Chemical Society.}
    \label{fig:3xiong2014}
\end{figure}

Xiong \textit{et al.}\cite{xiong2014active} reported one of the earliest demonstrations of integrating epitaxial BTO thin films with the silicon photonic platform. In their work, high-quality BTO was grown on silicon-on-insulator (SOI) substrates using reactive MBE Fig. \ref{fig:3xiong2014}. An 8 nm STO buffer layer was first deposited on Si(100) to enable perovskite growth, followed by an 80 nm BTO layer Fig. \ref{fig:3xiong2014}(d). The resulting BTO thin film exhibited an atomically smooth surface with a rms roughness of ~0.4 nm and a rocking curve FWHM of 0.5\degree  for the BTO (200) peak, as illustrated in Fig. \ref{fig:3xiong2014}(a). Based on this platform, the authors demonstrated EO modulators and extracted an effective electro-optic coefficient ($r_{\text{eff}}$) of 213 pm/V from device characterization. While the device details are discussed in Section \ref{section4}, this work established the feasibility of integrating epitaxial BTO on Si using MBE and provided early experimental validation of strong EO response in BTO thin films on a silicon photonic platform.

\begin{figure}
    \centering
    \includegraphics[width=1\linewidth]{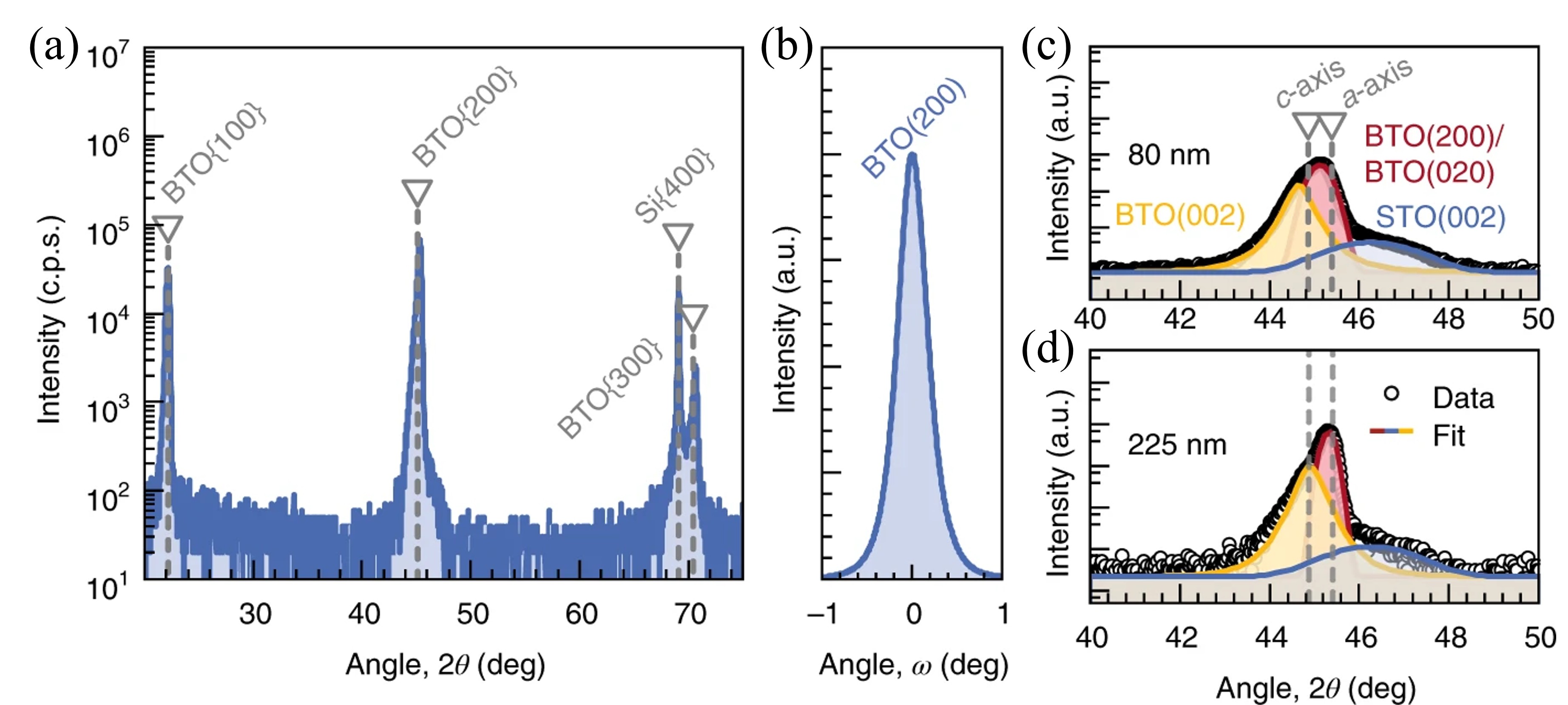}
    \caption{(a) XRD scan of the BTO thin film grown on SOI(100) with a STO buffer layer. (b) XRD rocking curve of BTO (200) peak, exhibiting a narrow FWHM of 0.3\degree. High-resolution XRD scans around the BTO (200) reflection for BTO films with thicknesses of (c) 80 nm and (d) 225 nm, showing the contributions from $a$- and $c$-axis-oriented domains. Reproduced with permission from Abel \textit{et al.}, Nature Materials 18, 42–47 (2019). Copyright 2018 Springer Customer Service Center GmbH\cite{abel2019large}.}
    \label{fig:3abel2019}
\end{figure}

Later, Abel \textit{et al.}\cite{abel2019large} reported one of the largest $r_{\text{eff}}$ to date for BTO integrated on the Si platform. In their work, high-quality epitaxial BTO thin films were first grown on SOI(100) using reactive MBE. An epitaxial 4 nm STO buffer layer was deposited to enable perovskite growth, followed by BTO layers with thicknesses between 80 and 225 nm grown at 500–600 \degree C. The deposited films exhibited excellent crystalline quality, with a rocking curve FWHM of ~0.3\degree, which is one of the narrowest values reported for MBE-grown BTO, as illustrated in Fig. \ref{fig:3abel2019}(a)-\ref{fig:3abel2019}(d). After growth, the BTO/STO/SOI stack was transferred onto a SiO$_2$-on-Si wafer using direct wafer bonding to form a platform suitable for device fabrication. Based on this platform, the authors fabricated EO modulators and extracted an exceptionally large $r_{42}$ of 923 pm/V. Details of the device design and performance will be discussed in Section \ref{section4}. This work demonstrated one of the highest-quality MBE-grown BTO films integrated on silicon, with crystallinity, surface morphology, and domain structure well suited for advanced EO photonic devices. These works establish MBE-grown BTO as a mature material platform for silicon-integrated photonics, enabling the high-performance EO devices which will be discussed in a subsequent section.

\begin{figure}
    \centering
    \includegraphics[width=1\linewidth]{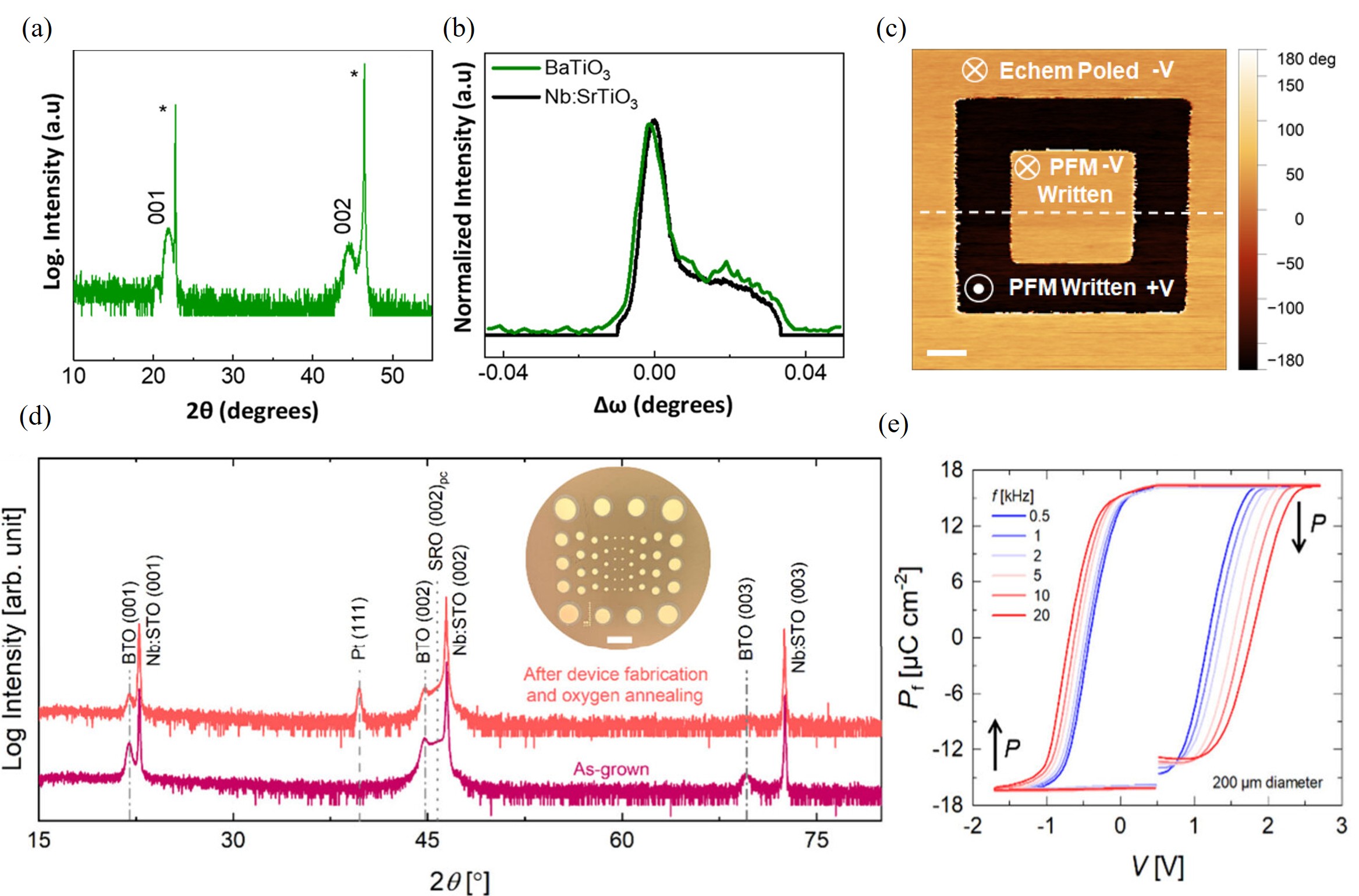}
    \caption{(a) XRD scan of the BTO thin film grown on Nb-doped STO (substrate peaks are denoted by $*$). (b) XRD rocking curve of thin film and substrate (002) peak. Electrochemical switching of ferroelectric polarization in BTO thin film, proved in PFM phase image (c) with scale bar 1 $\mu$m. (d) XRD scan of the BTO thin film grown on Nb-doped STO, with comparison of oxygen annealing process. (e) Hysteresis mapping of ferroelectric BTO thin film in frequency-dependent polarization with applied voltage. Reproduced with permission from Abbsai \textit{et al.}, Nano Letters 2022, 22, 10, 4276–4284, Copyright 2022 American Chemical Society\cite{abbasi2022ferroelectric}, and reproduced with permission from Manjeshwar \textit{et al.}, Nano Letters 25, 39, 14246–14255, Copyright 2025 Ameican Chemical Society\cite{manjeshwar2025ferroelectric}.}
    \label{fig:3abbasi2022}
\end{figure}

In addition, conductive substrates, such as Nb-doped STO, have also been demonstrated for switchable electronic properties from reversible ferroelectric polarization. Abbasi \textit{et al.} \cite{abbasi2022ferroelectric} demonstrated the growth of a 15 nm-thick \textit{c}-axis-oriented single-crystalline BTO film on a 0.5 wt\% Nb-doped STO substrate using reactive MBE. As illustrated in Fig. \ref{fig:3abbasi2022}(a)-\ref{fig:3abbasi2022}(b), the XRD scan shows clear diffraction peaks corresponding to the BTO thin film and Nb-doped STO substrate. The film was subsequently poled to reversibly switch the surface polarization state. The PFM image shown in Fig. \ref{fig:3abbasi2022}(c) shows clear local domain switching behavior. Manjeshwar \textit{et al.} \cite{manjeshwar2025ferroelectric} also reported a 40 nm-thick BTO film on a 0.5 wt\% Nb-doped STO substrate with an intermediate 16 nm-thick SRO layer using hybrid MBE. The resulting BTO film exhibited a rocking-curve FWHM of 0.71\degree, as shown in Fig. \ref{fig:3abbasi2022}(d). Through direct measurements of the polarization-voltage response, the ferroelectric nature of the BTO thin film was verified, as shown in Fig. \ref{fig:3abbasi2022}(e). The asymmetry in the switching behavior and leakage current was hypothesized to be associated with structural defects. These works demonstrate the feasibility of epitaxial growth of BTO on conductive substrates for future photonic and electronic applications.

\subsection{RF sputtering}
Sputtering is also a widely used technique for depositing BTO thin films due to its scalability, compatibility with large-area substrates, and relatively low equipment cost. In sputtering, energetic ions from a plasma bombard a target and eject atoms that subsequently condense onto a heated substrate. One of the earliest demonstrations of epitaxial sputtered BTO was reported by Kim \textit{et al.}\cite{kim1995structural}, who successfully deposited epitaxial BTO films on MgO(100). Although XRD indicated a well-defined $c$-axis orientation, TEM showed that thin $a$-domains (<10 nm) were embedded within the $c$-domain matrix, and the film exhibited a rocking-curve FWHM of 0.9\degree. Later, Kim \textit{et al.}\cite{kim2014ridge} demonstrated thicker BTO thin film on MgO(100) with a thickness ranging from 200 to 1000 nm using RF sputtering. They observed that the 1 µm thick film exhibited a broader rocking curve FWHM of 1.527\degree for the (100) peak, while a 750 nm film showed improved crystallinity with a FWHM of 0.81\degree and a smoother surface (rms $\approx$ 0.86 nm).

\begin{figure}
    \centering
    \includegraphics[width=1\linewidth]{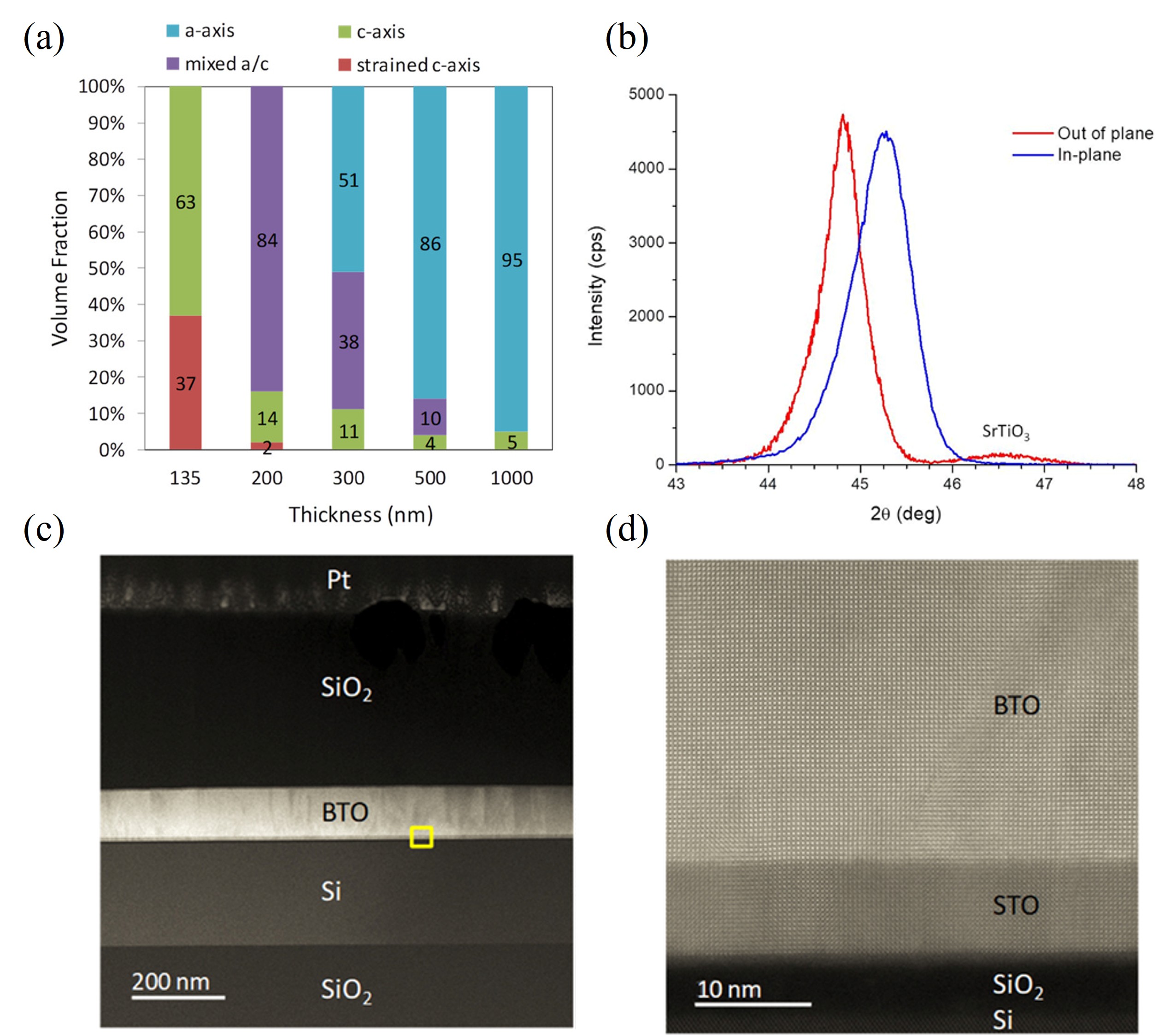}
    \caption{(a) Volume fraction of the crystallographic orientations in BTO films with different thicknesses. Reprinted with permission from Posadas \textit{et al.}, ACS Applied Materials \& Interfaces 13, 51230–51244 (2021)\cite{posadas2021thick}.  Copyright 2021 American Chemical Society. (b) XRD results of a $c$-axis oriented BTO film growth on a STO-buffered SOI subtrate. Cross-sectional images of (c) a sputtered BTO film growth on a SOI platform, and (d) a high magnification image of the BTO/STO/Si interface, revealing the presence of an interfical SiO$_2$ layer. Reproduced with permission from Posadas \textit{et al.}, Journal of Applied Physics 134, 073101 (2023) with the permission of AIP Publishing \cite{posadas2023rf}.}
    \label{fig:3posadas20212023}
\end{figure}

Sputtering is also a popular technique to integrate BTO onto the Si platform due to its wafer-scale uniformity and relatively high deposition rate, supported by intermediate buffer layers to mitigate the lattice mismatch between BTO and Si/SiO$_2$. Posadas \textit{et al.}\cite{posadas2021thick} demonstrated the growth of thick (up to 1 $\mu$m) epitaxial BTO films on SOI using off-axis RF magnetron sputtering combined with an MBE-grown STO buffer layer. They first investigated the effect of STO thickness and found that STO layers $\geq$6 nm provided the most stable template for epitaxial BTO growth. A 300 nm BTO film grown on 6-8 nm STO exhibited high crystalline quality with ~55\% $a$-axis orientation. Their thickness-dependent study showed that BTO films thinner than 135 nm remained fully $c$-axis oriented, while thicker films ($\geq$500 nm) gradually transitioned to $a$-axis orientation, as illustrated in Fig. \ref{fig:3posadas20212023}(a), consistent with strain-relaxation mechanisms reported by Hsu \textit{et al.}\cite{hsu2017controlled} For a 1 $\mu$m film, more than 95\% $a$-axis orientation was achieved with a rocking curve FWHM of 0.4\degree. In a subsequent work\cite{posadas2023rf}, the same group demonstrated the growth of $c$-axis–oriented BTO thin film for EO device fabrication. In that work, an 8 nm thick STO was first deposited on Si using MBE, followed by sputtering of a 110 nm thick BTO at 680 \degree C. Figure \ref{fig:3posadas20212023}(b) shows an XRD scan of the resulting BTO film, revealing a BTO(002) rocking-curve FWHM of 0.6\degree, with a surface roughness of approximately 0.4 nm. Figures \ref{fig:3posadas20212023}(c) and \ref{fig:3posadas20212023}(d) illustrate cross-sectional TEM images of the BTO films grown on STO-buffered SOI substrates and a magnified view of the BTO/STO/Si interface, respectively, indicating the presence of a thin interfacial SiO$_2$ layer between the STO buffer and the Si substrate. Based on this film platform, the authors fabricated BTO EO modulators onto SOI platform and extracted an EO coefficient ($r_{33}$ = 134.4 pm/V), demonstrating the feasibility of sputtered BTO for photonic applications.

More recently, Raju \textit{et al.}\cite{raju2025high} reported low-loss ridge waveguides fabricated directly in sputtering BTO thin films, In their work, a 5 nm thick STO layer was first grown oxidized SOI, followed by deposition of a 300 nm thick BTO layer by off-axis RF sputtering at temperatures exceeding 750 \degree C with a growth rate of 3.6 nm/min. The deposited films exhibited a rocking curve FWHM of 0.52\degree and a rms roughness less than 1 nm. This work demonstrates the feasibility of achieving low-loss waveguide configurations in BTO thin film using state-of-the-art sputtering and etching processes. Overall, these studies demonstrate that RF sputtering can produce high-quality epitaxial BTO thin films, offering a scalable and wafer-compatible route for integrated photonic applications.

\subsection{Chemical vapor deposition (CVD)}
In early development of thin-film BTO, chemical vapor deposition (CVD) was also considered a promising candidate for achieving epitaxial growth. One of the earliest demonstrations of ridge waveguides patterned in BTO thin films was reported by Gill \textit{et al.}\cite{gill1996thin}, who deposited the BTO thin film on (001)-oriented MgO using low-pressure MOCVD with metalorganic Ba and Ti precursors. The films were prepared at 725 \degree C with a growth rate of approximately 100 nm/h. XRD results indicated that the BTO films are $a$-axis oriented, and AFM showed rms surface roughness values of 7 nm and 9 nm for 300 and 200 nm thick films, respectively. These films were subsequently etched into ridge-waveguide geometries, and their propagation loss was characterized. Later, Tang \textit{et al.}\cite{tang2004low} demonstrated EO response in MOCVD-grown BTO films using a similar process. In their work, a 570 nm thick epitaxial BTO film was deposited on MgO(100), and a low-loss (<1 dB/cm) waveguide was realized by employing a strip-loaded Si$_3$N$_4$ configuration. These works demonstrated that MOCVD-grown BTO can support ridge or strip-loaded waveguide structures and confirmed that such epitaxial films possess measurable EO properties, despite the relatively high surface roughness and limited crystalline quality.

More recently, high-quality BTO films have been demonstrated using high-vacuum CVD (HV-CVD), as reported by Reinke \textit{et al.}\cite{reinke2017low} Compared to conventional CVD, the HV-CVD process enables epitaxial BTO growth at significantly lower temperatures, making it compatible with CMOS fabrication constraints. In their work, epitaxial BTO was grown at 400 \degree C on three substrates: MgO, STO, and STO-buffered Si. The deposited BTO films on STO and MgO exhibited rocking curve FWHMs of 1.3\degree and 3.2\degree, respectively. For BTO grown on Si, a 4 nm thick STO buffer layer was first deposited by MBE at 650 \degree C, followed by HV-CVD deposition of BTO at 400 \degree C. The resulting film exhibited $a$-axis orientation with a rocking curve FWHM of 1.24\degree and a rms roughness of 0.2 nm. A more recent improvement of the HV-CVD approach was reported by Szmyt \textit{et al.}\cite{szmyt2025high} In their work, the BTO thin film was deposited on a Si substrate with a 4 nm thick STO buffer layer at about 467 \degree C. The film exhibited improved quality, with a rocking curve FWHM of 0.77\degree and a rms roughness of 0.16 nm. This highlights the feasibility of low-temperature epitaxial BTO deposition via HV-CVD while remaining compatible with silicon integration. Although the traditional low-pressure CVD has generally provided BTO films with limited crystallinity and relatively high roughness, recent advancements in HV-CVD demonstrate that low-temperature epitaxial BTO films on silicon can achieve competitive quality, positioning CVD as a promising method for preparing BTO thin films for integrated photonics.

\subsection{Chemical solution deposition (CSD)}
Chemical solution deposition (CSD) has recently gathered attention due to its rapid, low-cost, low temperature, atmospheric-pressure approach for preparing BTO thin films. However, achieving high-quality epitaxial BTO by CSD remains challenging because crystallization occurs only during the post-deposition annealing step, in which grain growth, porosity, and defect formation must be simultaneously managed.

\begin{figure}
    \centering
    \includegraphics[width=1\linewidth]{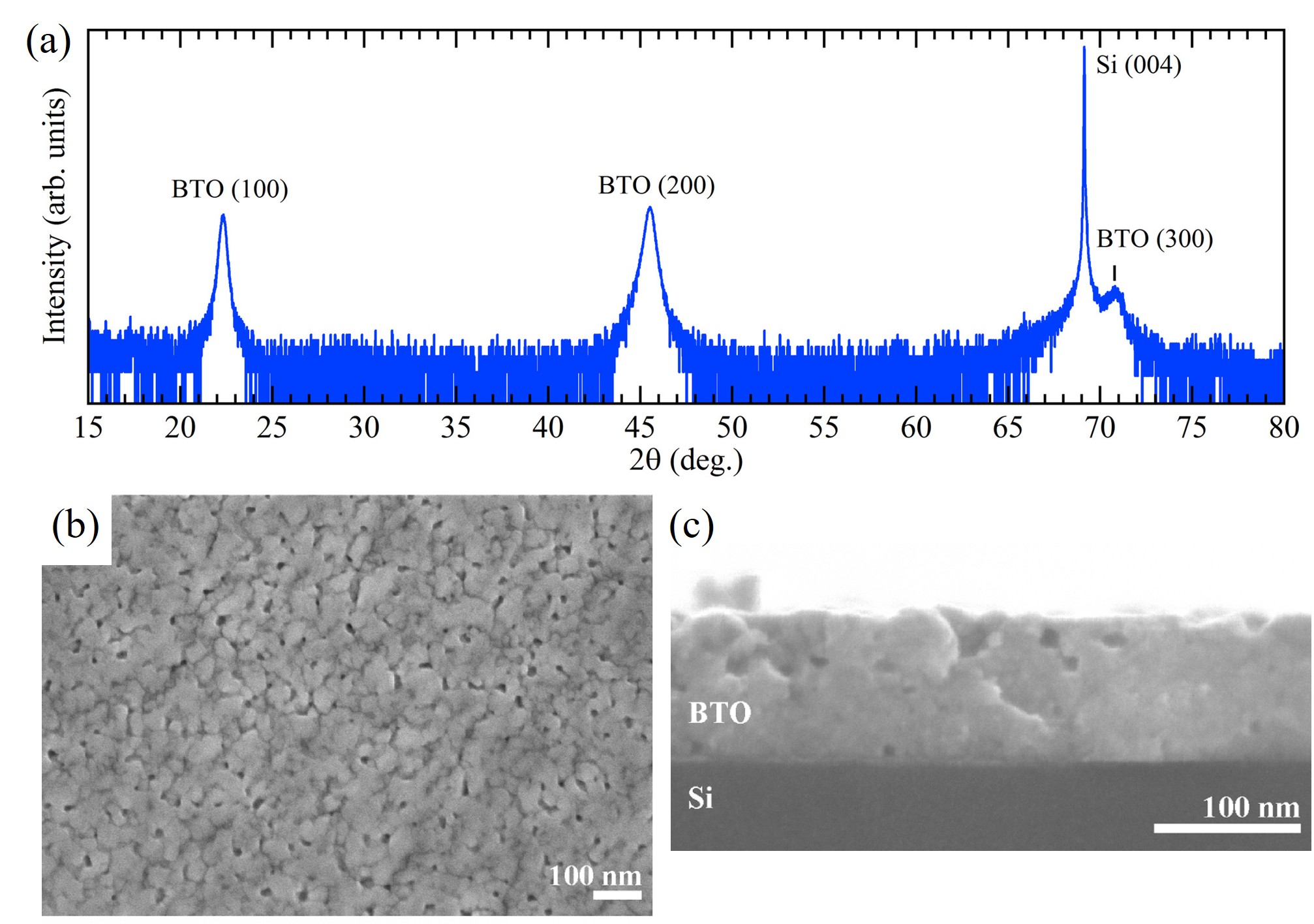}
    \caption{(a) Out-of-plane XRD scan of the BTO thin film deposited on a STO-buffered Si substrate using CSD method. (b) Top-view and (c) cross-sectional SEM images of the BTO film, showing the relatively rough top surface and the grains\cite{edmondson2020epitaxial}. Reproduced with permission from Edmondson \textit{et al.}, Journal of the American Chemical Society, \textbf{103}, 1209-1218 (2020). \copyright 2020 Wiley-VCH GmbH}
    \label{fig:3edmondson2019}
\end{figure}

Edmondson \textit{et al.}\cite{edmondson2020epitaxial} demonstrated that epitaxial and EO-active BTO thin films can be achieved on Si(001) using CSD. The authors employed STO-buffered Si substrates prepared by MBE for the growth of BTO. Then, the BTO precursor was spin-coated onto the substrate and annealed at 600\degree C for 1 hr, with the coating-annealing cycle repeated five times to reach an approximately 85 nm thick film. XRD measurements confirmed epitaxial growth with the $a$-axis oriented out-of-plane, exhibiting a rocking-curve FWHM of 1.23\degree for the BTO (200) reflection, as shown in Fig. \ref{fig:3edmondson2019}(a). SEM imaging showed grains with an average diameter of ~40 nm, as shown in Fig. \ref{fig:3edmondson2019}(b)-\ref{fig:3edmondson2019}(c). An AFM scan revealed local height variations on the order of 10 nm, though a relatively low rms roughness of 1.4 nm was obtained over a 5 $\times$ 5 $\mu$m area. EO characterization confirmed a measurable, though modest, EO response ($\sim$ 27 pm/V) in CSD-grown BTO on STO-buffered Si substrate. 

\begin{figure}
    \centering
    \includegraphics[width=1\linewidth]{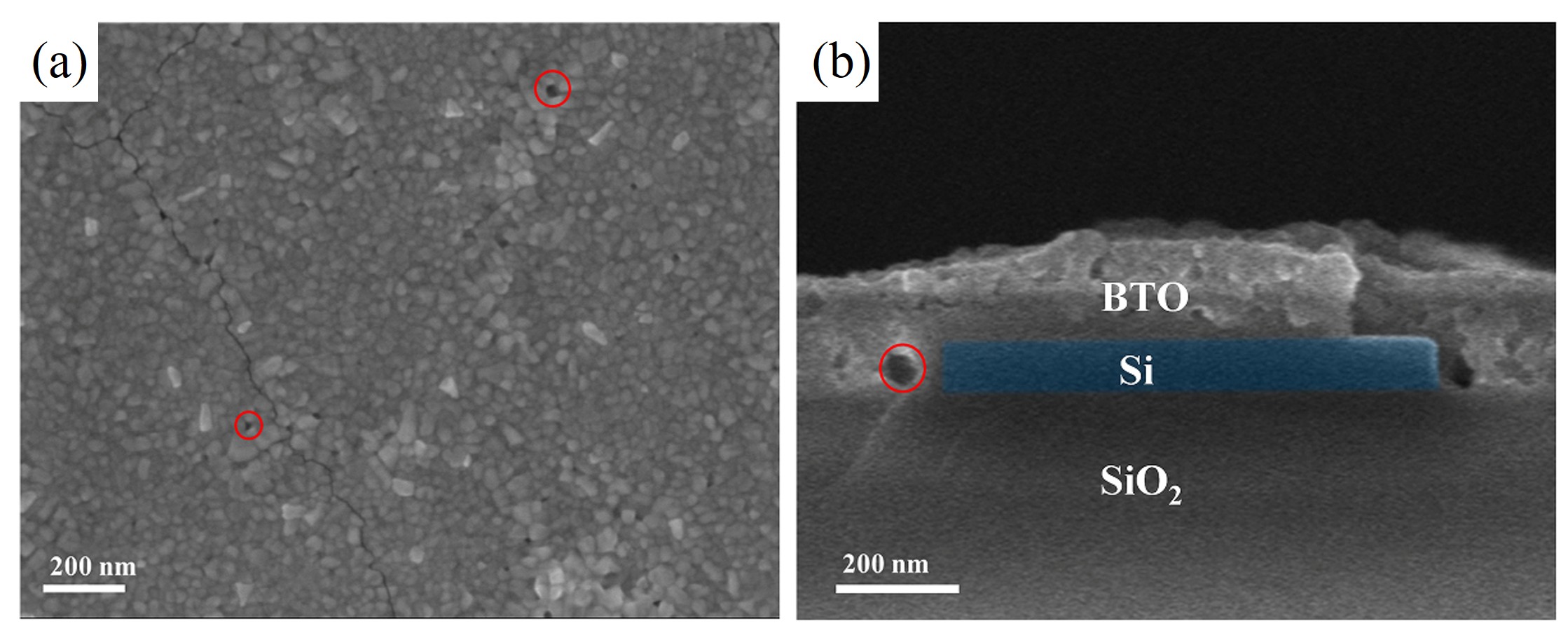}
    \caption{(a) Top-view SEM image of the BTO thin film prepared using the CSD, revealing surface cracks and structural defects (indicated by red circles). (b) Cross-sectional SEM image of the CSD-grown BTO layer deposited on a patterned SOI waveguide. Reproduced with permission from Zhang \textit{et al.}, Advanced Photonics Nexus 3, 066005–066005 (2024). Copyright 2024 authors, licensed under a Creative Common Attribution (CC BY) license\cite{zhang2024hybrid}.}
    \label{fig:3zhang2024}
\end{figure}

More recently, Zhang \textit{et al.}\cite{zhang2024hybrid} demonstrated a hybrid Si-BTO integrated photonics using BTO thin film prepared using CSD. In their work, SOI substrates were first patterned to form silicon ridge waveguides, followed by repeated BTO solution spin-coating at 4000 rpm and annealing cycles at 450 \degree C. In the final annealing step, the temperature was increased to 800 \degree C to enhance film crystallinity, as shown in Fig. \ref{fig:3zhang2024}. Although the resulting BTO films were polycrystalline with noticeable defects and surface non-uniformity, the authors successfully achieved EO modulation in racetrack resonators with an extracted effective refractive index of 27.2 pm/V, showing that CSD-grown BTO can function as an active cladding layer on SOI.

Overall, while CSD-grown BTO films typically exhibit higher defect densities, broader rocking-curve widths, and reduced EO response compared to vacuum-grown films, the method enables large-area, low-cost deposition of EO-active BTO on Si. These characteristics make CSD an attractive and scalable alternative for application when high crystalline quality is not necessary.

\subsection{Atomic layer deposition (ALD)}
Atomic layer deposition (ALD) has also been explored as a pathway for epitaxial BTO thin-film growth, due to its low deposition temperature, excellent thickness control, and scalability to large-area substrates. Compared with other techniques, such as PLD or MBE, ALD offers a unique advantage for CMOS-compatible integration because it can produce uniform films at temperatures far below those required by conventional epitaxial methods. 

\begin{figure}
    \centering
    \includegraphics[width=1\linewidth]{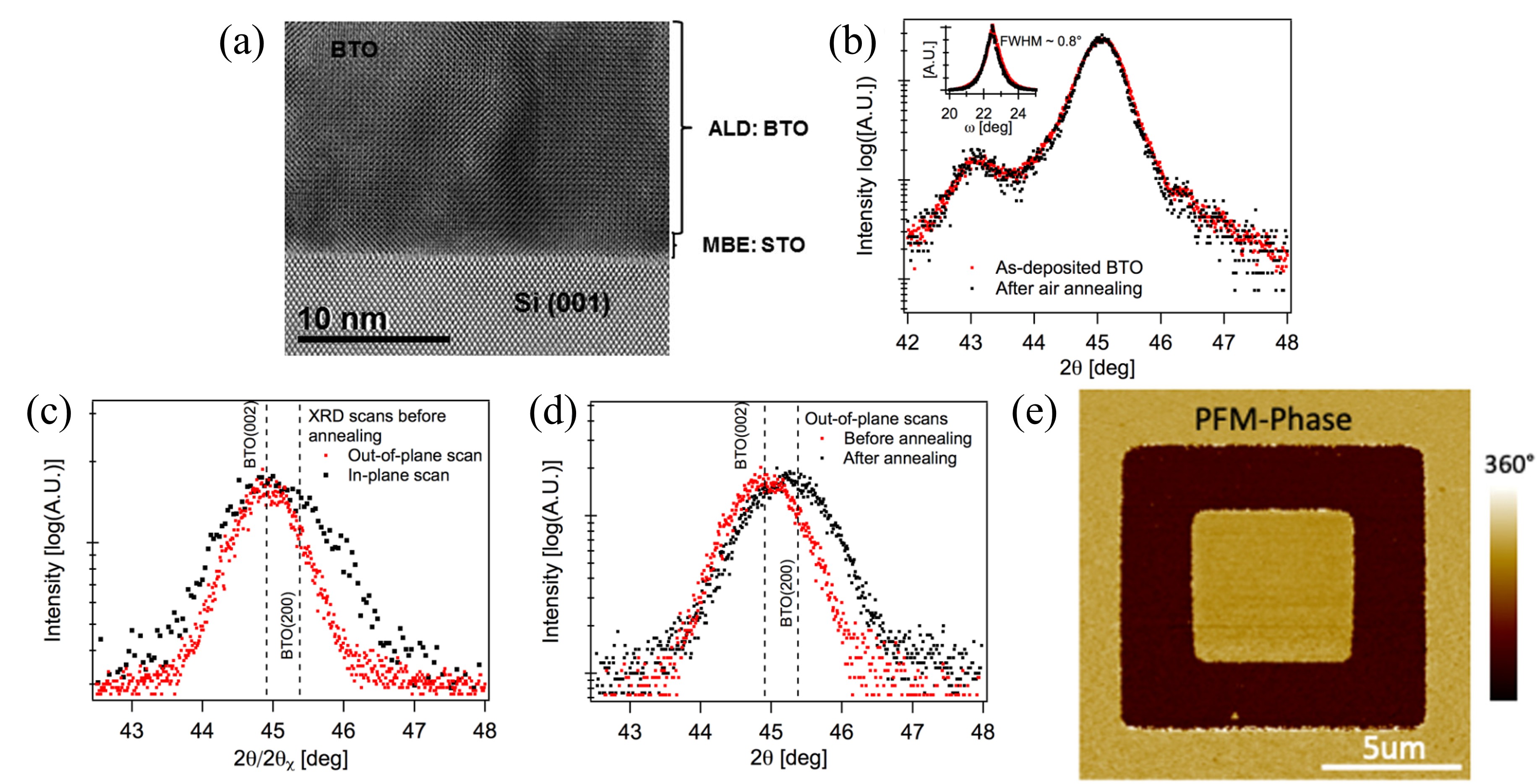}
    \caption{(a) Cross-sectional STEM image of an ALD-grown BTO film on a STO-buffered Si substrate after annealing at 600 \degree C, indicating not observable amorphous layer at the interface. Reproduced from Ngo \textit{et al.}, Applied Physics Letters 104 (2014), with the permission of AIP Publishing \cite{ngo2014epitaxial}. (b) XRD scan of a $c$-axis oriented 15 nm thick BTO film, showing no noticeable change after annealing at 300 \degree C. (c) Broad in-plane XRD scan of a 66 nm thick BTO film revealing the presence of both $a$- and $c$-axis reflections. (d) Out-of-plane XRD scans indicating strain relaxation and rotation of the BTO $c$-axis from out-of-plane to in-plane after annealing at 650 \degree C.  (e) PFM phase image after poling. Reproduced from Lin \textit{et al.}, Journal of Applied Physics 126 (2019), with the permission of AIP Publishing \cite{lin2019atomic}.}
    \label{fig:3ald}
\end{figure}

Ngo \textit{et al.}\cite{ngo2014epitaxial} reported the epitaxial growth of BTO thin film on Si(001) using ALD. In their work, a 1.6 nm thick STO buffer layer was first deposited on Si by MBE, followed by deposition of up to 20 nm thick BTO at 225 \degree C. After deposition, a 600 \degree C annealing for 5 minutes improved the rocking curve FWHM from 0.91\degree to 0.74\degree. Also, a rapid temperature ramp led to a transition from strong $c$-axis-orientation to weak $a$-axis orientation due to strain relaxation. Cross-sectional STEM imaging showed a clean STO/Si interface with no amorphous SiO$_2$, highlighting a key advantage of low-temperature ALD in suppressing interfacial oxidation, as illustrated in Fig. \ref{fig:3ald}(a). 

Lin \textit{et al.}\cite{lin2019atomic} later demonstrated both ferroelectric and EO responses in ALD-grown BTO thin film on Si substrate. In their study, an STO buffer layer was deposited on Si at 200 \degree C by MBE and crystallized by annealing at 600 \degree C. Subsequently, BTO films up to 66 nm thick were grown at 225 \degree C. Thin BTO films (15 nm) exhibited an out-of-plane $c$-axis orientation, with a rocking curve FWHM of approximately 0.8\degree for the BTO (002) peak, and showed no observable change after low-temperature air annealing at 300 °C, as shown in Fig. \ref{fig:3ald}(b). Figure \ref{fig:3ald}(c) shows a broad in-plane BTO peak covering both the $a$- and $c$-axis reflections, indicating the coexistence of (200) and (002) orientations in the 66 nm thick film. The authors also observed strain relaxation when the 66 nm thick films were annealed at 650 °C in O$_2$, resulting in a rotation of the BTO $c$-axis from out-of-plane to in-plane, as illustrated in Fig. \ref{fig:3ald}(d). Additionally, Fig. \ref{fig:3ald}(e) shows PFM measurements exhibiting clear 180° phase contrast after poling, confirming ferroelectric switching, and electro-optic characterization demonstrated a noticeable Pockels response.

While the overall crystalline quality of ALD-grown BTO remains moderate compared with films prepared by other approaches, the scalability, low-pressure deposition, and precise thickness control make ALD a promising and CMOS-compatible approach for epitaxial BTO thin films in integrated photonic applications.

\section{\label{section4}ELectro-Optic Photonics Based on BTO Thin Films}
In this section, we discuss devices, both passive and active, related to EO BTO-based integrated photonics. Here we classify the devices into two categories, monolithic and hybrid. In monolithic BTO devices, ridge waveguides are realized by patterning  BTO film, which results in a strong optical mode overlap with BTO itself. As for hybrid devices, strip-loaded ridge waveguides are typically achieved using other low-loss materials, such as Si or SiN. Such structures allow utilization of more mature manufacturing techniques of the ridge waveguide system, while still allowing exploitation of the favorable optical properties of BTO, allowing realization of EO modulators.

\subsection{\label{passive photonics}Passive photonics of BTO thin films}

\begin{table*}[t!]
\caption{\label{tab:BTO_passive_summary}
Summary of representative passive devices in BTO thin films}
\begin{ruledtabular}
\begin{tabular}{c c c c c}

\makecell{\textbf{Waveguide}\\ \textbf{type}}  &
\textbf{Year} &
\makecell{\textbf{BTO thickness}\\ \textbf{(nm)}} &
\makecell{\textbf{Propagation loss}\\ \textbf{(dB/cm)}} &
\makecell{\textbf{Loss-measurment}\\ \textbf{method}}
\\ \hline

Planar \cite{gill1997thin} & 1997 & 300  & 5 & -- \\
Planar \cite{buchal1998epitaxial} & 1998 & 600 & 2.9 & Scattering-loss imaging\\
Planar \cite{lisoni2001growth} & 2001 & 400 & 8 & Prism coupling\\
Planar \cite{petraru2002ferroelectric} & 2002 & 1000 &  2 & --\\
Planar \cite{leroy2013guided} & 2013 & 1200 & 5 & Fiber coupling\\
Planar \cite{posadas2023rf} & 2023 & 110 &  1.08 & Prism coupling\\
Ridge BTO \cite{gill1996thin} & 1996 & 200 & 4 & Scattering-loss imaging\\
Ridge BTO \cite{kim2014ridge} & 2014 & 1000  & 3-5 & --\\
Ridge BTO \cite{cao2021barium} & 2021 & 800  & 2 & Fabry–Perot\\
Ridge BTO \cite{dong2023monolithic} & 2023 & 500 & 3.17 & Cut-back\\
Ridge BTO \cite{Lin2024single} & 2024 & 500 & 12 & Cut-back\\
Ridge BTO \cite{raju2025high}& 2025 & 300  & 0.138/0.7 & Racetrack resonator/cut-back\\
Ridge BTO \cite{isti2025fabrication}& 2025 & 150 & 0.7 & Optical backscatter reflectometry\\
Ridge BTO \cite{kim2025lowlossmonolithicbarium}& 2025 & 340 & 0.32 & Racetrack resonator\\
Hybrid SiN/BTO \cite{tang2004low} & 2004 & 560 &  1 & Cut-back\\
Hybrid Si/BTO \cite{xiong2014active} & 2014 & 80 &  44 & Ring resonator\\
Hybrid Si/BTO \cite{abel2016hybrid} & 2016 & 50 &  47 & Cut-back\\
Hybrid Si/BTO \cite{eltes2016low} & 2016 & 50 &  6.3 & Cut-back\\
Hybrid Si/BTO \cite{jin2017monolithic} & 2017 & 500 &  4.2 & Cut-back\\
Hybrid Si/BTO \cite{abel2019large} & 2019 & 225 &  9 & Cut-back\\
Hybrid Si/BTO \cite{eltes2019batio3} & 2019 & 170 &  5.8 & Ring resonator\\
Hybrid SiN/BTO \cite{ortmann2019ultra} & 2019 & 80 &  9.4 & Cut-back\\
Hybrid Si/BTO \cite{eltes2020integrated} & 2020 & 225 &  2.9 & Cut-back\\
Hybrid Si/BTO \cite{geler-kremer2022ferroelectric} & 2022 & 225 &  4.8 & Racetrack resonator\\
Hybrid Si/BTO \cite{zhang2024hybrid} & 2024 & -- &  11.3 & Ring resonator\\
Hybrid SiN/BTO \cite{chrysostomidis2025ultra} & 2025 & 220 & 5.4 & Cut-back\\
Hybrid SiN/BTO \cite{psiquantum_team2025manufacturable}& 2025 & 135 &  0.54 & Cut-back\\

\end{tabular}
\end{ruledtabular}
\end{table*}

The development of low-loss waveguides in thin-film BTO is essential for realizing high-performance integrated photonic and EO devices that exploit its outstanding material properties. In recent decades, various waveguide implementations based on BTO thin films have been reported, exploring various approaches to achieve efficient optical confinement and low propagation loss. As summarized in Table \ref{tab:BTO_passive_summary}, the reported propagation losses span more than an order of magnitude, reflecting not only variations in film quality but also the diversity of waveguide configurations and fabrication processes. Here we classify passive waveguides in BTO thin films into three categories: planar (slab) waveguides, monolithic BTO ridge waveguides, and heterogeneous hybrid strip-loaded waveguides.

\begin{figure}
    \centering
    \includegraphics[width=1\linewidth]{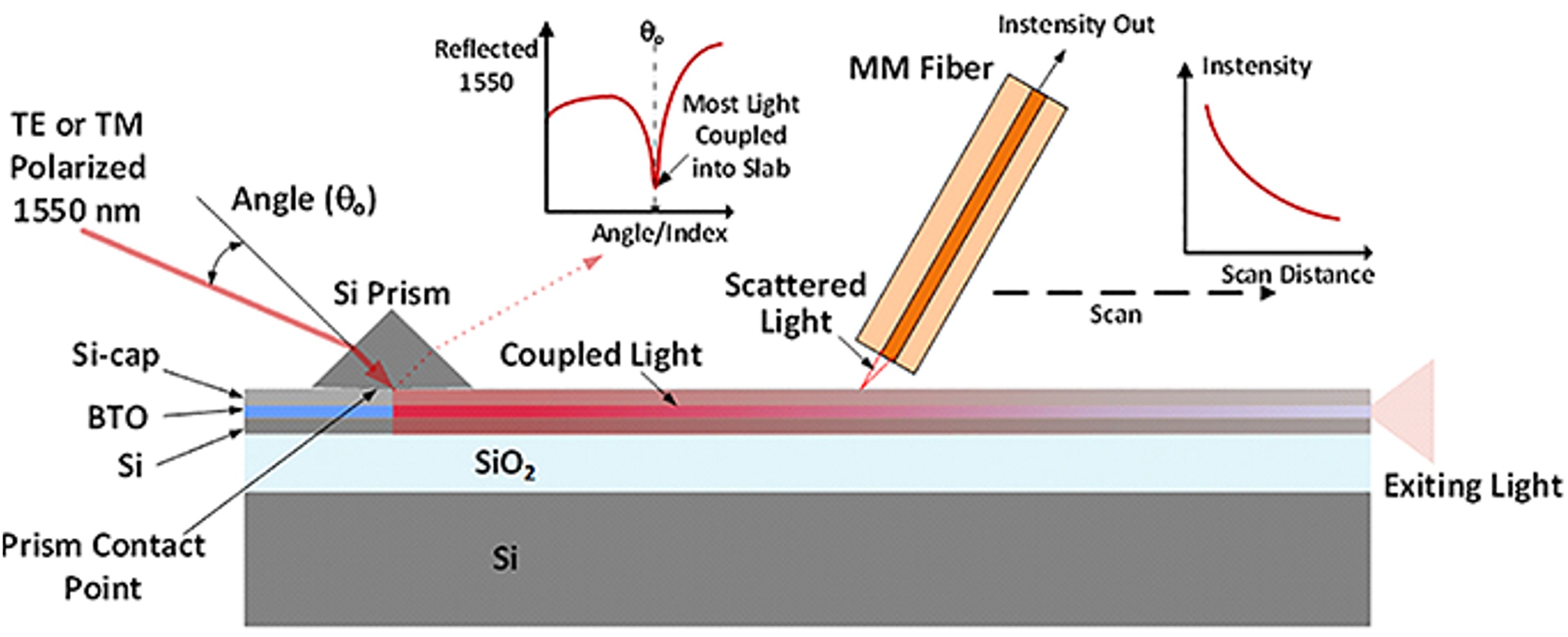}
    \caption{Schematic of prism coupling method for propagation loss measurement of planar waveguides. Reproduced from Posadas \textit{et al.}, Journal of Applied Physics 134, 073101 (2023), with the permission of AIP Publishing\cite{posadas2023rf}.}
    \label{fig:4prismcouplling}
\end{figure}

Planar or slab waveguides utilize the BTO thin film itself as a guiding layer without lateral patterning. These studies were motivated to understand the intrinsic material quality, optical transparency, and scattering mechanisms without introducing additional loss from etching processes. For planar waveguides, BTO thin films are typically grown on substrates with a lower refractive index than BTO and are cladded by air or oxide layers, which have lower refractive indices than the BTO itself. Based on this vertical refractive-index contrast, optical modes can be guided within the BTO thin film, and by employing different characterization methods, the propagation loss of the films themselves can be measured, as shown in Fig. \ref{fig:4prismcouplling}. In planar BTO waveguides, propagation losses are most commonly characterized using prism coupling \cite{lisoni2001growth, posadas2023rf} or scattering-loss imaging methods \cite{buchal1998epitaxial}. The measured losses in planar waveguides are strongly correlated with film thickness, surface roughness, crystalline quality, and the choice of substrate. Thus, planar-waveguide loss measurements provide valuable insights into material-limited losses and serve as an important indicator of thin-film quality. However, due to the lack of lateral confinement, planar BTO waveguides have limited applicability in photonic device integration on their own.

To enable compact integration, it is essential to introduce lateral refractive-index contrast. As a result, researchers have been attracted to explore monolithic BTO ridge waveguides, where the BTO thin film is directly etched into a ridge geometry to achieve lateral mode confinement. In principle, this monolithic configuration offers several advantages, such as the optical mode being mainly confined within the ferroelectric material, which then later maximizes the efficiency of the EO response. This waveguide geometry is widely adopted in other ferroelectric photonic platforms, such as lithium niobate (LN) \cite{chen2021integrated,li2023high, hu2025integrated} and lithium tantalate (LT)\cite{zhang2025ultrabroadband, wang2024lithium}, enabling compact components including interferometers \cite{hassanien2022compact,wang2018nanophotonic} and resonators \cite{yu2024tunable,yin2021electro,wang2019monolithic}.

\begin{figure}
    \centering
    \includegraphics[width=1\linewidth]{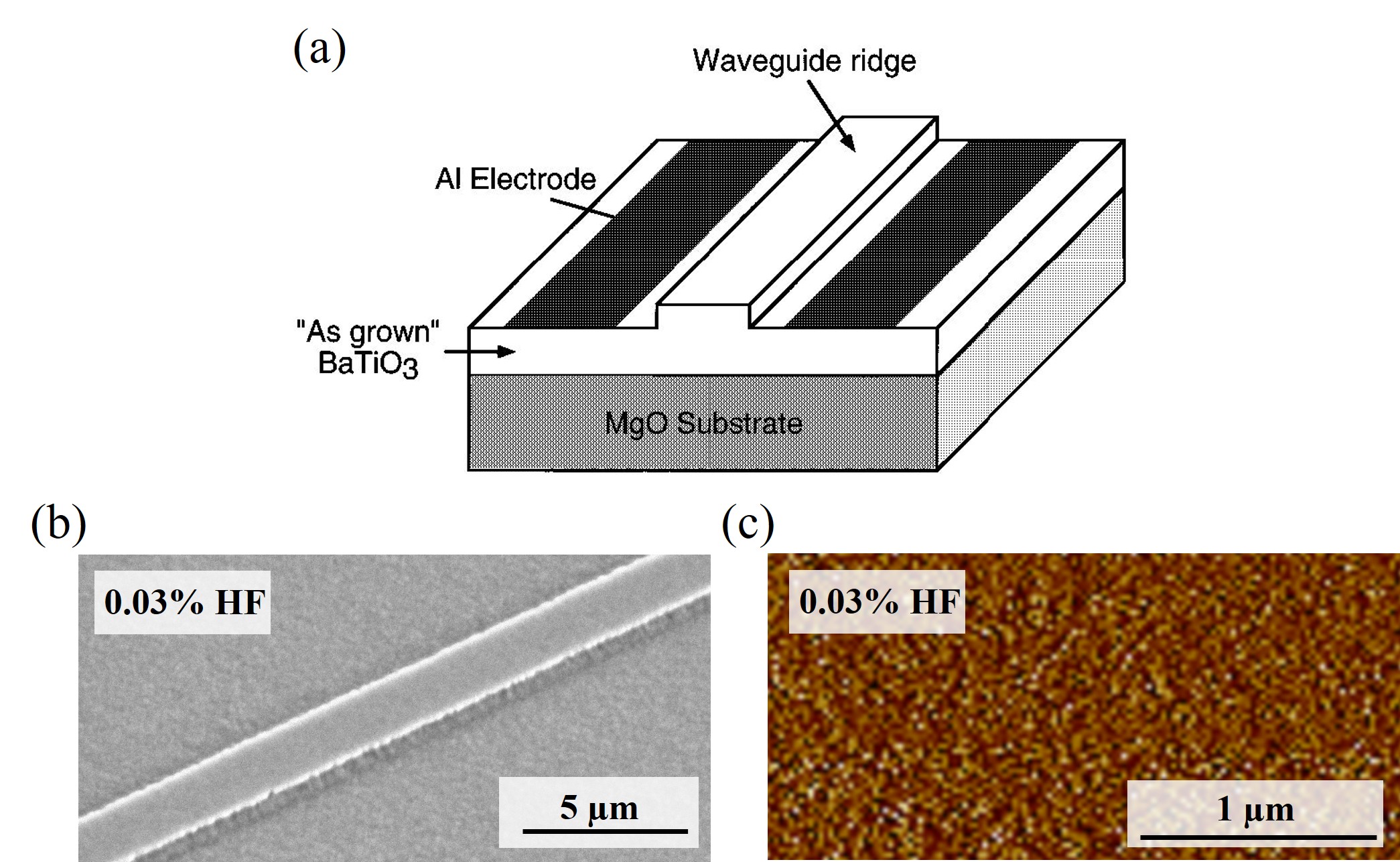}
    \caption{(a) Schematic of BTO ridge waveguides on MgO substrate with lateral electrodes. Reproduced from Gill \textit{et al.}, Applied Physics Letters 71, 1783–1785 (1997), with the permission of AIP Publishing\cite{gill1997thin}. (b)-(c) SEM image and surface topography images of wet-etched thin-film BTO\cite{Lin2024single}.}
    \label{fig:4btoridge}
\end{figure}

However, due to the chemically and mechanically stable nature of BTO, direct etching of BTO remains challenging. As summarized in Table \ref{tab:BTO_etching_summary}, both wet-etching and dry-etching approaches have been explored to pattern BTO thin films. Wet etching with selected mask materials has offered excellent etching selectivity and can enable relatively deep etch depths.
However, due to its inherently isotropic nature, wet etching typically results in sloped sidewalls and consequently degraded lateral mode confinement. For instance, the ion-milled BTO waveguides reported by Raju \textit{et al.} \cite{raju2025high} exhibited sidewall angles of 75\degree, while the wet-etched BTO waveguides reported by Lin \textit{et al.} \cite{Lin2024single} exhibited a relatively low sidewall angle of approximately 45\degree.
Gill \textit{et al.} \cite{gill1996thin,gill1997thin} reported one of the earliest demonstrations of BTO ridge waveguides fabricated on MgO substrates using a 1\% HF solution and successfully demonstrated amplitude modulation by applying an electric field across the resulting BTO ridge waveguide, as shown in Fig. \ref{fig:4btoridge}(a).

More recently, Lin \textit{et al.} \cite{Lin2024single} investigated the impact of HF concentration (ranging from 2\% to 0.03\%) on the surface roughness of BTO thin film after wet etching. The results showed that lower HF concentrations lead to reduced etch rates and improved surface roughness. Figures \ref{fig:4btoridge}(b) and \ref{fig:4btoridge}(c) illustrate the BTO thin film after wet etching using a 0.03\% HF solution. A surface RMS roughness of 2.02 nm was achieved, compared to the as-grown RMS roughness of 0.8 nm. Due to the isotropic nature of the wet etching process, a sidewall slope of 45\degree\ was observed. By using this approach, propagation losses of 14 and 12 dB/cm were reported for ridge waveguides with widths of 2.5 $\mu$m and 3 $\mu$m, respectively. Although the propagation losses achieved by wet etching are still higher than those required for low-loss waveguides, a functional MZI structure with an extinction ratio of 25 dB was demonstrated in their subsequent work \cite{Lin2025mach} through optimization of the waveguide geometry.

\begin{table*}[t!]
\caption{\label{tab:BTO_etching_summary}
Summary of etching approaches for monolithic BTO ridge waveguides and the propagation losses measured from the ridge waveguides}
\begin{ruledtabular}
\begin{tabular}{c c c c c}

\textbf{Etch methods} &
\textbf{Year} &
\textbf{Film thickness (nm)} &
\textbf{Etch depth (nm)} &
\textbf{Propagation loss (dB/cm)} 
\\ \hline

HF wet-etching \cite{gill1996thin} & 1996 & 200 & 15 & 4 \\
HF wet-etching \cite{gill1997thin} & 1997 & 320 & 40 & -- \\
Ion beam etching \cite{petraru2003integrated} & 1997 & 1000 & 70 & -- \\
HCl/H$_2$O$_2$ wet-etching \cite{kim2014ridge} & 2014 & 1000 & 150 & 3-5 \\
ICP dry-etching\cite{cao2021barium} & 2021 & 800 & 170 & 2 \\
Ion milling \cite{dong2023monolithic} & 2023 & 500 & 175 & 3.17 \\
HF/HNO$_3$ wet-etching\cite{posadas2023rf} & 2023 & 105 & 75 & -- \\
HF wet-etching \cite{Lin2024single} & 2024 & 500 & 250 & 12\\
Ion milling \cite{raju2025high}& 2025 & 300 & 100 & 0.138/0.7 \\
Ion beam milling \cite{isti2025fabrication}& 2025 & 150 & 150 & 0.7\\
Ion beam milling \cite{kim2025lowlossmonolithicbarium}& 2025 & 340 & 180 & 0.32\\
\end{tabular}
\end{ruledtabular}
\end{table*}

Compared to wet-etching, dry-etching approaches, including ion-beam milling and inductively coupled plasma (ICP)-based etching, are more desirable for achieving compact design due to their anisotropic etching characteristics, which allows higher-aspect-ratio patterning. However, due to low etch selectivity with the available hard masks, dry-etching in BTO typically results in limited etch depths\cite{li2014comparative,chen2021analysis}. It is worth noting that, although the anisotropic nature of dry etching, achieving highly vertical sidewalls remains challenging because of the limited etch selectivity.

\begin{figure}
    \centering
    \includegraphics[width=1\linewidth]{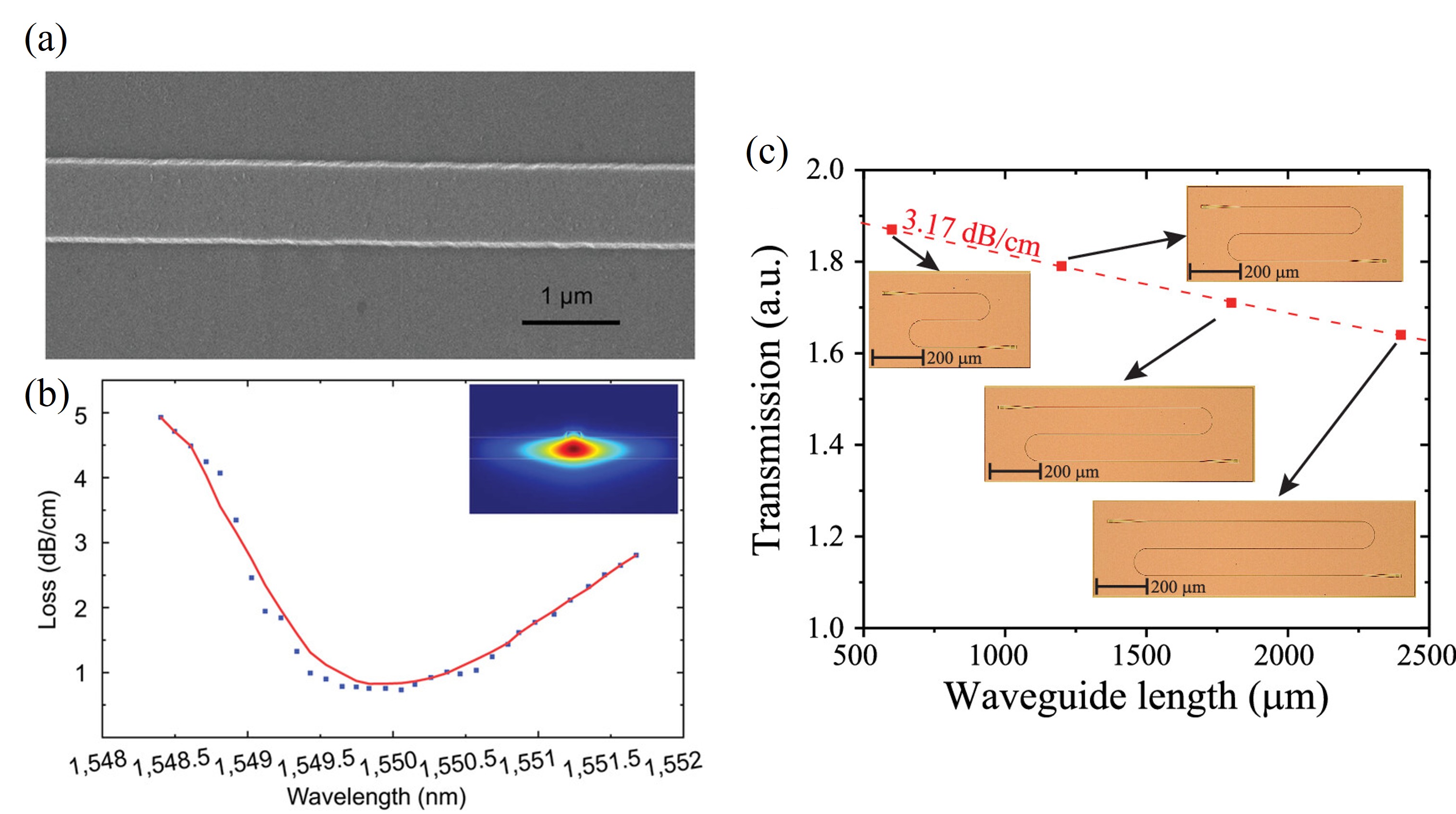}
    \caption{(a) Helium-ion microscopy image of the ICP-etched BTO ridge waveguide and (b) the propagation loss measurement. Reproduced with permission from Cao \textit{et al.}, Advanced Materials 33, 2101128 \copyright 2021 Wiley-VCH GmbH\cite{cao2021barium}. (c) Cut-back methods for propagation loss measurement on BTO ridge waveguides with various lengths\cite{dong2023monolithic}.  Reprinted with permission from Dong \textit{et al.}, ACS Photonics, \textbf{10}, 4367-4376. Copyright 2023 American Chemical Society.}
    \label{fig:4earlydryetching}
\end{figure}

Cao \textit{et al.}\cite{cao2021barium} reported a demonstration of thick BTO film growth on DSO substrates, followed by ICP dry etching to a depth of 170 nm in an 800 nm thick film. Using the Fabry-Perot method, a propagation loss of about 2 dB/cm was measured as shown in Fig. \ref{fig:4earlydryetching}(a) and \ref{fig:4earlydryetching}(b). Another demonstration of monolithic BTO ridge waveguides using ion milling was reported by Dong \textit{et al.} \cite{dong2023monolithic}. In their work, ridge geometries were defined by etching 175 nm in a 500 nm thick BTO film. By using the cut-back method, a propagation loss of 3.17 dB/cm was extracted, as shown in Fig. \ref{fig:4earlydryetching}(c). 

These etching-related challenges have limited the passive performance of monolithic BTO ridge waveguides and motivated the adoption of hybrid waveguide configurations, in which lateral confinement is provided by a different material while the BTO layer remains unetched. Nevertheless, recent advances in thin-film quality and etching processes have enabled some demonstrations of monolithic BTO ridge waveguides with significantly reduced propagation loss. 

\begin{figure}
    \centering
    \includegraphics[width=1\linewidth]{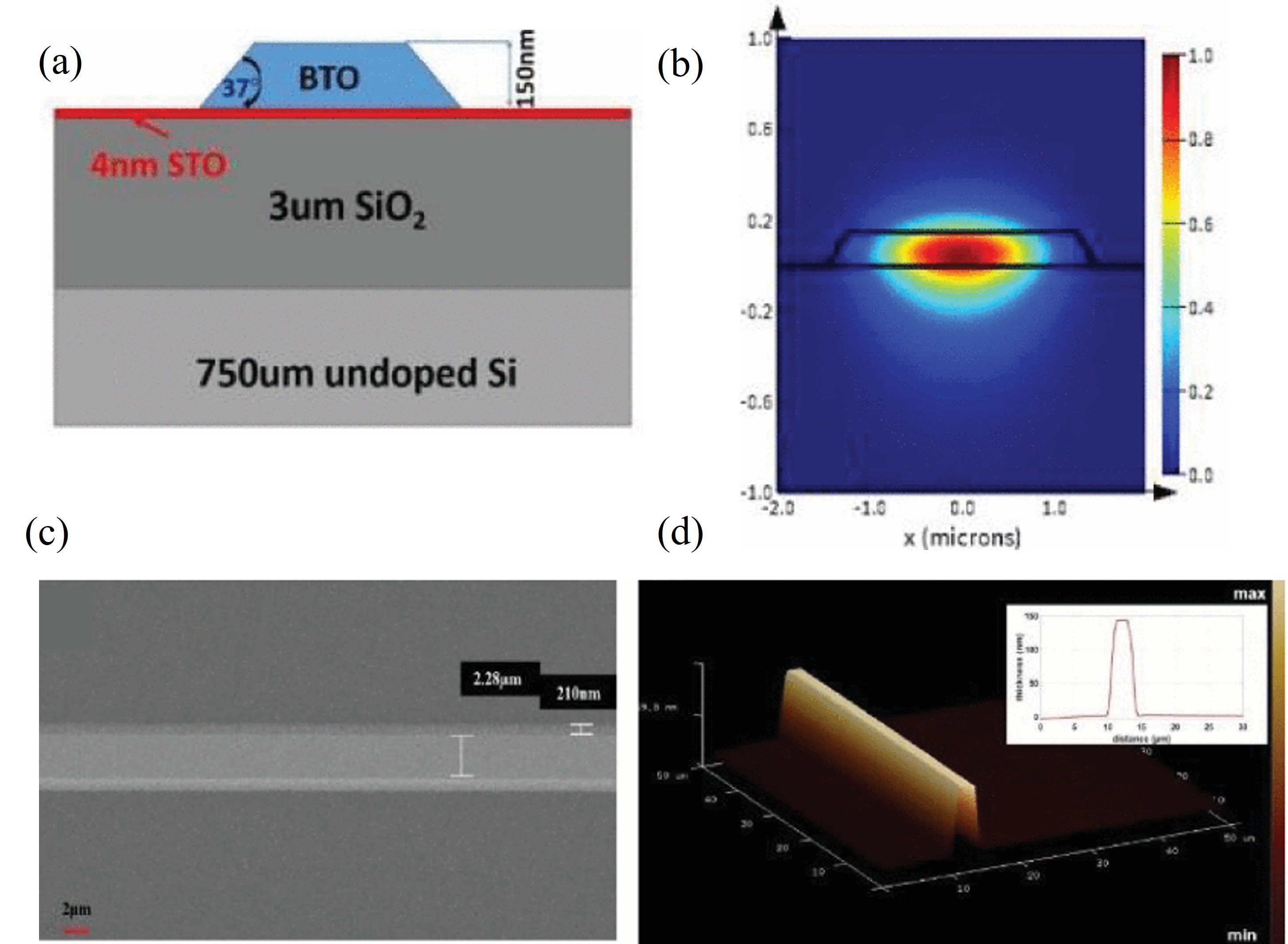}
    \caption{(a) Schematic of BTO ridge waveguide fabricated on SiO$_2$. (b) Optical mode simulation indicating that most of the mode overlap with BTO layer. (c) SEM and (d) AFM images of the ion-beam milled BTO ridge waveguide.\cite{isti2025fabrication} Reprinted with permission from IEEE Photonics Technology Letters. \copyright 2025 IEEE.
    }
    \label{fig:4isti2025}
\end{figure}

Isti \textit{et al.} \cite{isti2025fabrication} demonstrated low-loss BTO ridge waveguides fabricated using ion-beam milling. In their work, epitaxial BTO thin films were prepared on an SOI substrate with an STO buffer layer, and the BTO layer was fully etched to form ridge waveguides, as illustrated in Fig. \ref{fig:4isti2025}(a) and \ref{fig:4isti2025}(b). The resulting BTO ridge waveguide exhibit a sidewall angle of 37\degree and a sidewall rms roughness less than 2.5 nm. Using optical backscatter reflectometry, a propagation loss of 0.7 dB/cm was characterized, which is one of the lowest propagation losses reported for BTO-based waveguides.

\begin{figure}
    \centering
    \includegraphics[width=1\linewidth]{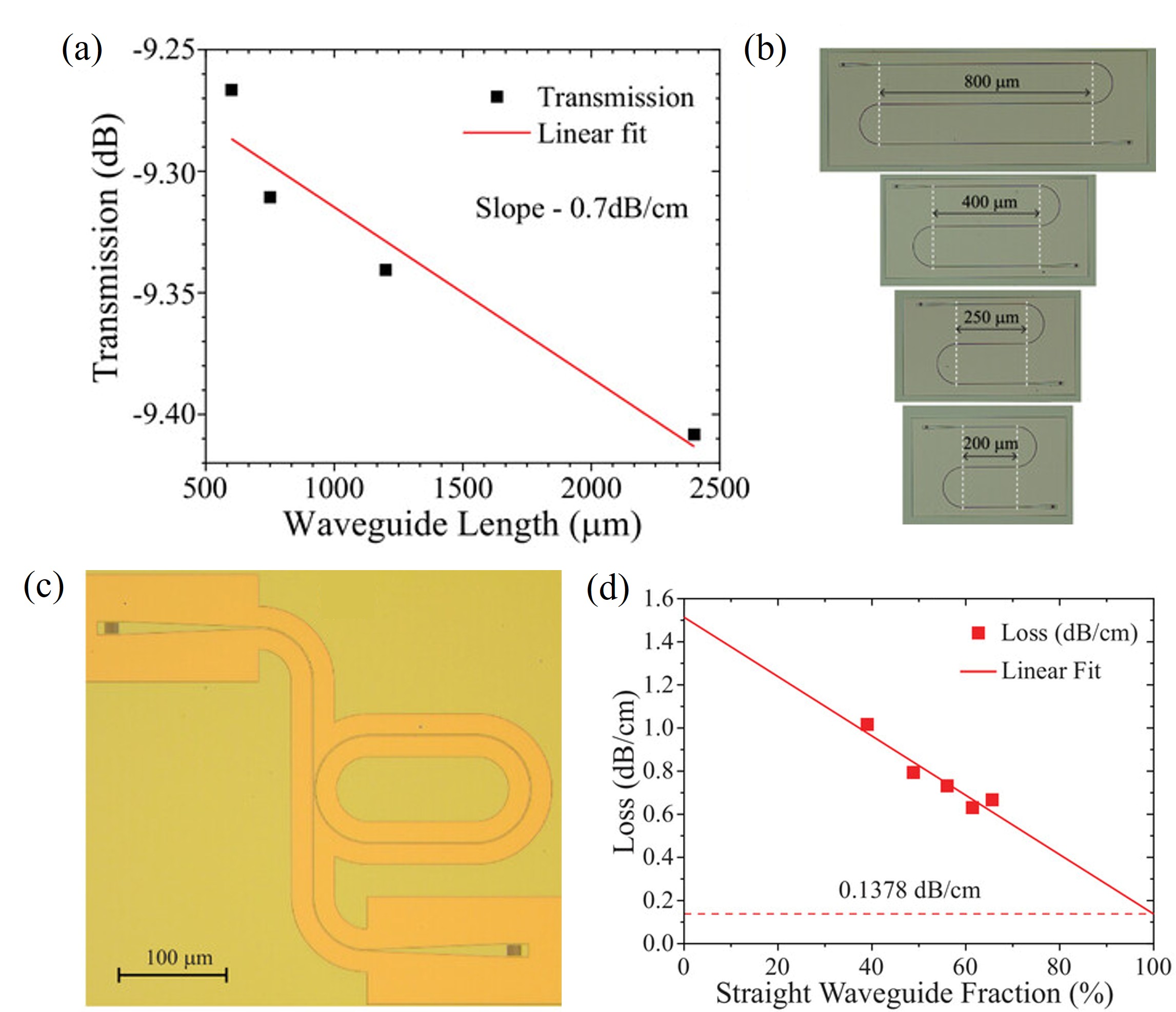}
    \caption{(a) Propagation loss measured using cut-back method. (b) Optical microscope images of ion-milled BTO ridge waveguides with different lengths. (c) Optical microscope image of the fabricated racetrack resonator. (d) Propagation loss extracted from the racetrack resonator structures. Reproduced with permission from Raju \textit{et al.}, Laser \& Photonics Reviews 19, 10 (2025). \copyright Wiley\cite{raju2025high}.}
    \label{fig:4raju2025}
\end{figure}

Another exceptional demonstration of monolithic BTO waveguide was reported by Raju \textit{et al.} \cite{raju2025high}. In their work, epitaxial BTO films were grown by RF sputtering on STO-buffered SOI substrates, and subsequently patterned using Ar-based ion milling with an etch depth of 100 nm in a 300 nm thick film. Using the cut-back method, a propagation loss of 0.7 dB/cm was obtained, as shown in Fig. \ref{fig:4raju2025}(a) and \ref{fig:4raju2025}(b). The ion-milled waveguides exhibited relatively steep sidewalls with an angle of approximately 75\degree. Furthermore, high-Q ring and racetrack resonators were demonstrated. For the racetrack resonators, an unloaded Q-factor of 5 $\times$ 10$^5$ was measured, and by varying the straight waveguide fraction, a propagation loss of 0.1378 dB/cm was obtained, as illustrated in Fig. \ref{fig:4raju2025}(c) and \ref{fig:4raju2025}(d). To the best of our knowledge, this represents the lowest propagation loss reported to date for both monolithic and hybrid BTO-based waveguides. 

Furthermore, more recently, Kim \textit{et al.} \cite{kim2025lowlossmonolithicbarium} reported the highest intrinsic Q-factor demonstrated to date in a monolithic thin-film BTO platform. In their work, a 340 nm-thick $a$-axis-oriented BTO-on-insulator film was patterned using Ar-based ion milling with a small fraction of Cl$_2$ included to mitigate sidewall redeposition during dry etching. The ridge waveguides had an etch depth of 180 nm at an etch rate of approximately 17.5 nm/min. The resulting waveguides exhibited a surface RMS roughness of 0.29 nm and sidewall angles of approximately 60\degree. As a results, an intrinsic Q-factor of 1.35 $\times$ 10$^{6}$ was measured in a micro-racetrack resonator, corresponding to a propagation loss of 0.32 dB/cm. Although this propagation loss is not the lowest reported among monolithic and hybrid BTO waveguides, the achieved Q-factor is the highest reported to date, which was attributed to the improved sidewall smoothness and verticality and optimized micro-racetrack resonator design.
Overall, these recent works indicate that low-loss monolithic BTO photonics is achievable under optimized growth and etching conditions.

Despite recent demonstrations of low-loss monolithic BTO ridge waveguides, most prior art were featured hybrid waveguide configurations to circumvent the difficulties associated with direct BTO patterning and etching. In such heterogeneous approaches, lateral confinement is achieved by adopting a different material to actually create strip-loaded waveguides, most commonly Si, amorphous Si, or SiN, while the BTO layer remains unpatterned. This strategy exploits the exceptional properties of BTO while leveraging well-established patterning techniques of the strip-loaded materials. However, in these strip-loaded geometries, the optical mode is primarily guided by the patterned strip and overlaps evanescently with the BTO film, which typically reduces the effective EO modulation efficiency. Here, the optical overlap factor refers to the fraction of the guided optical mode residing within the BTO layer, whereas the EO overlap factor additionally accounts for the overlap between the optical mode and the applied electric field within the BTO. As a result, the EO overlap factor more directly indicates the achievable modulation efficiency in practical device geometries.

In contrast, monolithic BTO ridge waveguides reported by Cao \textit{et al.}\cite{cao2021barium} and Lin \textit{et al.}\cite{Lin2024single} obtained optical overlap factors of 92\% and 75\%, respectively; most hybrid BTO-based waveguides exhibit overlap factors less than 50\%. Although EO overlap factors accounting for the electric-field distribution are not always reported, the difference in optical overlap alone indicates the typical advantage of monolithic waveguides over hybrid configurations. Nevertheless, based on the mature fabrication techniques of Si and SiN platforms, early hybrid waveguides typically enabled significantly lower propagation loss compared to monolithic BTO waveguides. Table \ref{tab:BTO_hybrid_overlap} summarizes representative hybrid BTO-based waveguides reported in the literature together with their corresponding optical or EO overlap factors.  It should also be mentioned that in addition to the geometric mode cross-sectional overlap with the BTO film, the uniformity and strength of the applied modulating field inside the BTO also play an important role in effective EO modulation. The field penetration into the BTO is highly dependent on the waveguide structure.

\begin{table*}[t!]
\caption{\label{tab:BTO_hybrid_overlap}
Summary of reported optical or EO overlap factors for hybrid strip-loaded waveguides. Reported propagation losses and resonator Q-factor are also included for comparison.
}
\begin{ruledtabular}
\begin{tabular}{c c c c c c}

\makecell{\textbf{Strip-loaded}\\ \textbf{material}} &
\textbf{Substrate / Integration Platform} &
\textbf{Year} &
\textbf{Overlap factor (\%)} &
\makecell{\textbf{Propagation loss}\\ \textbf{(dB/cm)}} &
\textbf{Q-factor} 
\\ \hline

SiN\cite{tang2004low} & BTO on MgO & 2004 & 59 (EO) & 1 & -- \\
Si\cite{xiong2014active} & BTO on SOI & 2014 & 22 (EO) & 44 & 7,000 \\
Si\cite{abel2016hybrid} & BTO on SOI & 2016 & 19.9 (O) & 47 & 4,600 \\
Si\cite{eltes2016low} & BTO on SOI & 2016 & 14 (O) & 6.3 & 20,000 \\
Si\cite{abel2019large} & BTO on SOI & 2019 & TE/TM : 39/55 (O) & 9 & 5,000 \\
Si\cite{eltes2019batio3} & BTO on SOI & 2019 & 38 (O) & 5.8 & 50,000 \\
SiN\cite{ortmann2019ultra} & BTO on SOI & 2019 & 18 (O) & 9.4 & -- \\
SiN\cite{eltes2020integrated} & BTO on SOI & 2020 & 18 (O) & 5.6 & 1,800 \\
Si\cite{zhang2024hybrid} & SOI with BTO cladding & 2024 & 15 (O) & 11.3 & 48,000 \\
SiN\cite{deng2026self-buffered} & BTO on LSAT & 2026 & 45 (EO) & -- & -- \\

\end{tabular}
\end{ruledtabular}
\footnotetext{O denotes optical mode overlap; EO denotes EO overlap reported from literature.}
\end{table*}

\begin{figure}
    \centering
    \includegraphics[width=0.9\linewidth]{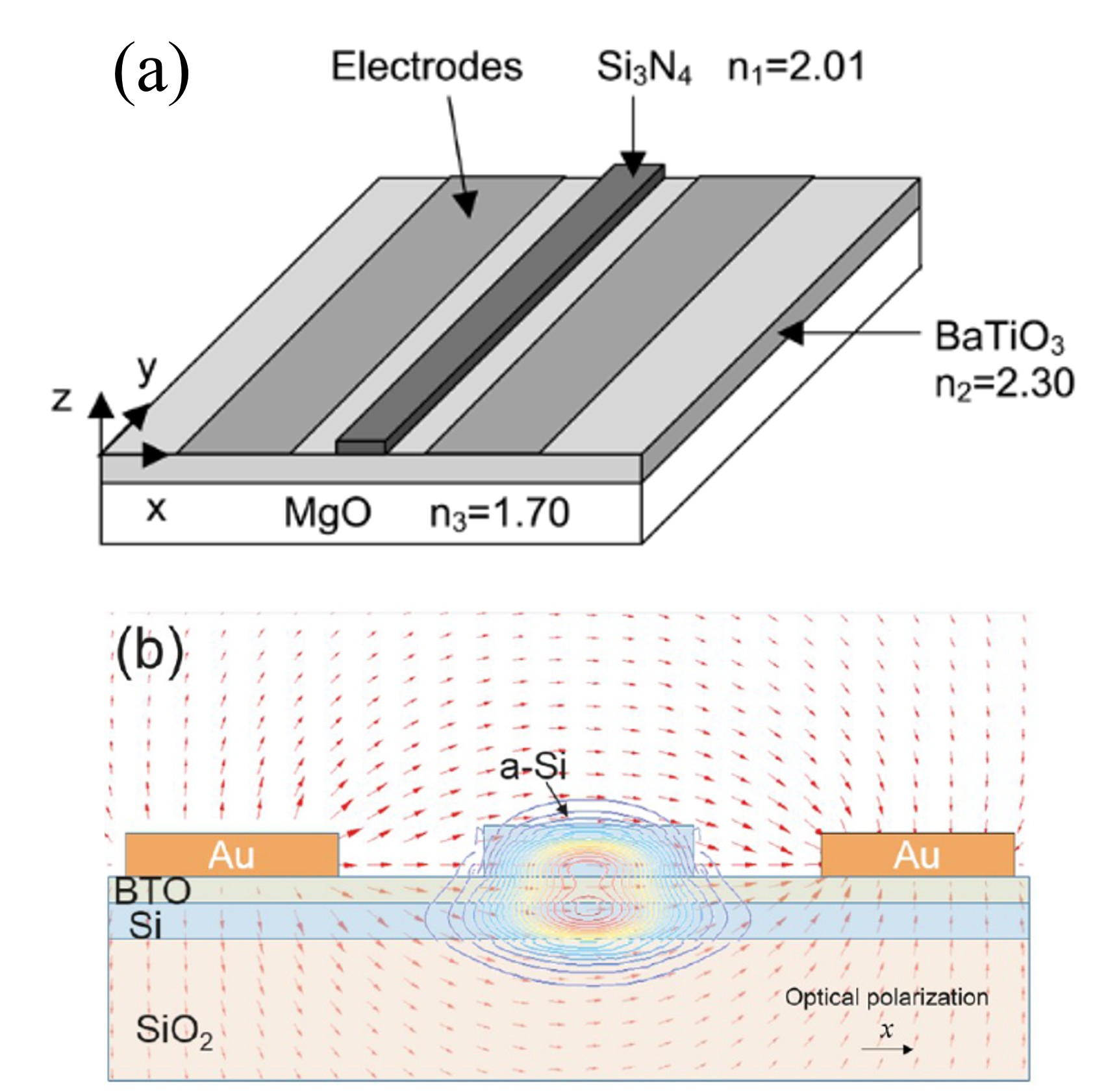}
    \caption{(a) Schematic of a SiN strip-loaded BTO waveguide on an MgO substrate \cite{tang2004low}. Reprinted with permission from IEEE Photonics Technology Letters. \copyright 2004 IEEE. (b) Schematic of a Si strip-loaded BTO waveguide on an SOI platform, in which a thin BTO layer is sandwiched between a bottom Si layer and a top Si strip to enhance EO overlap \cite{xiong2014active}. Reprinted with permission from Xiong \textit{et al.}, Nano letters 14, 1419–1425 (2014). Copyright 2014 American Chemical Society.
    }
    \label{fig:4tangxiong}
\end{figure}

To the best of our knowledge, Tang \textit{et al.} \cite{tang2004low} reported the first demonstration of a heterogeneously integrated hybrid BTO-based waveguide in 2004. In their work, a SiN strip with a width of 4 µm and a thickness of 125 nm was patterned on a 560 nm thick BTO film grown on an MgO substrate, as illustrated in Fig. \ref{fig:4tangxiong}(a). The resulting hybrid SiN–BTO waveguide exhibited propagation losses of 1 and 1.1 dB/cm for TE and TM polarizations, respectively. In addition, the authors analyzed the overlap between the optical mode and the applied electric field, which is 59\%, an important parameter that determines the effective EO response of the device. Later, Xiong \textit{et al.} \cite{xiong2014active} demonstrated strip-loaded Si ridge waveguides on BTO thin films integrated on an SOI platform. Because silicon has a much higher refractive index than BTO, direct strip-loading without careful design can significantly reduce the optical mode overlap with the BTO layer. To address this issue, the authors designed a sandwich configuration in which a thin BTO layer was placed between a bottom Si layer and a top Si strip, as illustrated in Fig. \ref{fig:4tangxiong}(b). As a result, an EO overlap factor of 22\% was achieved, and resonators fabricated using this hybrid Si–BTO waveguide exhibited a loaded Q factor of 7000, from which a propagation loss of 44 dB/cm was extracted.

\begin{figure}
    \centering
    \includegraphics[width=1\linewidth]{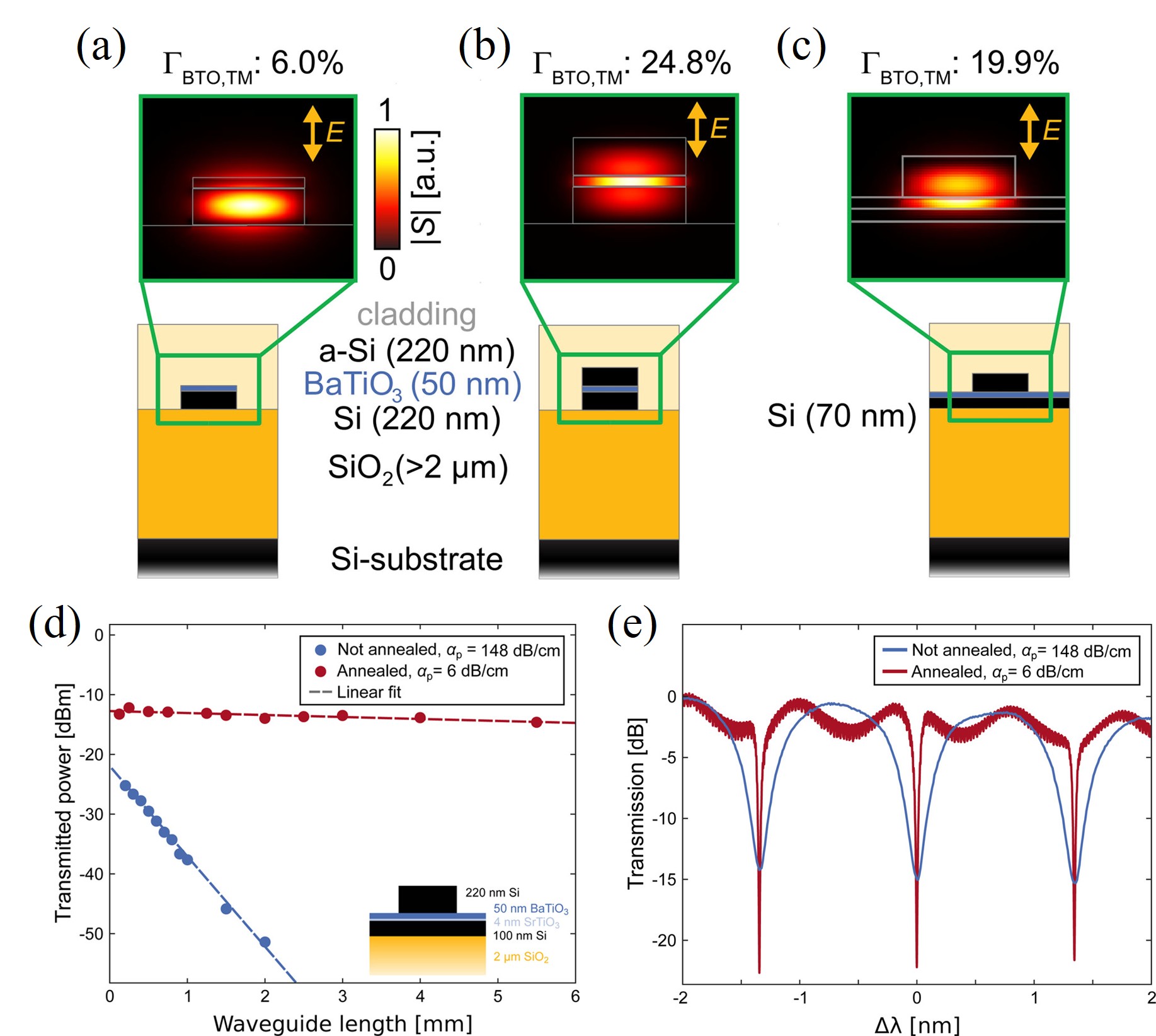}
    \caption{(a)-(c) Schematic illustrations of different hybrid Si–BTO waveguide configurations investigated by Abel \textit{et al.}\cite{abel2016hybrid}. Reprinted with permission from Journal of Lightwave Technology. \copyright 2016 IEEE. (d) Propagation loss as a function of waveguide length before and after annealing (e) Transmission spectra of ring resonators fabricated using the hybrid Si–BTO waveguides\cite{eltes2016low}. 
    }
    \label{fig:4abelelteshybrid}
\end{figure}

Abel \textit{et al.}\cite{abel2016hybrid} also reported demonstration of hybrid Si-BTO. In this study, the authors explored different waveguide configurations to identify a desirable trade-off between effective optical confinement in the BTO layer and fabrication practicality, as shown in Fig. \ref{fig:4abelelteshybrid}(a)-\ref{fig:4abelelteshybrid}(c). A Si ridge waveguide with BTO used as the cladding exhibits a relatively low optical overlap with the BTO layer, while a fully etched Si–BTO–Si sandwich configuration provides a significantly enhanced optical overlap factor of 24.8\%. However, this fully etched geometry requires direct BTO patterning, which can lead to rough sidewalls and increased scattering loss. Additionally, the configuration increases the difficulty of placing electrodes parallel to the waveguide.

To address these challenges, the authors adopted a configuration similar to that reported by Xiong \textit{et al.} \cite{xiong2014active}, in which a Si strip waveguide is patterned on top of an unetched BTO thin film integrated on an SOI platform, as illustrated in Fig. \ref{fig:4abelelteshybrid}(c). This configuration achieves an optical overlap of approximately 20\% within the BTO layer while avoiding direct BTO etching. Using this hybrid waveguide platform, ring resonators were fabricated, exhibiting a loaded Q factor of 4,600 and a propagation loss of 47 dB/cm for the TE polarization. From the same group, Eltes \textit{et al.} \cite{eltes2016low} further investigated the cause of the high propagation loss observed in such hybrid Si–BTO waveguides. They identified significant optical absorption in the underlying STO seed layer induced by hydrogen exposure during processing. By introducing a post-fabrication annealing process, the authors mitigated this absorption, resulting in a substantially reduced propagation loss of approximately 6 dB/cm and an increased resonator Q factor exceeding 20,000, as shown in Fig. \ref{fig:4abelelteshybrid}(d) and \ref{fig:4abelelteshybrid}(e).

\begin{figure}
    \centering
    \includegraphics[width=1\linewidth]{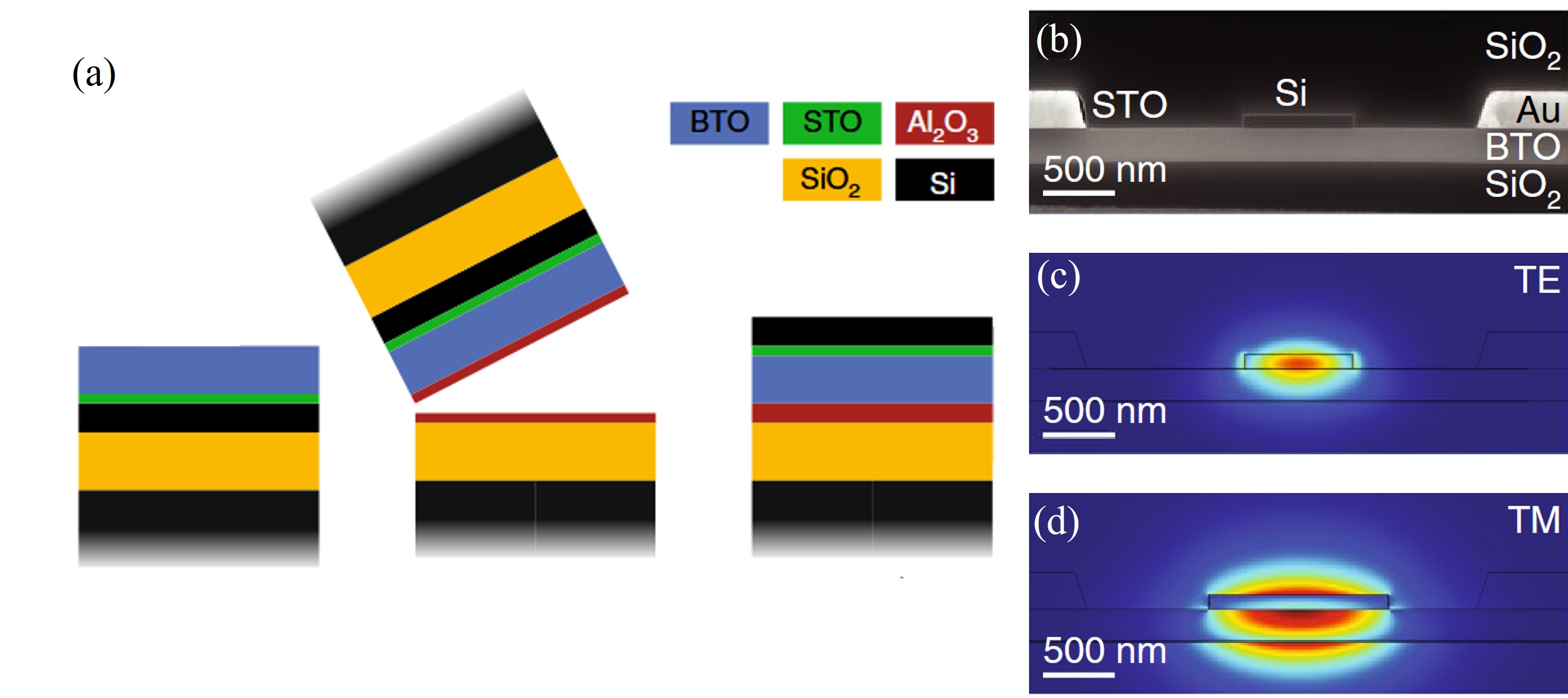}
    \caption{(a) Schematic of the wafer bonding and substrate removal processes used to fabricate a hybrid Si–BTO platform, resulting in a Si/STO/BTO/Al$_2$O$_3$/SiO$_2$ stack. (b) Cross-sectional image of the fabricated hybrid Si–BTO waveguide. Simulated optical mode profiles for the fundamental (c) TE and (d) TM polarizations, respectively. Reproduced with permission from Abel \textit{et al.}, Nature Materials 18, 42–47 (2019). Copyright 2018 Springer Customer Service Center GmbH\cite{abel2019large}.
    }
    \label{fig:4able2019hybrid}
\end{figure}

Later, Abel \textit{et al.}\cite{abel2019large} reported a hybrid Si strp-loaded waveguide platform on single-crystalline BTO integration on silicon. In this approach, epitaxial BTO thin films were first grown on STO-buffered SOI substrates using MBE, followed by wafer bonding to transfer the BTO/STO/Si stack onto a SiO$_2$ carrier wafer using an Al$_2$O$_3$ bonding interface. After removal of the donor wafer by grinding and selective wet etching, the final stack consisted of Si/STO/BTO/Al$_2$O$_3$/SiO$_2$/Si, as illustrated in Fig. \ref{fig:4able2019hybrid}(a) and \ref{fig:4able2019hybrid}(b). Silicon strip waveguides were then patterned on top of the BTO layer to form hybrid waveguides without direct BTO etching. In this configuration, the optical mode overlap within the BTO layer reached approximately 39\% and 55\% for TE and TM polarizations, respectively, as shown in Fig. \ref{fig:4able2019hybrid}(c) and \ref{fig:4able2019hybrid}(d). Ring resonators fabricated using this platform exhibited a loaded Q factor of about 5,000, corresponding to propagation losses of 9 dB/cm for TM polarization.
In a follow-up study, Eltes \textit{et al.}\cite{eltes2019batio3} employed a similar hybrid Si–BTO waveguide design but with further optimization of the Si waveguide fabrication process which allowed an improvement in sidewall roughness and propagation loss. A ring resonator with a radius of 30 $\mu$m exhibited a Q factor of up to 50,000, corresponding to an extracted propagation loss of 5.8 dB/cm.

More recently, Alexander \textit{et al.} reported a manufacturable hybrid BTO-based photonic platform for large-scale quantum photonic integration \cite{psiquantum_team2025manufacturable}. In this platform, EO phase shifting is realized by integrating a thin BTO film with SiN strip waveguides embedded in a SiO$_2$ cladding. Benefiting from the low-loss SiN platform, this hybrid configuration demonstrated, to the best of our knowledge, the lowest reported propagation loss among hybrid BTO-based waveguides, reaching approximately 0.53 dB/cm.

\begin{figure}
    \centering
    \includegraphics[width=0.8\linewidth]{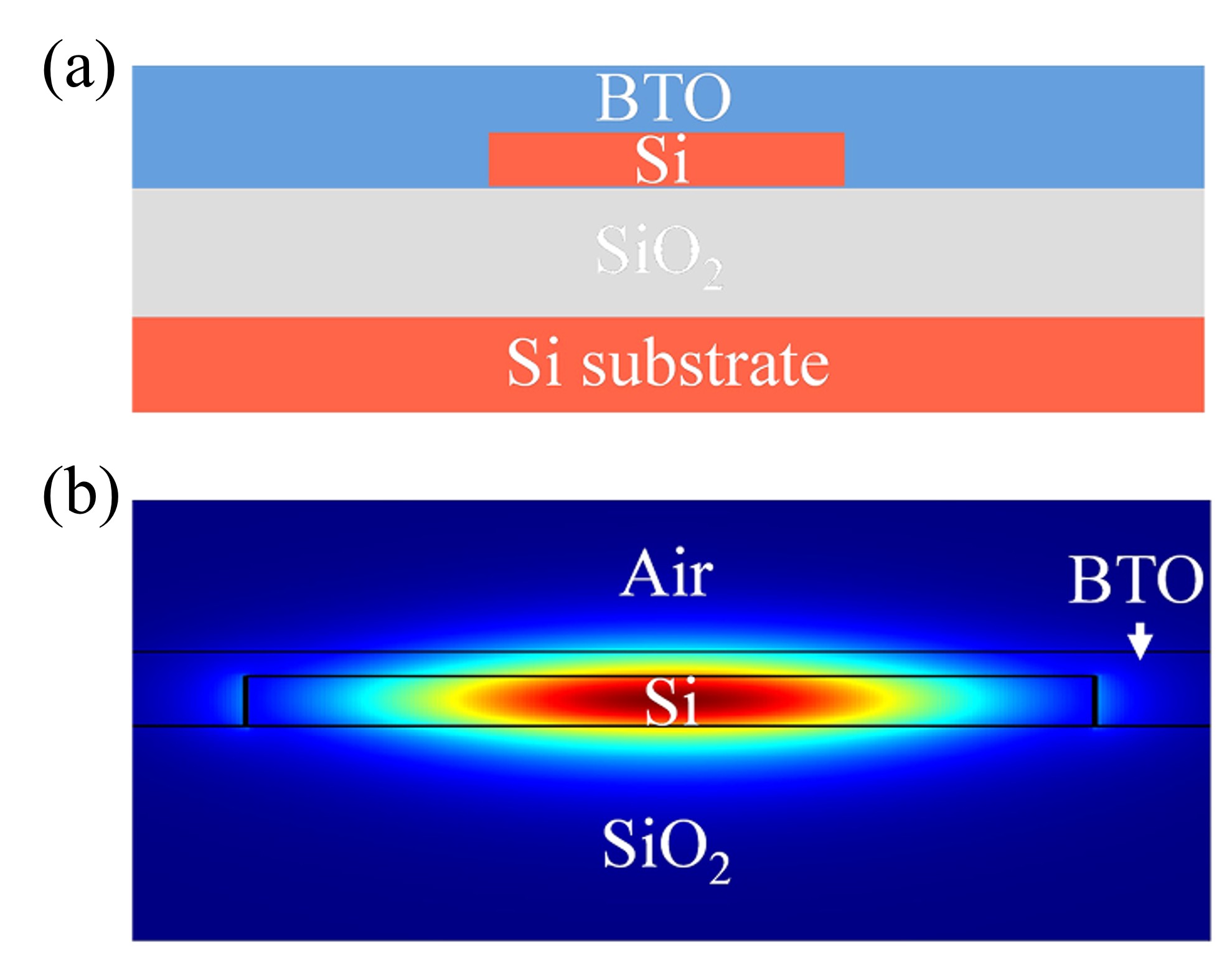}
    \caption{(a) Cross-sectional schematic of a hybrid Si-BTO waveguide fabricated by CSD, where the BTO thin film serves as the upper cladding of an SOI waveguide. (b) Simulated optical mode profile of the hybrid waveguide, indicating that most of the optical mode is confined within the Si core. Reproduced with permission from Zhang \textit{et al.}, Advanced Photonics Nexus 3, 066005–066005 (2024). Copyright 2024 authors, licensed under a Creative Common Attribution (CC BY) license\cite{zhang2024hybrid}.}
    \label{fig:4zhanghybrid}
\end{figure}

In contrast to vacuum-deposited BTO films, Zhang \textit{et al.}\cite{zhang2024hybrid} reported a hybrid waveguide configuration in which BTO served as a cladding layer on Si ridge waveguides using the CSD method. In this approach, BTO thin films were spin-coated and crystallized by thermal annealing to cover the patterned SOI waveguides. Due to the higher refractive index of Si compared to BTO, the optical mode overlap within the BTO layer was limited to approximately 15\%, as illustrated in Fig. \ref{fig:4zhanghybrid}(a) and \ref{fig:4zhanghybrid}(b). The resulting hybrid waveguides exhibited a propagation loss of 11.3 dB/cm, higher than that of bare silicon waveguides, which can be attributed to the relatively lower crystalline quality and increased scattering from the CSD-grown BTO layer. Nevertheless, racetrack resonators fabricated on this platform achieved Q-factors up to 48,000.

\subsection{\label{active EOM}Active electro-optic modulators based on BTO thin films}

\begin{table*}[t!]
\caption{\label{tab:BTO_EO_summary}
Summary of representative active devices in BTO thin films}
\begin{ruledtabular}
\begin{tabular}{c c c c c c}

\textbf{Implementation} &
\textbf{Platform} &
\textbf{BTO orientation} &
\makecell{\textbf{EO coefficient}\\ \textbf{ (pm/V)}}  &
\textbf{$V_\pi L$ (V$\cdot$cm)} &
\textbf{3dB EO bandwidth}
\\ \hline

Film\cite{cao2023characterization} & BTO/SRO/DSO & c & $r_{13}$ = 4.2; $r_{23}$ = 4.6; $r_{33}$ = 9 & -- & -- \\

Film\cite{leroy2013guided} & BTO/ITO/MgO & polycrystalline & $r_{13}$ = 18; $r_{33}$ = 23 & -- & -- \\

Film\cite{abel2013strong} & BTO/SOI & a & $r_{\text{eff}}$ = 148; $r_{42}$ = 105 & -- & -- \\

Film\cite{edmondson2020epitaxial} & BTO/Si & a & $r_{\text{eff}}$ = 27 & -- & -- \\

Film\cite{lin2019atomic} & BTO/Si & c & $r_{\text{eff}}$ = 26 & -- & -- \\

Monolithic\cite{gill1997thin} & BTO/MgO & a & $r_{\text{eff}}$ = 50 & 1.35 & -- \\

Monolithic\cite{petraru2002ferroelectric} & BTO/MgO & \makecell{{a}\\{c}} &
\makecell{{$r_{\text{eff}}$ = 734 @ 632 nm}\\ {$r_{\text{eff}}$ = 80 @ 632 nm}} & \makecell{{ 1.89 @ 632 nm; 2.85 @ 1550 nm}\\{2.4 @ 632 nm; 4.5 @ 1550 nm}} & 
\makecell{{--}\\{1 MHz @ 1550 nm}} \\

Monolithic\cite{petraru2003integrated} & BTO/MgO & polycrystalline & $r_{\text{eff}}$ = 22 & -- & 1 MHz \\

Monolithic\cite{kim2014ridge} & BTO/MgO & c & $r_{51}$ = 110 & -- & -- \\

Monolithic\cite{cao2023active} & BTO/DSO & c & $r_{42}$ = 600 & -- & -- \\

Monolithic\cite{dong2023monolithic} & BTO/STO/SiO$_2$ & a & $r_{\text{eff}}$ = 89 & 2.32 & -- \\

Monolithic\cite{posadas2023rf} & BTO/STO/SOI & c & $r_{33}$ = 134.4 & 10.55 & 10 MHz \\

Monolithic\cite{winiger2024pld} & BTO/MgO & mixed a/c & $r_{42}$ = 1030 & -- & -- \\

Monolithic\cite{chelladurai2025barium} & BTO/SiO$_2$ & a & $r_{42}$ = 481; $r_{33}$ = 125 & -- & -- \\

Monolithic\cite{kim2025lowlossmonolithicbarium} & BTO/SiO$_2$ & a & $r_{eff}$ = 162 & 0.54 & -- \\

Hybrid\cite{tang2004low} & SiN/BTO/MgO & c & $r_{\text{eff}}$ = 38 & 4.5 & -- \\

Hybrid\cite{tang2004electrooptic} & SiN/BTO/MgO & c & $r_{\text{eff}}$ = 150 & 1.25 & 3.7 GHz \\

Hybrid\cite{tang2004low-voltage} & SiN/BTO/MgO & c & $r_{\text{eff}}$ = 162 & 1.1 & -- \\

Hybrid\cite{xiong2014active} & Si/BTO/SOI & a & $r_{\text{eff}}$ = 213 & 1.5 & 4.9 Ghz \\

Hybrid\cite{abel2016hybrid} & Si/BTO/SOI & mixed a/c & $r_{\text{eff}}$ = 300 & 1.35 & -- \\

Hybrid\cite{abel2019large} & Si/BTO/SiO$_2$ & a & $r_{42}$ = 923; $r_{42}$ = 342 & 0.45 & 
\makecell{{65 GHz (Plasmonic)}\\{30 GHz (Photonic)}}\\

Hybrid\cite{eltes2019batio3} & Si/BTO/SiO$_2$ & a & $r_{\text{eff}}$ = 380 & 0.23 & 2 GHz \\

Hybrid\cite{ortmann2019ultra} & SiN/BTO/SiO$_2$ & a & $r_{\text{eff}}$ = 343 & 0.3 & -- \\

Hybrid\cite{eltes2020integrated} & Si/BTO/SiO$_2$ & a & $r_{\text{eff}}$ = 700 & 5 & -- \\

Hybrid\cite{posadas2021thick} & SiN/BTO/SOI & a & $r_{\text{eff}}$ = 183 & 0.421 & -- \\

Hybrid\cite{reynaud2023si} & BTO/SOI & a & $r_{\text{eff}}$ = 119 & -- & -- \\

Hybrid\cite{eltes2023thin-film} & SiN/BTO/SiO$_2$ & a & -- & 0.48 & -- \\

Hybrid\cite{lievens2024integration} & BTO/SOI & a & $r_{\text{eff}}$ = 170 & 9.9 & -- \\

Hybrid\cite{yu2024tuning} & BTO/LSAT & a & $r_{\text{eff}}$ = 175 & -- & -- \\

Hybrid\cite{wen2024enhanced} & BTO/STO & a & $r_{\text{eff}}$ = 286 & -- & -- \\

Hybrid\cite{zhang2024hybrid} & BTO/SOI & polycrystalline & $r_{\text{eff}}$ = 27.2 & 4.51 & -- \\

Hybrid\cite{chrysostomidis2025ultra} & BTO/SOI & mixed a/c & $r_{\text{eff}}$ = 783 & 2.52 & -- \\

Hybrid\cite{psiquantum_team2025manufacturable} & SiN/BTO & mixed a/c & $r_{\text{eff}}$ = 1000 & 0.77 & 6.9 GHz \\

Hybrid\cite{reynaud2022microstructural} & SiO$_2$/BTO/SOI & mixed a/c & $r_{\text{eff}}$ = 352 & 1.5 & -- \\

Hybrid\cite{kohli2025plasmonic} & SiO$_2$/BTO/SiN & a & $r_{\text{eff}}$ $\sim180$ & 0.0027 & 110 GHz \\

Hybrid\cite{deng2026self-buffered} & SiN/BTO/LSAT & mixed a/c & $r_{42}$ = 358 & 0.7 & 12 GHz \\

\end{tabular}
\end{ruledtabular}
\end{table*}

The ultimate performance of integrated photonic devices is determined by both the passive performance, and more critically, their EO response. Due to the extraordinary nonlinear optical properties of BTO, it has attracted significant interest as a platform for achieving superior EO performance compared with other state-of-the-art integrated photonic materials. While a variety of devices exploit the EO effect, this section focuses specifically on optical modulators, in which EO-induced refractive-index changes are used to modulate optical signals.

Table \ref{tab:BTO_EO_summary} summarizes representative EO modulators demonstrated on thin-film BTO-based platforms. In EO modulators, the voltage-length product ($V_\pi L$) is a key figure of merit that indicate modulation efficiency. Here, $V_\pi$ is the have-wave voltage required to induce a $\pi$ phase shift, and $L$ is the interaction length of the modulation region, typically defined by the electrode length. Although $V_\pi$ is dependent on the device length, the product $V_\pi L$ provides a figure of merit that enables a direct comparison between different modulator designs. A lower $V_\pi L$ indicates a stronger EO interaction and more efficient modulation. In addition to modulation efficiency, high-speed operation is a critical requirement for practical EO devices. The EO bandwidth is mainly characterized by the 3-dB EO bandwidth, defined as the frequency at which the modulated output drops to half  its low-frequency value.


In thin-film BTO-based devices, it is crucial to retain bulk-like EO properties, particularly the excellent $r_{42}$ or $r_{51}$ parameters, which are also included in Table \ref{tab:BTO_EO_summary}. For device-level optical modulation demonstrations, the EO parameters are typically extracted from modulation performance. For comparison, we also include EO coefficients reported from thin-film material studies, in which EO parameters are more directly measured from the films themselves. As introduced in Section \ref{EOC}, the EO coefficients extracted from BTO strongly depend on the crystallographic orientation of the film, the polarization of the optical mode, and the direction of the applied electric field. Therefore, controlling the film orientation is crucial, as the accessible EO coefficients and modulation differ significantly between $a$-axis- and $c$-axis-oriented films. 

Unlike EO coefficients such as $r_{13}$ and $r_{33}$, which only induce refractive index changes along the principal crystal axes, the off-diagonal components indicated by $r_{42}$ and $r_{51}$ lead to changes in the off-diagonal components of the permittivity tensor. As a result, EO modulation based on these coefficients can induce not only phase modulation but also polarization rotation.  This occurs because diagonalization of a permittivity tensor with non-zero off-diagonal components is effectively a rotation of the optic axes.  Propagating light which is well-polarized along a (non-rotated) basis vector may generally have field components that overlap more than one of the principal axes in the diagonal (rotated) basis; the light effectively sees a birefringent crystal in such a situation.

In $a$-axis-oriented BTO thin films, the crystallographic $c$-axis lies within the plane of the film. This geometry is attractive for integrated photonics because it enables the $r_{42}$ coefficient by applying an in-plane electric field using lateral electrodes, leading to changes in the in-plane effective refractive index for TE-polarized light. In this configuration, optic axis rotation induced by $r_{42}$ remains confined within the film's plane and does not result in TE-TM mode conversion. (The basis vectors rotate about the $a$ axis for a $y$-oriented applied field.) However, $a$-axis-oriented BTO films grown on common substrates typically exhibit multi-domain in-plane structures, in which the local $c$-axis orientation varies across the film. As a result, the in-plane crystallographic axes consist of mixed $\pm$$a$ and $\pm$$c$ domains, as reported in several studies.

For such multi-domain films, maximizing the EO response is crucial, as the EO contributions from differently oriented domains can partially cancel each other, leading to a significantly reduced net modulation efficiency. To address this issue, numerous studies have demonstrated that the highest EO efficiency is achieved when the in-plane optical propagation direction is aligned at 45\degree with respective to the crystallographic $a$- and $c$-axes\cite{abel2013strong,abel2019large,chelladurai2025barium}. In this configuration, pre-poling or a constant bias is typically required to align the in-plane domains, after which the applied electric field can effectively utilize the $r_{42}$ coefficient. With lateral electrodes along such a waveguide, the modulating field strength is effectively reduced by a factor of $\sqrt{2}$, as is the poling field which uses the same electrodes. As a result, the performance of such EO phase shifters or modulators strongly depends on the poling or bias voltage used to align the initially disordered in-plane domains. Under this condition, EO contribution from the poled domains add constructively.

It is worth noting that, in this configuration, the EO coefficient extracted from the device measurements represents an effective value, $r_{\text{eff}}$, which incorporates contributions from multiple tensor elements and is also usually stated with respect to the applied field, i.e., $r_{\text{eff}}$ acts on $E_{\textit{applied}}$. This effective coefficient treats the net EO response as a pure phase modulation and neglects the impact of polarization rotation arising from off-diagonal tensor components. This 45\degree alignment strategy has therefore been widely adopted in in-plane multi-domain BTO-based EO devices to effectively exploit the large intrinsic $r_{42}$ coefficient despite the presence of domain disorder.  Care must be taken in comparing various devices in literature, however, because although $r_{42}$ can be mathematically extracted from $r_{\text{eff}}$, the device geometry leading to the measurement may not permit the full $r_{42}$ to be exploited.  In such cases, $r_{42}$ can be thought of as a proxy of the BTO's material quality, and $r_{\textit{eff}}$ represents what a modulator can actually use.

In contrast, $c$-axis-oriented BTO thin films, in which the $c$-axis is normal to the film plane, exhibit a fundamentally different EO response compared to the $a$-axis-oriented films. In this orientation, for a single-domain BTO film, an in-plane electric field only activates the $r_{42}$ (or equivalently, $r_{51}$) coefficient, and the induced off-diagonal components of the permittivity tensor inescapably leads to both refractive index modulation and polarization rotation. Consequently, an input TE-polarized optical mode can be partially or fully converted to a TM-polarized mode during propagation in a waveguide upon application of an electric field from coplanar waveguides, or vice versa. While this configuration can fully utilize the large $r_{42}$ coefficient without degradation from in-plane domain disorder, the accompanying polarization rotation can complicate device operation, particularly in phase shifters and interferometric modulator configurations where pure phase modulation is preferred. It is worth noting that for EO devices driven solely by the $r_{42}$ coefficient, the simultaneous occurrence of optic axis rotation and direction-sensitive refractive index change means that a well-defined $V_\pi$ is generally not applicable, since the EO modulation is not purely based on a simple change in refractive index.  In other words, applying a voltage $V$ can induce birefringence rather than a simple index change.

With the orientation-dependent EO mechanisms of BTO clarified, we now review some representative experimental demonstrations of EO response in thin-film BTO, starting from film-level measurements to integrated EO modulators based on both $c$-axis- and $a$-axis-oriented films.



\begin{figure}
    \centering
    \includegraphics[width=1\linewidth]{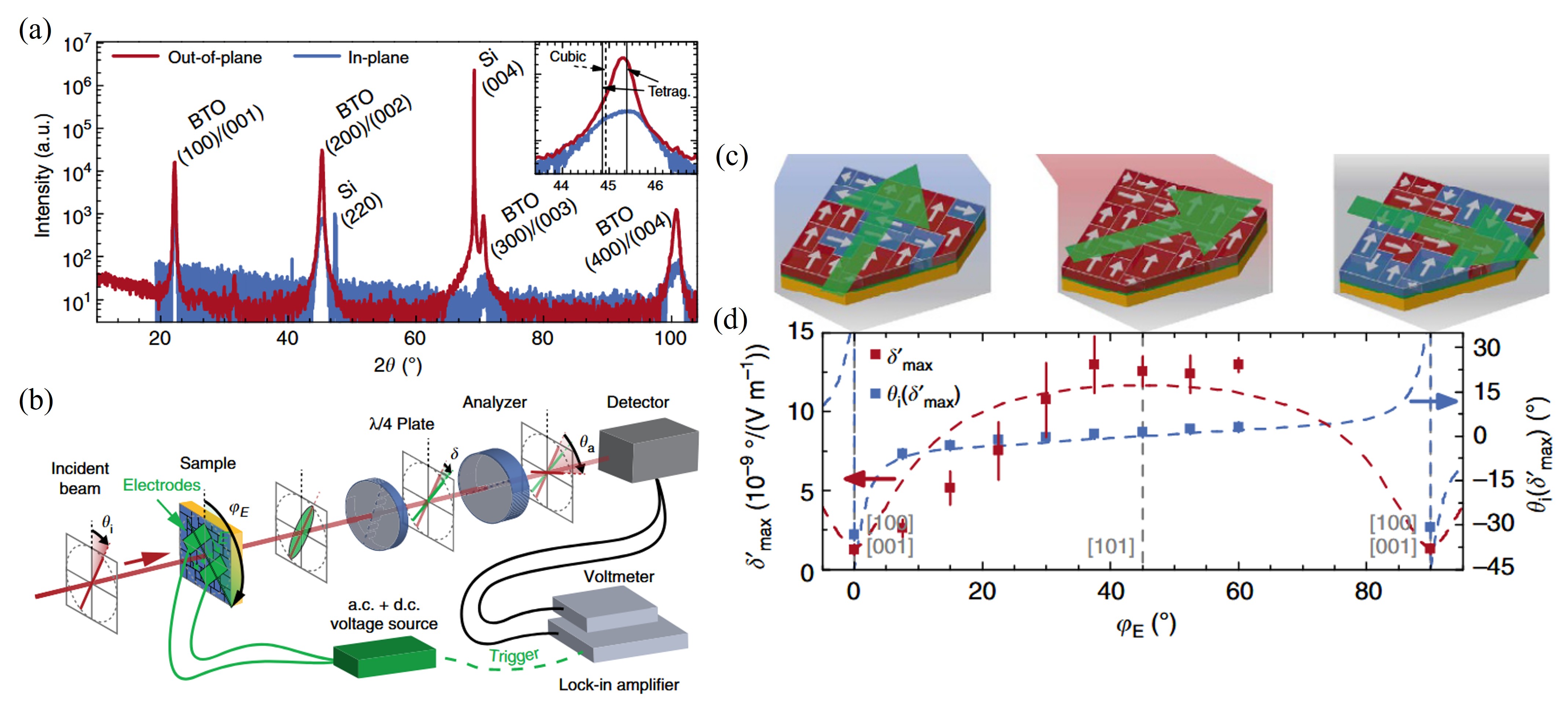}
    \caption{(a) Out-of-plane and in-plane XRD scans of an MBE-grown BTO thin film on Si, indicating a multi-domain ferroelectric structure. (b) Schematic of the free-space optical setup used for EO coefficient characterization of the BTO thin film, incorporating polarization analysis to detect the EO-induced birefringence changes. (c) Schematic illustration of different orientations of the applied electric field relative to the crystal axes of the multi-domain BTO film, indicating that when the electric field is aligned with a principal crystal axis, only a fraction of domains can be flipped (poled). (d) Maximum EO response quantified by the derivative of the EO-induced rotation angle ($\delta\;'$), and the corresponding incident polarization angle ($\theta_i$) as a function of orientation of the electric field ($\varphi_E$), indicating that the largest EO response is achieved when the electric field is oriented at 45\degree with respect to the in-plane crystal axes. Reproduced with permission from Abel et al., Nature Communications 4, 1671 (2013), Springer Nature \cite{abel2013strong}.}
    \label{fig:4abel2013}
\end{figure}

For film-level EO characterization, Abel \textit{et al.} \cite{abel2013strong} reported one of the earliest and comprehensive demonstrations of a strong EO response in epitaxial BTO thin films integrated on Si, providing an important framework for understanding and extracting EO coefficients in multi-domain ferroelectric thin films. In this work, epitaxial BTO films were grown on STO-buffered Si substrates by MBE. In-plane XRD measurements indicated a multi-domain ferroelectric structure in the BTO films, as illustrated in Fig. \ref{fig:4abel2013}(a).

To extract the EO coefficients of the thin-film BTO, the authors employed a free-space optical measurement setup that accounts for the multi-domain ferroelectric nature and domain flipping under an applied electric field, as shown in Fig. \ref{fig:4abel2013}(b). In this setup, linearly polarized light propagates through the BTO film with electrodes, where the Pockels effect induces a field-dependent change in birefringence. As a result, the transmitted light becomes slightly elliptically polarized. The beam then passes through a quarter-wave plate to convert the polarization state back to linear polarization, followed by a rotating analyzer and photodetector to record the optical intensity variation. By monitoring the intensity modulation as a function of analyzer angle, the EO-induced birefringence changes can be recorded.

The EO response is expressed in terms of the field-dependent rotation angle derivative, $\delta'$, which reflects the change in optical anisotropy induced by the applied electric field. Prior to measurement, the multi-domain BTO films were pre-poled at 40 V for 1 min, and the induced ferroelectric domain alignment was shown to be stable over several days. Two distinct electric-field orientations were investigated, as illustrated in Fig. \ref{fig:4abel2013}(c). When the electric field was align along the (001)$_{\textit{BTO}}$ direction, the EO response was primarily attributed to the $r_{13}$ and $r_{33}$ coefficients. In contrast, when the electric field was oriented along the (101)$_{\textit{BTO}}$, the EO response was dominated by the $r_{42}$ coefficient, and an $r_{42}$ value of 105 pm/V was extracted. Additionally, the authors also extracted an effective EO coefficient $r_{\text{eff}}$ of 148 pm/V by treating the modulation as a pure refractive-index change and neglecting the tensorial nature of EO response.

Figure \ref{fig:4abel2013}(d) further shows that in multi-domain $a$-axis-oriented BTO thin films, aligning the electric field ($\varphi_E$) at 45\degree relative to the in-plane crystal axes enables the maximum EO response when the incident optical polarization angle ($\theta_i$) is close to 0\degree. In contrast, alignment along a principal crystal axis leads to partial cancellation of the net EO response due to competing contributions from differently oriented domains.

Overall, this work by Abel \textit{et al.} demonstrated a robust free-space methodology for extracting EO coefficients in ferroelectric thin films and showed that the measured EO response depends on the relative orientation between the incident optical polarization, the applied electric field, and the crystal axes. It provided one the earliest explanations for the now widely adopted 45\degree alignment strategy in multi-domain BTO-based EO devices. Compared with other early studies that focused on the effective EO coefficient by analyzing only net phase shifts and neglecting the tensor nature of the EO response, this work clearly distinguished between intrinsic tensor elements and effective coefficients, offering a comprehensive framework for the design of thin-film BTO EO modulators.

Such a free-space measurement setup has been adopted in subsequent studies. For instance, Posadas \textit{et al.}\cite{posadas2021thick} reported thick BTO films (up to $\sim$ 1 $\mu$m) grown on STO-buffered SOI substrates, which exhibit multi-domain in-plane ferroelectric structures. The authors adopted the free-space EO characterization approach introduced by Abel \textit{et al.}\cite{abel2013strong} to evaluate the EO response of the deposited BTO films. Using this method, an effective EO coefficient of up to 183 pm/V was extracted. In addition, the authors fabricated hybrid a SiN/BTO MZI to extract EO parameters at the device level. From the devices, a comparable effective EO coefficient of 157.5 pm/V and a $V_{\pi}L$ of 0.421 V$\cdot$cm were obtained. This work demonstrates that the free-space EO characterization approach introduced by Abel \textit{et al.} provides a transferable and reliable method for evaluating EO performance across different BTO growth techniques.

\begin{figure}
    \centering
    \includegraphics[width=1\linewidth]{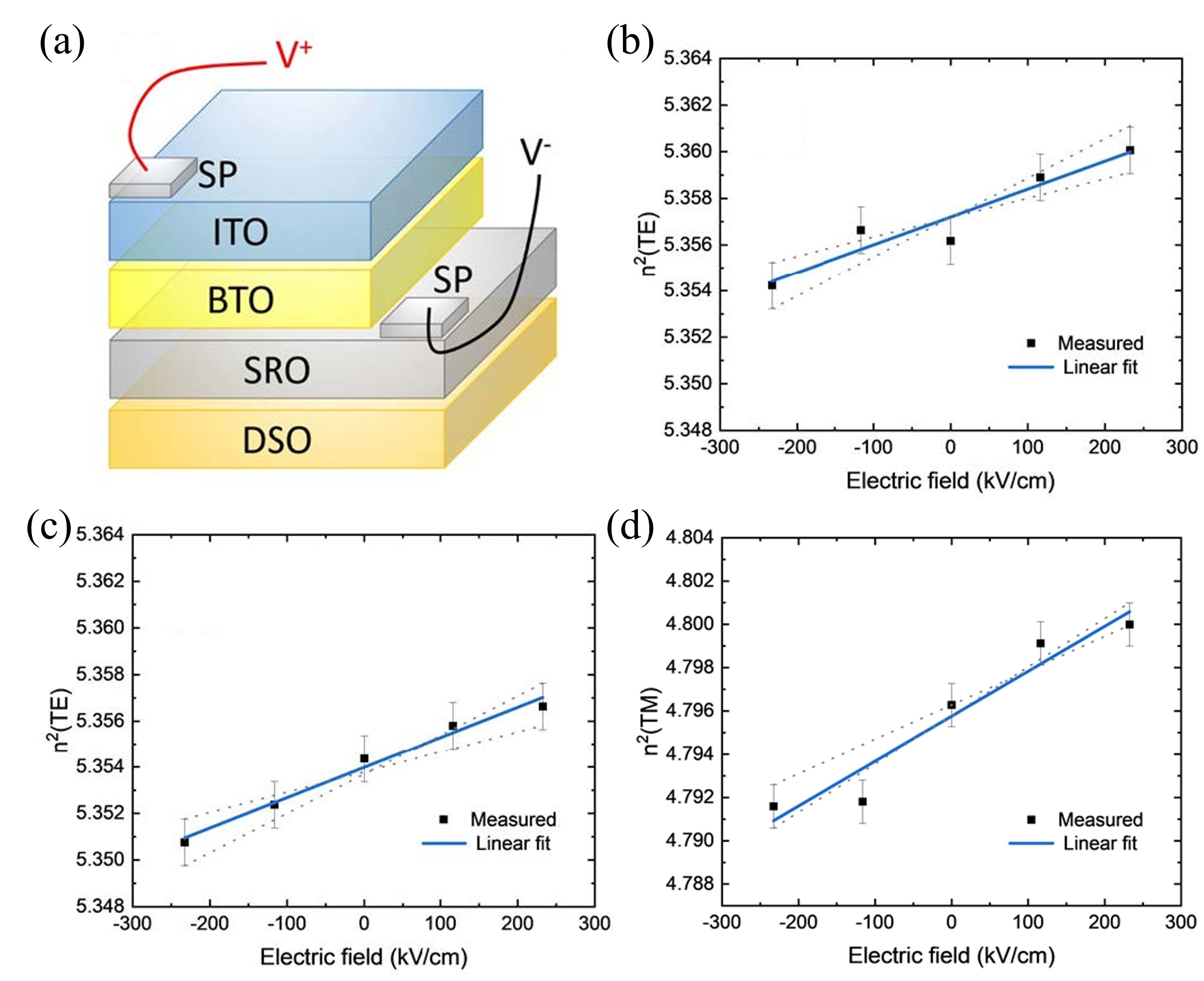}
    \caption{(a) Schematic of the vertical ITO/BTO/SRO/DSO structure used for EO characterization. (b-c) 
    Electric-field-induced refractive-index changes measured using TE-polarized light propagating along the two in-plane crystallographic directions, corresponding to the $r_{13}$ and $r_{23}$ EO coefficients, respectively. (d) Refractive-index modulation measured using TM-polarized light, corresponding to the $r_{33}$ EO. Reproduced with permission from Cao \textit{et al.}, Optical Materials Express 13, 152–160 (2022). Copyright 2022 authors, licensed under a Creative Common Attribution (CC BY) license\cite{cao2022characterization}.}
    \label{fig:4cao2022}
\end{figure}

Besides effective EO coefficients dominated by the $r_{42}$ parameters and extracted from multi-domain thin films, Cao \textit{et al.}\cite{cao2022characterization} reported a experimental characterization of the EO coefficients $r_{13}$, $r_{23}$, and $r_{33}$ in single-crystal BTO thin films grown epitaxially on DSO substrates. Unlike the studies that focus on extracted the $r_{42}$ coefficient in multi-domain films, this work directly measured the pure refractive-index modulation ($r_{\textit{i3}}$ values), which does not induce polarization rotation. The authors employed a vertical capacitor structure consisting of ITO/BTO/SrRuO$_3$(SRO)/DSO, enabling application of a vertical electric field across the BTO film, as illustrated in Fig. \ref{fig:4cao2022}(a). The BTO thin film was deposited on DSO by PLD, with a thin SRO layer serving as the bottom electrode, resulting in a single-domain, $c$-axis-oriented BTO film.

Owning to the single-domain nature of the film, this study is free from the random in-plane domain distribution commonly observed in $a$-axis-oriented BTO thin films. This allowed the authors to directly measured intrinsic EO coefficients rather than extracting effective values from effective net coefficient. Using the prism-coupling technique at a wavelength of 636.6 nm, with the applied electric field along the crystal $c$-axis, the EO coefficients were determined by measuring polarization-dependent refractive-index changes under DC bias. As a result, $r_{13}$ = 4.2 pm/V, $r_{23}$ = 4.6 pm/V, and $r_{33}$ = 9 pm/V were obtained, as shown in Fig. \ref{fig:4cao2022}(b) - \ref{fig:4cao2022}(d). The slight difference between the in-plane coefficients $r_{13}$ and $r_{23}$ was attributed to a small anisotropy in BTO mediated from the DSO substrate.  The coefficients are less than expected from bulk values.

To investigate further, the authors then repeated the measurements on BTO films grown on La$_{0.67}$Sr$_{0.33}$MnO$_3$ (LSMO) electrodes. Due the larger lattice mismatch between LSMO and BTO compared to DSO, the film quality is degraded and correspondingly exhibited reduced $r$ values. These results indicate that higher crystalline quality correlates with stronger EO response in BTO thin films. Although the measured coefficients are smaller than bulk BTO values, this work provides a clear and direct demonstration of polarization-preserving phase modulation in thin-film BTO under a $c$-oriented electric field, serving as an important benchmark for film-level EO characterization.

At the device level, Cao \textit{et al.}\cite{cao2023active} reported a clear experimental demonstration of polarization rotation purely induced by the off-diagonal Pockels coefficient $r_{42}$ in thin-film BTO. In this work, an active polarization rotator was realized using a single-domain $c$-axis-oriented BTO thin film with coplanar electrodes. Unlike most BTO-based EO modulators, which exploit the large $r_{42}$ coefficient for pure refractive-index modulation, this work directly exploits the polarization-rotation effect associated with the off-diagonal EO tensor elements, enabling controlled conversion between TE and TM modes in a simple straight waveguide geometry.

\begin{figure*}
    \centering
    \includegraphics[width=0.8\linewidth]{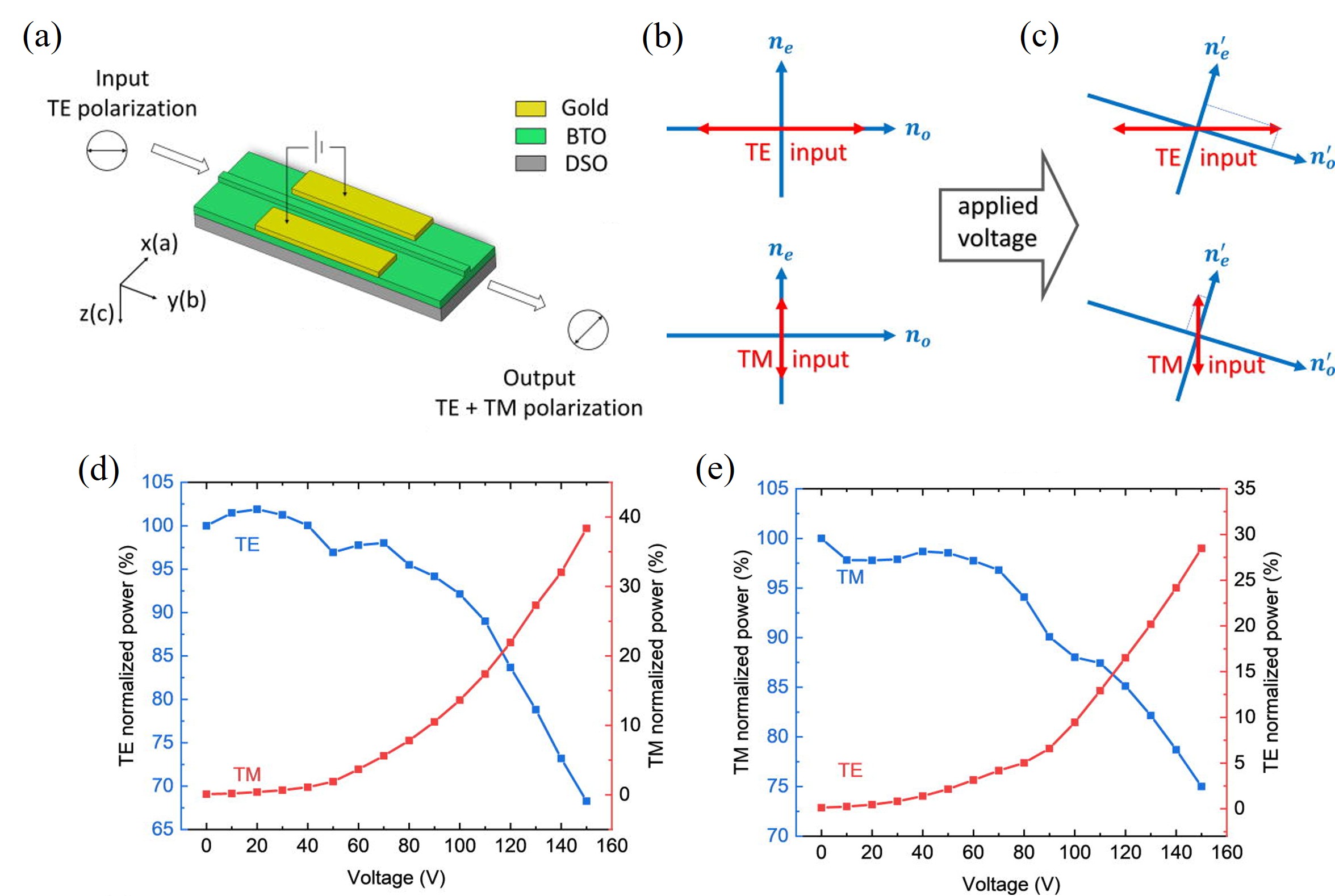}
    \caption{(a) Schematic of the polarization rotator structure fabricated in a $c$-axis-oriented BTO thin film on a DSO substrate with lateral electrodes. (b) Orientation fo the principle optical axes of BTO in the absence of an external electric field. (c) Rotation of the principle optical axes under an applied in-plane electric field, resulting in polarization conversion via the off-diagonal Pockels coefficient $r_{42}$. Normalized optical power in the TE and TM polarizations as a function of applied voltage for (d) TE- and (e) TM-polarized input light, respectively. Reproduced from Cao \textit{et al.}, Applied Physics Letters 122, 031106 (2023), with the permission of AIP Publishing\cite{cao2023active}.}
    \label{fig:4cao2023}
\end{figure*}

The polarization rotator was fabricated in a single-crystalline BTO thin film grown on a DSO substrate. The BTO ridge waveguide was fabricated by dry etching, followed by lateral electrode patterning adjacent to the waveguide, as shown in Fig. \ref{fig:4cao2023}(a). Figures \ref{fig:4cao2023}(b) and \ref{fig:4cao2023}(c) illustrate the polarization rotation mechanism based on the $r_{42}$ coefficient. In the absence of an external electric field, the guided optical modes are aligned with the principle optic axes, with TE and TM polarizations corresponding to the ordinary ($n_o$) an extraordinary ($n_e$) refractive indices, respectively. When an in-plane electric field is applied, the off-diagonal EO tensor elements introduce a rotation of the optic axes, resulting in two new principal axes $n'_o$ and $n'_e$. As a result, the original input polarization contains vector components projecting onto both rotated axes, leading to polarization conversion during propagation.

Experimentally, with TE-polarized input light, up to 31\% of the optical power was converted to the TM mode, corresponding to a TM extinction ratio of 26.1 dB with an applied voltage of 150 V, as shown in Fig. \ref{fig:4cao2023}(d). A similar result was observed for TM-to-TE conversion, with a maximum conversion efficiency of 25\% and a TE extinction ratio of 23.8 dB, as shown in Fig. \ref{fig:4cao2023}(e). It is worth noting that the total guided optical power remains conserved. The applied voltage redistributes the power between orthogonal polarizations rather than introducing additional propagation loss.

Overall, compared to most EO devices based on $a$-axis-oriented thin-film BTO, which typically suffer from random distributed in-plane $a$- and $c$-axis domains, this work utilizes a single-domain $c$-axis-oriented film, allowing direct access to the intrinsic $r_{42}$ coefficient without averaging effects from multiple domain orientations. This study therefore provides a clear experimental demonstration of polarization rotation purely from the off-diagonal Pockels response in BTO and demonstrates an alternative EO modulation mechanism to conventional phase modulation.

\begin{figure*}
    \centering
    \includegraphics[width=1\linewidth]{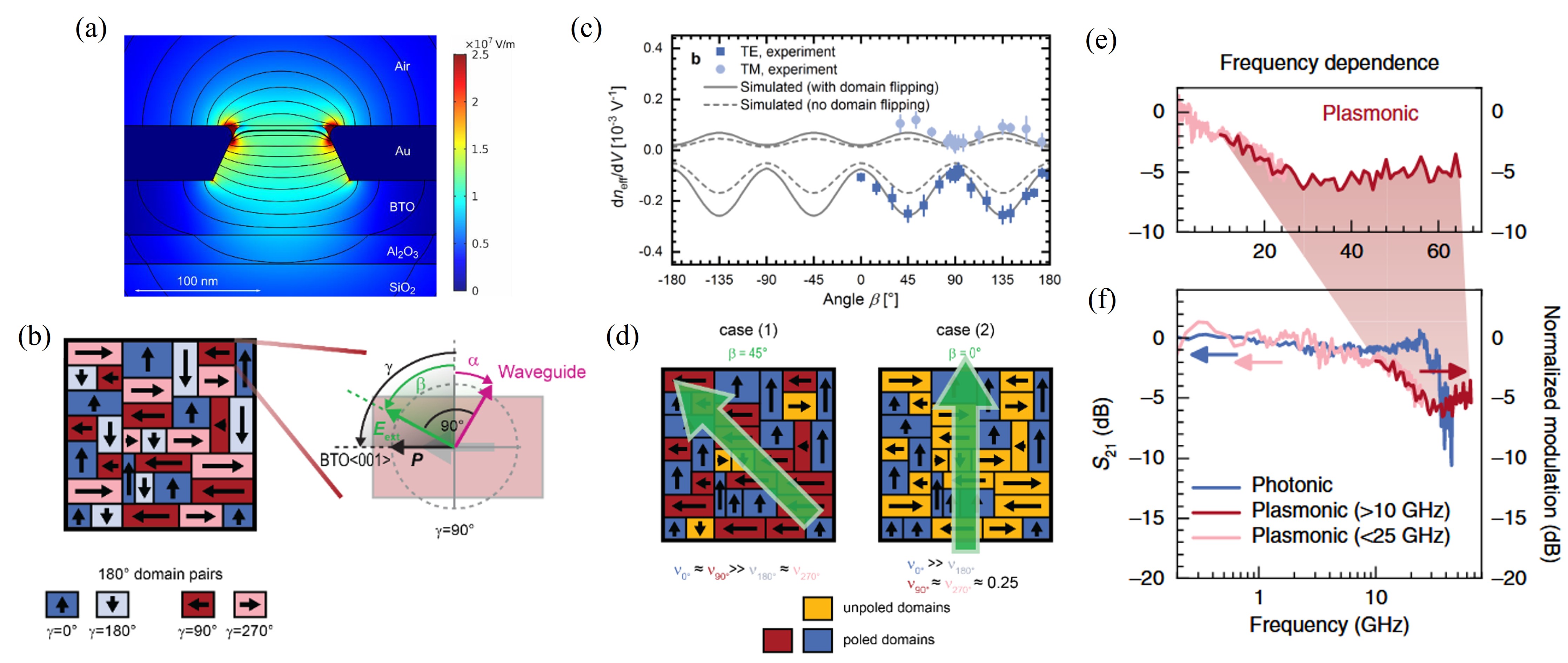}
    \caption{(a) Schematic of the plasmonic structure and the simulated static electric-field distribution. (b) Illustration of multi-domain structure in $a$-axis-orients BTO thin films, showing the random distribution of four in-plane ferroelectric domain variants ($\gamma$ denotes the domain orientation). The angle $\beta$ represents the orientation between the applied electric field and the in-plane crystallographic axes, while $\alpha$ denotes the waveguide propagation direction, with $\alpha + \beta = 90\degree$. (c) Effective EO response obtained from both simulations and experiments as a function of $\beta$, showing that maximum EO efficiency is achieved at $\beta$ = 45 \degree. 
    (d) Schematic comparison of domain structures for $\beta$ = 45\degree and $\beta$ = 0\degree, illustrating that fewer domains can be poled at $\beta$ = 0\degree, leading to reduced EO efficiency. (e) Close-up frequency response highlighting the reduction in EO modulation efficiency in the plasmonic device. (f) Measured high-frequency EO response of both photonic and plasmonic devices based on thin-film BTO, demonstrating operation up to tens of gigahertz. Reproduced with permission from Abel \textit{et al.}, Nature Materials 18, 42–47 (2019). Copyright 2018 Springer Customer Service Center GmbH\cite{abel2019large}.}
    \label{fig:4abel2019sum}
\end{figure*}

For EO devices based on $a$-axis-oriented BTO thin films, Abel \textit{et al.}\cite{abel2019large} reported one of the largest EO coefficients to date in a heterogeneously integrated BTO-on-Si platform, showing the potential of BTO for integrated EO phonics while retaining bulk-like nonlinear properties. As introduced previously in Section \ref{passive photonics} and as shown in Fig. \ref{fig:4able2019hybrid}, this platform was realized by wafer bonding an MBE-grown epitaxial BTO thin film onto an SOI substrate. After removing the donor substrate, the top Si layer was then patterned into photonic waveguides, enabling systematic investigation of EO modulation mechanism in a hybrid photonic configuration. In addition to photonic waveguides, plasmonic phase shifters were also fabricated directly on the BTO film without strip-loaded Si waveguides, as illustrated in Fig. \ref{fig:4abel2019sum}(a).

For photonic devices, racetrack resonators were employed to study EO-induced refractive-index changes via resonance shifts under an applied electric field. Due to the multi-domain nature of $a$-axis-oriented BTO films, as shown in Fig. \ref{fig:4abel2019sum}(b), the in-plane polarization consists of four ferroelectric domain variants. This random distribution of different oriented domains leads to a strong dependence of the EO response on the relative orientation between the optical propagation direction, the applied electric field, and the crystallographic axes. The authors systematically investigated the effective EO coefficient as a function of the angle $\beta$, defined as the angle between the applied electric field and the in-plane crystal axes, as illustrated in Fig. \ref{fig:4abel2019sum}(c). The maximum EO response was observed at $\beta$ = 45\degree. Furthermore, an improved net EO response was observed when domain flipping (poling) was induced by an external bias. Figure \ref{fig:4abel2019sum}(d) compares the domain configurations for $\beta$ = 45\degree and $\beta$ = 0\degree. At $\beta$ = 45\degree, the diagonal electric field orientation allows a large fraction of domains to be polarized under sufficient bias, resulting in constructive EO contributions, while at $\beta$ = 0\degree, only domains initially parallel to the field can be switched, leading to reduced EO efficiency. By combining angle-dependent measurements and modeling, the authors extracted EO coefficients of approximately $r_{42}$ = 923 pm/V and $r_{33}$ = 342 pm/V, representing the largest reported $r_{42}$ value in thin-film BTO to date. Nevertheless, compared to an ideal single-domain film, the fully poled multi-domain film still exhibits a reduced EO response due to incomplete domain alignment. An optimized racetrack resonator design achieved a voltage-length product of approximately 0.45 V$\cdot$cm and a flat EO response up to 30 GHz.

In addition to photonic waveguides, the authors also demonstrated plasmonic BTO phase shifters, in which the optical confinement factor within the BTO layer was increased to approximately 50\%, while eliminating potential plasma-dispersion contributions from Si. The compact device footprint further aids high-frequency operation. As a result, under a bias voltage of 2.5 V, EO modulation was demonstrated up to 65 GHz, as shown in Fig. \ref{fig:4abel2019sum}(f). The plasmonic devices exhibited a reduction in EO response in the frequency range from 2 to 30 GHz, as illustrated in Fig. \ref{fig:4abel2019sum}(e), which was likely attributed to deformation of the BTO film induced by piezoelectric effects, Although the plasmonic configuration reached a superior modulation efficiency and extended bandwidth, uncertainties such as possible contribution from residual $c$-axis-oriented regions with in the nominally $a$-axis-oriented film prevented precise extraction of EO coefficients from the measurements.

The work provided a detailed investigation of multi-domain ferroelectric behavior in $a$-axis-oriented BTO thin films. By accounting for domain populations, domain switching under bias, and hysteretic effects, the authors established a quantitative framework for understanding and optimizing EO performance in thin-film BTO. The results clearly demonstrated optimized domain alignment with near-bulk-like EO coefficient in thin films, enabling excellent EO modulation with operation frequencies extending to several tens of gigahertz, and establishing a strong foundation for future high-speed integrated BTO devices.

\begin{figure}
    \centering
    \includegraphics[width=1\linewidth]{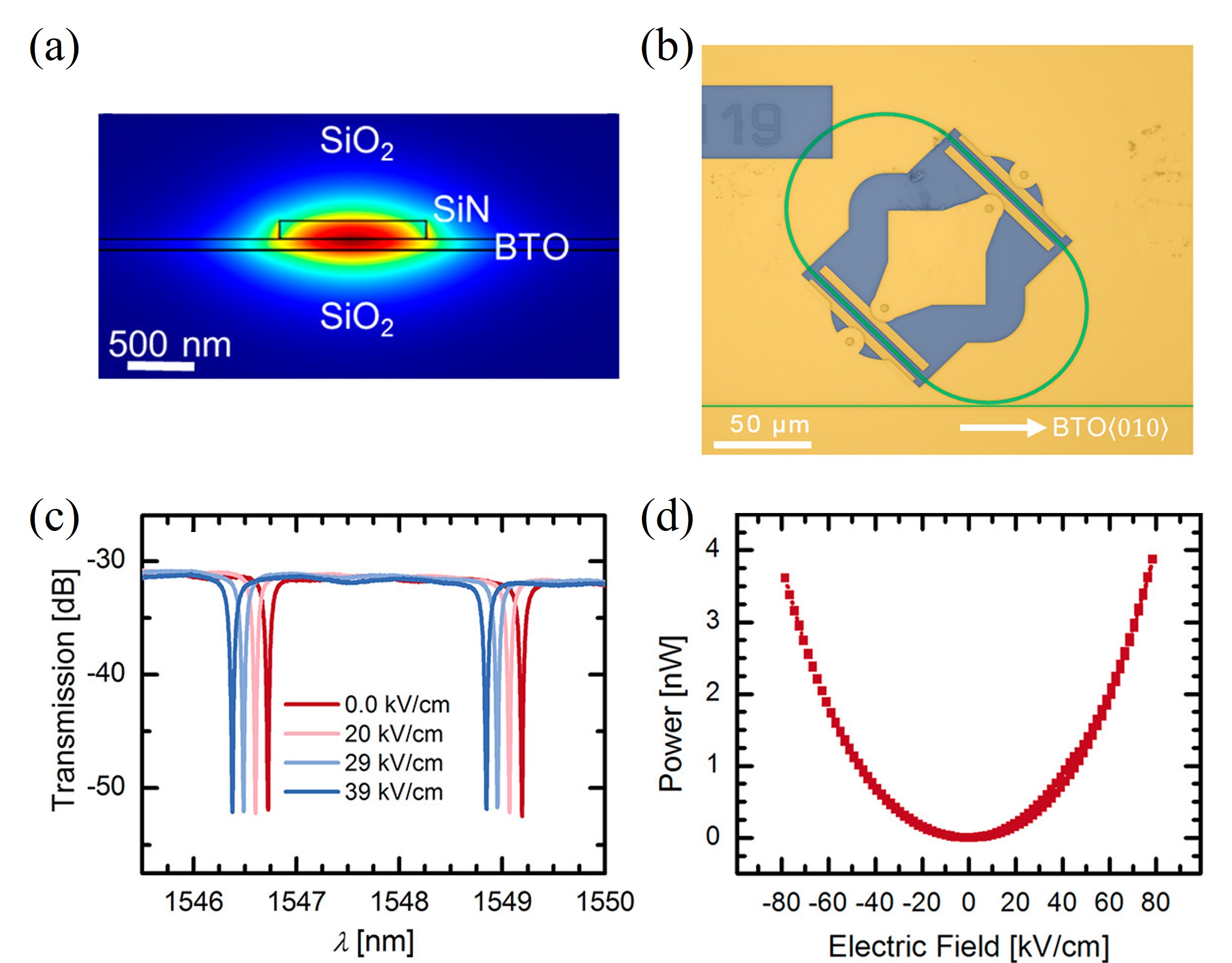}
    \caption{(a) Simulated optical mode profile of a SiN strip-loaded waveguide on a BTO thin film. (b) Optical microscope image of the fabricated racetrack resonator with integrated electrodes. (c) Transmission spectra of EO response of racetrack resonator under different electric fields. (d) Electrical power consumption as a function of the applied electric field during the EO tuning of the resonator.Reprinted with permission from Ortmann \textit{et al.}, ACS Photonics 6, 2677–2684 (2019). Copyright 2019 American Chemical Society\cite{ortmann2019ultra}.}
    \label{fig:4ortmann2019}
\end{figure}

From the same group, Ortmann \textit{et al.}\cite{ortmann2019ultra} demonstrated a closely related hybrid waveguide platform with a practical focus on achieving ultra-low-power EO modulation in BTO-based devices. In this work, a similar BTO thin-film transfer approach was employed to realized a BTO-on-SiO$_2$ stack, followed by the fabrication of a strip-loaded SiN waveguide on top of the BTO film, as shown in Fig. \ref{fig:4ortmann2019}(a). This configuration forms a hybrid SiN/BTO waveguide geometry with an optical overlap factor of approximately 18\% within the BTO active layer and a propagation loss of 9.4 dB/cm, as illustrated in Fig. \ref{fig:4ortmann2019}(b).

EO modulation was implemented using racetrack resonators, with the electrodes oriented at 45\degree with respect to the in-plane crystal axes to maximize the effective EO response. Figure \ref{fig:4ortmann2019}(c) shows  transmission spectra of the racetrack resonator under different applied electric fields. From the EO-induced resonance shifts, an effective Pockels coefficient of approximately 343 pm/V was extracted. It is worth noting that, owning to the carrier-free Pockels effect in BTO, the static tuning power is limited primarily by the leakage current through the ferroelectric material. As a result, an ultra-low static power consumption of approximately 106 nW was achieve to tune the resonance by one free spectral range (FSR), as illustrated in Fig. \ref{fig:4ortmann2019}(d).

The authors further demonstrated compensation of a thermal-induced index drift over a 20\degree C temperature range, requiring less than 1 nW of electrical power to maintain resonance alignment. This result indicating a key advantage of BTO-based EO tuning for Si photonics, where thermal drift is a major concern due to the relatively large thermo-optic coefficient of Si. Combining both ultra-low-power consumption and efficient EO modulation, this work establishes thin-film BTO as a promising material platform for energy-efficient integrated photonic systems.

\begin{figure}
    \centering
    \includegraphics[width=1\linewidth]{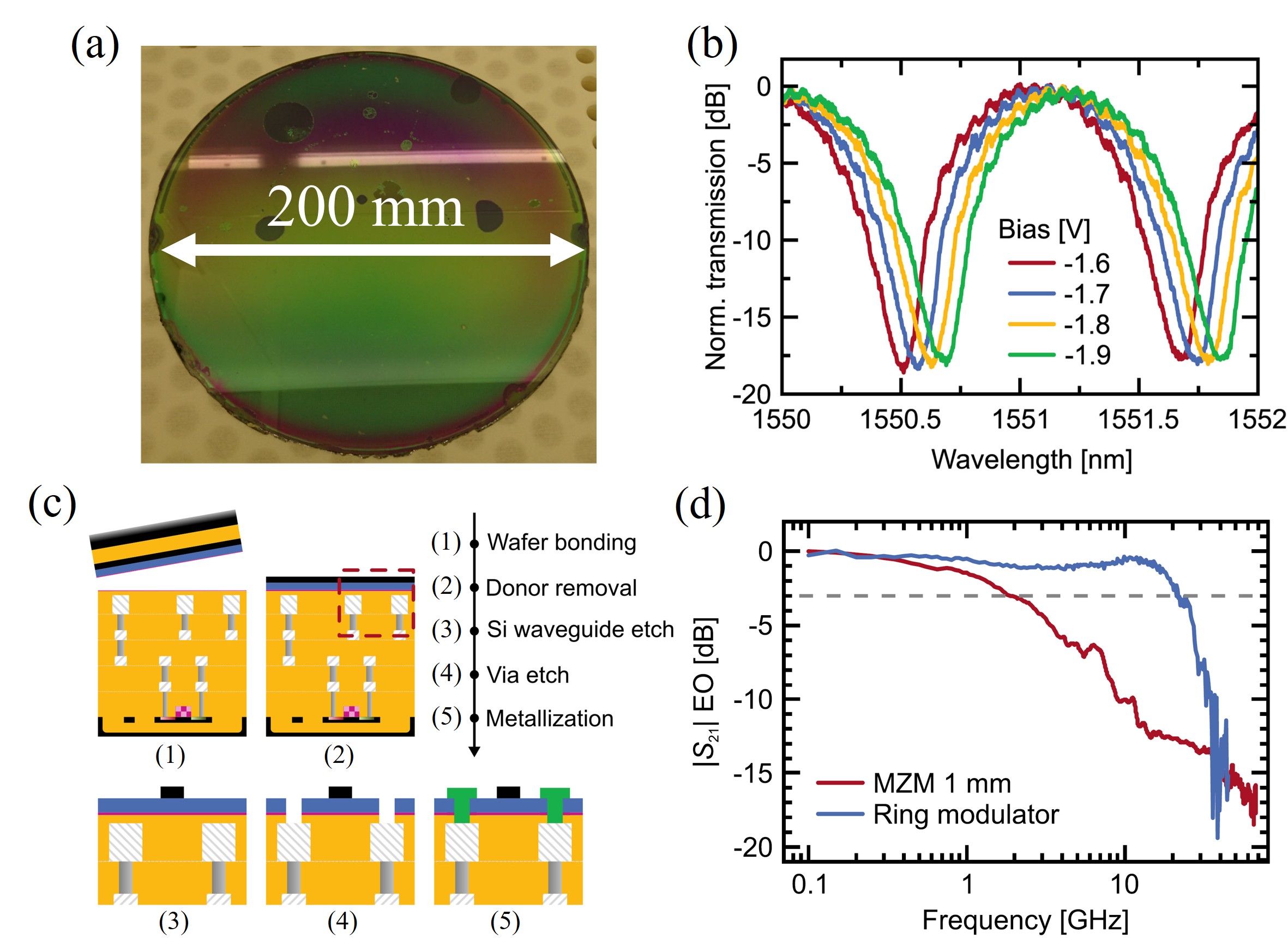}
    \caption{(a) Photo of a 200-mm wafer after transfer of epitaxial thin-film BTO onto a Si photonic integrated platform. (b) Schematic of the BEOL integration flow for thin-film BTO. (c) Transmission spectra of an EO MZI under various bias voltage, indicating voltage-induced refractive-index modulation. (d) High-frequency EO response measurement for a MZI and a ring modulator with a radius 10-$\mu$m. Reproduced with permission from Eltes \textit{et al.}, Journal of Lightwave Technology 37, 1456–1462 (2019). Copyright 2026 authors, licensed under a Creative Common Attribution (CC BY) license\cite{eltes2019batio3}.}
    \label{fig:4eltes2019}
\end{figure}

Eltes \textit{et al.}\cite{eltes2019batio3} demonstrated a wafer-scale and foundry-compatible integration of thin-film BTO into a Si photonics platform by incorporating EO modulators into the back-end-of-line (BEOL) of a photonic integrated circuit (PIC) process. This work represents an extension of the BTO thin-film transfer approach previously introduced by Abel \textit{et al.}\cite{abel2019large}, scaling it to 200-mm wafers, as illustrated in Fig. \ref{fig:4eltes2019}(a). Importantly, the BTO integration was performed after completion of the front-end photonic devices, enabling compatibility with industrial CMOS fabrication flows.

In the process flow (Fig. \ref{fig:4eltes2019}(b)), the Si photonics BEOL was interrupted at the fourth metallization after interlayer dielectric planarization. The epitaxial BTO thin film was then transferred onto the planarized PIC wafer via oxide bonding, followed by patterning the Si layer to form strip-loaded hybrid waveguides. After this, the remaining BEOL processes were resumed to define the electrodes.

Using this platform, the authors demonstrated both MZIs and ring resonators. Passive ring resonators with a radius of 30 $\mu$m exhibited a Q-factor of approximately 50,000, corresponding to a propagation loss of 5.8 dB/cm. For EO MZIs (Fig. \ref{fig:4eltes2019}(c)), a voltage-length $V_{\pi}L$ of 0.23 V$\cdot$cm was measured, from which an effective EO coefficient of 380 pm/V was extracted, accounting for an optical mode overlap of approximately 38\% with the BTO layer. High-speed measurements were also performed, as shown in Fig. \ref{fig:4eltes2019}(d). The MZI modulators exhibited a 3-dB EO bandwidth of about 2 GHz. To extend the modulation bandwidth, the authors reduced the ring resonator radius to 10 $\mu$m, lowering the Q-factor to approximately 15,000 but improving EO bandwidths up to 20 GHz.

This work demonstrated a scalable BEOL integration route for thin-film BTO within an advanced Si photonics platforms. By achieving low $V_{\pi}L$ and compatibility with large-wafer scale CMOS processing, Eltes \textit{et al.} showed that thin-film BTO can deliver high EO efficiency while meeting the integration and scalability requirements necessary for the wafer-scale BTO-based EO devices.

\begin{figure*}[t]
    \centering
    \includegraphics[width=\textwidth]{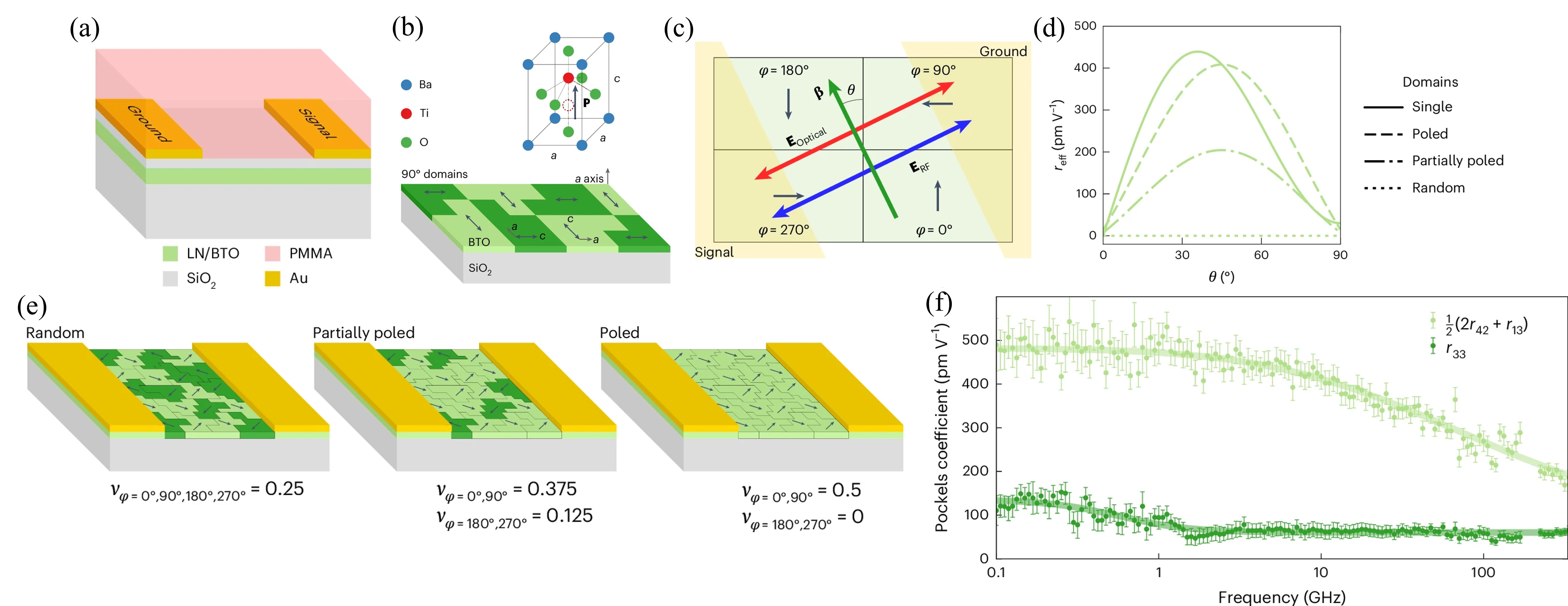}
    \caption{(a) Schematic of the plasmonic slot phase shifter fabricated on a BTO thin film. (b) Illustration of the multi-domain structure in an $a$-axis oriented film. Black arrows indicate the orientation of the in-plane ferroelectric domains. (c) Schematic showing the relative orientation between ferroelectric domains, optical polarization, and the applied electric field. (d) Numerical simulation of the effective Pockels coefficient $r_{\text{eff}}$ for single-domain, fully poled, partially poled, randomly oriented BTO films, showing that the maximum EO efficiency can be achieve when the optical propagation direction is aligned at 45\degree with the respect to the crystallographic axes. (e) Schematics of randomly oriented, partially poled, and fully poled $a$-axis-oriented BTO thin film with lateral electrodes. (f) Measured frequency dependence of the Pockels coefficients of BTO thin films. Since $r_{42}$ is typically much larger than $r_{13}$, therefore $\tfrac{1}{2}$($r_{13}$+2$r_{42}$) $\approx$ $r_{42}$. Reproduced with permission from Chelladurai \textit{et al.}, Nature Materials 24, 868-875 (2025). Copyright 2025 authors, licensed under a Creative Common Attribution (CC BY) license\cite{chelladurai2025barium}.}
    \label{fig:4chelladurai}
\end{figure*}

Recently a comprehensive investigation of the frequency-dependent EO response of thin-film BTO was reported by Chelladurai \textit{et al.}\cite{chelladurai2025barium}, particularly focusing on the extraction of EO coefficients and dielectric permittivity at high frequencies up to sub-terahertz regime. In this work, the authors employed a nanoscale plasmonic slot phase-shifter configuration using thin-film BTO as the active material, as illustrated in Fig. \ref{fig:4chelladurai}(a). This compact structure offers advantages for high-frequency characterization, including a very short interaction length ($\sim$25 $\mu$m), which minimize the velocity-mismatch effects, as well as low total insertion loss, enabling characterization of EO coefficients despite the weak EO response at high frequencies. 

Using these integrated phase shifters, the authors extracted both the permittivity and Pockels coefficients of BTO thin films over an exceptionally broad frequency range from 100 MHz to 330 GHz. The high-frequency permittivity results have been discussed previously in Section \ref{fig:chap2-2}. Due to the multi-domain nature of the $a$-axis-oriented BTO films, as illustrated in Fig. \ref{fig:4chelladurai}(b), the authors first analyzed the effective EO coefficients related to the electrode orientation and the crystal domain. Simulations, as shown in Fig. \ref{fig:4chelladurai}(d), indicate that for a partially or fully poled film (schematically illustrated in Fig. \ref{fig:4chelladurai}(e)), the effective EO coefficient is maximized when the electrode orientation is aligned at 45\degree with respect to the in-plane crystal axes. However, achieving a fully poled state requires a sustained external bias, and due to instrumental limitations, the film in this work could only be partially poled by applying a DC bias prior to the high-frequency EO measurements.

Under these conditions, the authors extracted a frequency-dependent $r_{42}$ coefficient of approximately 481 pm/V at 100 MHz, which decreases to about 191 pm/V at 330 GHz, as shown in Fig. \ref{fig:4chelladurai}(f). In contrast, the $r_{33}$ coefficient was found to be 125 pm/V at 100 MHz and remained relatively constant above 1 GHz. This behavior suggests that the observed reduction in $r_{42}$ at high frequencies is associated with a decrease in the permittivity along the $a$-axis, whereas the permittivity along the $c$-axis remains relatively stable beyond 1 GHz, consistent with the frequency-independent behavior of $r_{33}$. Importantly, this study establishes a clear quantitative link between crystallographic orientation, ferroelectric domain structure, and the effective EO response in multi-domain thin-film BTO. Consistent with earlier reports, the authors confirm that optimal exploitation of $r_{\text{eff}}$ in $a$-axis-oriented films requires domain polarization, while partial poling leads to reduces EO efficiency and a randomly polarized (unpoled) film can results in near-zero net EO response. 

Finally, this work represents, to the best of our knowledge, the only experimental demonstration of EO characterization of thin-film BTO extending into the sub-terahertz regime. While most reported EO modulators based on thin-film BTO have operated at a frequency of several tens of gigahertz, the frequency-dependent measurements presented in this work provide a critical insight into intrinsic material limitations at ultra-high frequencies and potential challenges for realizing broadband EO modulation in thin-film BTO-based platforms.

\begin{figure*}[t]
    \centering
    \includegraphics[width=\textwidth]{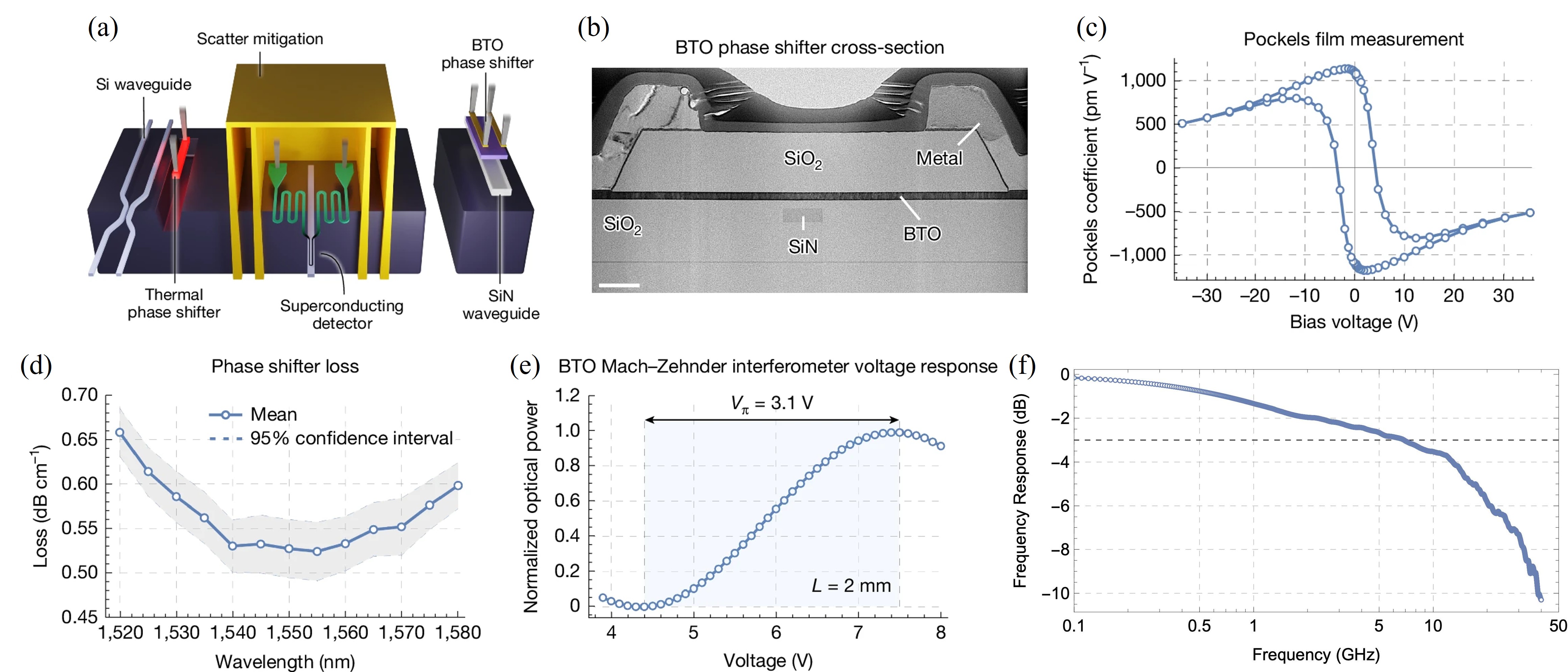}
    \caption{(a) Schematic of key components in a Si-photonic-based quantum computing system. (b) Cross-sectional SEM image of the hybrid BTO/SiN phase shifter. (c) Effective EO coefficient extracted from free-space Pockels measurements as a function of applied bias voltage.(d) Propagation loss of the hybrid phase shifter measured using cut-back methods. (e) Static EO response of a 2 mm long MZI, indicating a $V_\pi$ of 3.1 V, and a corresponding $V_{\pi}L$ = 0.62 V$\cdot$cm. (f) High-speed EO characterization of a heterogeneously integrated EO MZI modulator, showing a 3-dB EO bandwidth of 6.9 GHz. Reproduced with permission from Alexander \textit{et al.}, Nature 641, 876–883 (2025). Copyright 2025 authors, licensed under a Creative Common Attribution (CC BY) license\cite{psiquantum_team2025manufacturable}.}
    \label{fig:4PsiQuantum}
\end{figure*}

Alexander \textit{et al.}\cite{psiquantum_team2025manufacturable} recently reported a manufacturable, wafer-scale Si photonic platform targeting large-scale photonic quantum computing. Due to the requirements of fault-tolerant quantum systems, this platform demands optical components with performance beyond the current state of the art. Figure \ref{fig:4PsiQuantum}(a) shows the key components in such a platform. Among these components, high-performance EO phase shifters were realized by integrating thin-film BTO with SiN photonics. In this work, epitaxial BTO thin films were grown by MBE and subsequently heterogeneously integrated with SiN waveguides via oxide bonding, as shown in Fig. \ref{fig:4PsiQuantum}(b), enabling scalable EO functionality on a mature silicon photonics platform.

The MBE-grown BTO thin films were first characterized using a free-space measurement setup, where a 1550 nm laser was incident normal to the film surface. From this measurement, an exceptionally large effective EO coefficient of approximately 1000 pm/V was extracted, representing one of the largest reported effective EO responses in thin-film BTO to date, as shown in Fig. \ref{fig:4PsiQuantum}(c). It is worth noting that this extracted value was obtained from a free-space measurement with optical propagation normal to the film surface, whereas in the integrated devices the optical mode propagates in the plane of the film. In addition, the propagation loss of the hybrid BTO/SiN waveguides was measured to be approximately 0.5 dB/cm under a 14 V bias, as illustrated in Fig. \ref{fig:4PsiQuantum}(d). To better emulate practical device operation, the waveguides were pre-poled at 40 V prior to cut-back loss measurements, with a bias voltage of 14 V applied during measurement, ensuring stable ferroelectric domain alignment.

The EO modulation performance was subsequently characterized using both static and high-frequency electrical signals applied to fabricated MZIs and phase shifters. Figure \ref{fig:4PsiQuantum}(e) shows the DC EO response of a 2 mm long MZI incorporating an integrated BTO phase shifter, from which a voltage-length product $V_{\pi}L$ = 0.62 V$\cdot$cm was extracted. The device was poled at 40 V prior to measurement to align the in-plane ferroelectric domains. High-speed measurements were performed on a 3 mm long MZI modulator, yielding a 3 dB EO bandwidth of 6.9 GHz, as shown in Fig. \ref{fig:4PsiQuantum}(f).

Overall, this work successfully demonstrates the integration of thin-film BTO into an advanced and scalable Si photonic platform for potential quantum computing applications. The combination of low $V_{pi}L$, high EO efficiency, multi-gigahertz bandwidth, low optical loss, and wafer-scale manufacturability of hybrid BTO/SiN phase shifters meets several requirements necessary for future photonic quantum systems. By exploiting the large Pockels response of BTO within a silicon-compatible process flow, this platform establishes BTO as a promising EO material not only for quantum photonics but also for future large-scale integrated photonic circuits.

In addition to hybrid devices on a Si-based platform described above, Deng \textit{et al.}\cite{deng2026self-buffered} recently reported a SiN/BTO waveguide on a (LaAlO$_3$)$_{0.3}$-(Sr$_2$TaAlO$_6$)$_{0.7}$ (LSAT) substrate. In contrast to Si-based platforms, LSAT offers less lattice mismatch with BTO while it has a lower refractive index, making it a promising insulating substrate for thin-film BTO-based integrated photonics.

The BTO thin films grown on LSAT in this work exhibit a laterally periodic $a$/$c$ multi-domain configuration extending throughout the film thickness. The authors first characterized the EO response at the film level using a free-space EO measurement setup. Using this approach, the BTO film on LSAT exhibited a maximum effective EO coefficient of 253 pm/V when an in-plane electric field was applied along the <110> crystal direction. From this effective value, an intrinsic $r_{42}$ coefficient of 358 pm/V was extracted. In contrast, when the electric field was aligned along the <100> direction, a significantly reduced effective coefficient of 34 pm/V was obtained. This dependence of the EO response on the electric-field orientation is consistent with other EO devices fabricated in multi-domain BTO films and reflects the tensorial nature of the EO effect in BTO generally.

\begin{figure}
    \centering
    \includegraphics[width=1\linewidth]{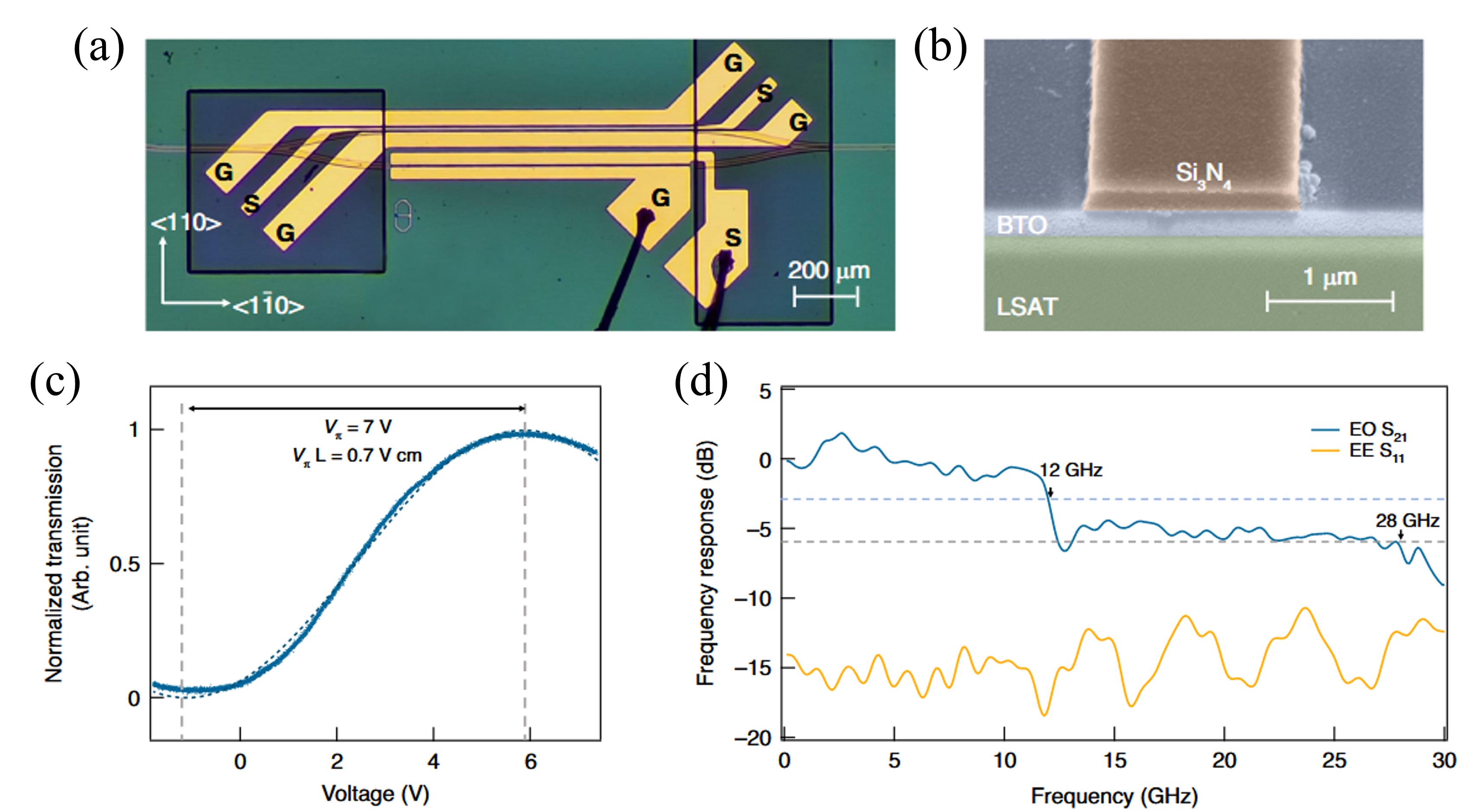}
    \caption{(a) Optical microscope image of a fabricated hybrid SiN/BTO MZI on a LSAT substrate, with coplanar electrodes aligned along the <110> direction. (b) Cross-sectional image of a strip-loaded SiN waveguide on BTO thin film grown on LSAT. (c) EO transmission response of the MZI as a function of applied voltage, indicating that $V_\pi$ = 7 V and a corresponding $V_{\pi}L$ = 0.7 V$\cdot$cm. (d) Frequency response of the corresponding EO MZI modulator, showing a 3 dB bandwidth of approximately 12 GHz. Reproduced with permission from Deng \textit{et al.}, Light: Science \& Applications 15, 21 (2026). Copyright 2026 authors, licensed under a Creative Common Attribution (CC BY) license\cite{deng2026self-buffered}.}
    \label{fig:4deng2026}
\end{figure}

Based on this material platform, Deng \textit{et al.} fabricated a hybrid SiN strip-loaded BTO MZI with a 1-mm long phase shifter, as shown in Fig. \ref{fig:4deng2026}(a) and \ref{fig:4deng2026}(b). The coplanar electrodes were aligned along the <110> crystallographic direction to maximize the EO efficiency. As a result, the fabricated EO MZI exhibited a half-wave voltage $V_\pi$ of 7 V, corresponding to a voltage-length product $V_{\pi}L$ = 0.7 V$\cdot$cm, as shown in Fig. \ref{fig:4deng2026}(c). Additionally, high-frequency measurements showed a 3 dB EO bandwidth of approximately 12 GHz, and a 6 dB bandwidth extending to 28 GHz, as illustrated in Fig. \ref{fig:4deng2026}(d). The bandwidth limitation was likely attributed to the group-velocity mismatch of the optical modes and RF signals.

This work demonstrates an alternative route toward high-performance BTO EO devices that does not rely on integration with Si. By combining epitaxial BTO growth on an oxide-insulator substrate with a hybrid SiN/BTO waveguide configuration, this platform achieves competitive modulation efficiency while simplifying material integration. Compared with the more widely explored BTO-on-Si approaches, this work demonstrates the viability of BTO-on-insulator platforms for integrated photonic systems and broadens the design space for future high-performance EO devices.

\section{Applications}
Owing to the most prominent EO characteristics and excellent linear and nonlinear optical properties of BTO materials, along with recent advances in the techniques of thin-film BTO growth and microfabrication of optical devices, the thin-film BTO integrated photonic platform has emerged in increasingly broad application fields such as high-speed optical communications, metasurface-based free-space optical manipulation, optical sensing, nonlinear optics, and optical neuromorphic computing. Here, we highlight selected application areas of this emerging platform as well as future opportunities.

\subsection{High-speed optical communications}
Optical communication is a technology that encodes digital signals (i.e., 0 and 1) onto the degrees of freedom (such as intensity and phase) of optical signals, enabling the transmission of digital information through optical signal modulation. Such intensity or phase modulation of an optical signal is highly suitable for implementation using an EO modulator. Due to the outstanding EO characteristics and integration capabilities of EO modulators based on BTO integrated photonic platforms as demonstrated in Section \ref{active EOM}, these devices have emerged as promising candidates for future optical communications applications. 

In optical communications, data rate is one of the most critical performance metrics, with ongoing efforts aimed at maximizing its value to achieve high-speed optical communications. This corresponds to a requirement that EO modulators possess a broad EO bandwidth. In principle, a larger EO bandwidth directly enables a higher maximum data rate before intensity or phase modulation distortion leads to an unacceptable bit error rate. Moreover, a lower half-wave voltage $V_\pi$-dependent applied peak-to-peak voltage $V_{\text{pp}}$ of an EO modulator implies lower driving power consumption. Also important are having low optical insertion and propagation losses, as well as small fabrication footprint.  It is worth noting that while the half-wave voltage-length product ($V_\pi L$) serves as a key parameter for evaluating the EO modulation capability, practical optical communication applications place greater emphasis on the individual $V_{\text{pp}}$ and the footprint related to the modulation length $L$. 

In recent years, significant advancements have been achieved in thin-film BTO-based EO modulators for high-speed optical communications, as summarized in Table \ref{tab:BTO_opt commun_summary}. All have compact footprints at the millimeter-squared (mm$^2$) level. Currently, both photonic \cite{eltes2023thin-film} and plasmonic \cite{kohli2025plasmonic} device architectures have achieved high-speed data transmission. With respect to the photonic device in Ref. \cite{eltes2023thin-film}, the driving voltage $V_{\text{pp}}$ in low-power-consumption mode reached a sub-1 V level of as low as 0.83 V, while an increase in $V_{\text{pp}}$ leading to the enhanced voltage distinguishability enables a high data rate of 262 Gb/s. More recently, Qiu \textit{et al.} demonstrated a BTO dual-polarization in-phase quadrature modulator (DP-IQM) for coherent optical transmission \cite{qiu2026barium}. Using 120 GBaud DP 64-QAM signaling, the system achieved a net data rate of 1.15 Tb/s over an 80 km fiber link, representing a state-of-the-art system-level demonstration of BTO-based high-speed optical communications. In addition to the transmission performance, this work investigated the influence of poling bias on the system response and showed that stronger domain alignment can improve the effective EO response and transmission performance. These results highlight both the potential of thin-film BTO modulators for coherent communication systems and the importance of poling dynamics and bias control in practical BTO devices. Moreover, Qiu \textit{et al.} recently also demonstrated a thin-film BTO 4-lane data-center-reach optical link (DR4) chip operating in the O-band with a net data rate of 1.6 Tb/s (4 × 448 Gbps PAM-4) through monolithic integration on a commercial silicon photonics platform, highlighting the growing maturity of BTO photonics for next-generation optical interconnects\cite{Qiu26barium}.
On the other hand, plasmonic devices can more readily achieve a broader EO bandwidth, and as a result, they can also support high data rates; up to 448 Gb/s \cite{kohli2025plasmonic} was achieved in a BTO slot plasmonic modulator on a SiN platform. However, due to metal absorption, these devices exhibit significantly higher optical losses ($\sim$20 dB \cite{kohli2025plasmonic} fiber-to-fiber) compared to photonic devices ($\sim$2 dB \cite{eltes2023thin-film}). A plasmonic racetrack microresonator-based configuration with a $Q$ = 1931 quality factor was found to have improved transmission losses \cite{kohli2025plasmonic}; plasmonic racetrack modulators in BTO are expected to have low temperature sensitivity with stable modulation in a small form factor.

\begin{table*}[t!]
\caption{\label{tab:BTO_opt commun_summary}
Summary of representative thin-film BTO EO modulators for high-speed optical communications.}
\begin{ruledtabular}
\begin{tabular}{c c c c c c c}

\textbf{Configuration} &
\textbf{Year} &
\textbf{Data rate} &
\textbf{Modulation scheme} &
\textbf{$V_{\text{pp}}$ (V)} &
\textbf{Optical insertion loss (dB)} &
\textbf{Footprint (mm$^2$)}
\\ \hline

Photonic \cite{xiong2014active} & 2014 & 300 Mb/s & NRZ\footnote{Non-return-to-zero (NRZ) intensity modulation scheme.} & 6.6 & $\sim3.3$\footnote{Estimated using propagation loss $\alpha$ of 44 dB/cm and modulation arm length of 750 $\mu$m.} & $<1\times1$ \\

Photonic \cite{abel2019large} & 2019 & 40 Gb/s & NRZ & N.A. & N.A. & $\sim0.1\times0.1$ \\

Plasmonic \cite{abel2019large} & 2019 & 50 Gb/s & BPSK\footnote{Binary phase-shift keying (BPSK) phase modulation scheme.} & 1.6 & $\sim14$\footnote{Estimated using propagation loss $\alpha$ of $\sim$1.4 dB/$\mu$m and modulation length of 10 $\mu$m.} & $<0.1\times0.1$ \\

Photonic \cite{eltes2019batio3} & 2019 & 25 Gb/s & NRZ & 4 & $\sim$0.58\footnote{Estimated using propagation loss $\alpha$ of 5.8 dB/cm and modulation length of 1 mm.} & $\sim1\times1$ \\

Photonic \cite{eltes2020integrated} & 2020 & 20 Gb/s & NRZ & 1.7 & N.A. & $\sim0.1\times0.1$ \\

Photonic\cite{eltes2023thin-film} & 2023 & 262 Gb/s & PAM-6\footnote{Pulse-Amplitude Modulation 6 (PAM-6) intensity modulation scheme.} & 3 & 2 & $0.4\times1.6$ \\

 &  & 200 Gb/s & PAM-4\footnote{Pulse-Amplitude Modulation 4 (PAM-4) intensity modulation scheme.} & 0.83 & 2 & $0.4\times1.6$ \\

Photonic\cite{zhang2024hybrid} & 2024 & 50 Gb/s & NRZ & N.A. & N.A. & $<0.1\times0.1$ \\

Plasmonic\cite{kohli2025plasmonic} & 2025 & 340 Gb/s & PAM-4 using MZ\footnote{PAM-4 intensity modulation scheme using Mach-Zehnder (MZ) modulator.} & 1.13 & 20.3 & $<0.1\times0.1$ \\

 &  & 448 Gb/s & QAM-4 using IQ\footnote{Quadrature Amplitude Modulation 4 (QAM-4) intensity and phase modulation scheme using In-phase/quadrature (IQ) modulator.} & 1.13 & 23.9 @1550 nm & $0.75\times2.15$ \\

 &  & 200 Gb/s & PAM-2 using RT\footnote{Pulse-Amplitude Modulation 2 (PAM-2) intensity modulation scheme using racetrack (RT) modulator.} & 1.13 & 2 & $\sim0.25\times0.125$ \\

 Photonic\cite{qiu2026barium} & 2026 & 1.15 Tb/s & DP QAM-64 using IQ\footnote{Dual-polarization (DP) Quadrature Amplitude Modulation 64 (QAM-64) phase modulation scheme using In-phase/quadrature (IQ) modulator.} & N.A. & 14.5\footnote{Fiber-to-fiber insertion loss, with on-chip loss of 3.5 dB.} & $\sim4.5\times1$ \\

 Photonic\cite{Qiu26barium} & 2026 & 1.6 Tb/s & DR4 PAM-4 using MZ\footnote{4-lane data-center-reach optical link (DR4) PAM-4 intensity modulation scheme using Mach-Zehnder (MZ) modulator.} & N.A. & 12.5-14\footnote{Fiber-to-fiber insertion loss, with on-chip loss of 3 dB.} & $5.1\times3.5$\footnote{For 4 MZ modulators.} \\

\end{tabular}
\end{ruledtabular}
\end{table*}

\subsection{Metasurface-based applications}
On the integrated thin-film photonic platform, artificially designed two-dimensional planar structures at sub-optical wavelength scales can be fabricated, which are referred to as "metasurfaces" \cite{neshev2018optical,abdelraouf2022recent,kuznetsov2024roadmap}. These not only allow for the precise design and manipulation of degrees of freedom of free-space light such as the amplitude, phase, and polarization, thereby breaking through the functional limitations of traditional optical devices, but also possess features of ultra-thinness and high integration capability, thus surpassing the volume constraints of traditional optical devices. Crucially, the performance of such metasurfaces are fundamentally governed by their constituent materials. In other words, the material properties serve as the cornerstone of their design, directly dictating the operating principles and potential applications. For these reasons, BTO, renowned for its superior EO properties and considerable optical nonlinearity, coupled with advances in high-quality thin film growth and device micro-fabrication, has attracted substantial attention as a promising material platform for metasurfaces.

\begin{figure*}
    \centering
    \includegraphics[width=0.78\linewidth]{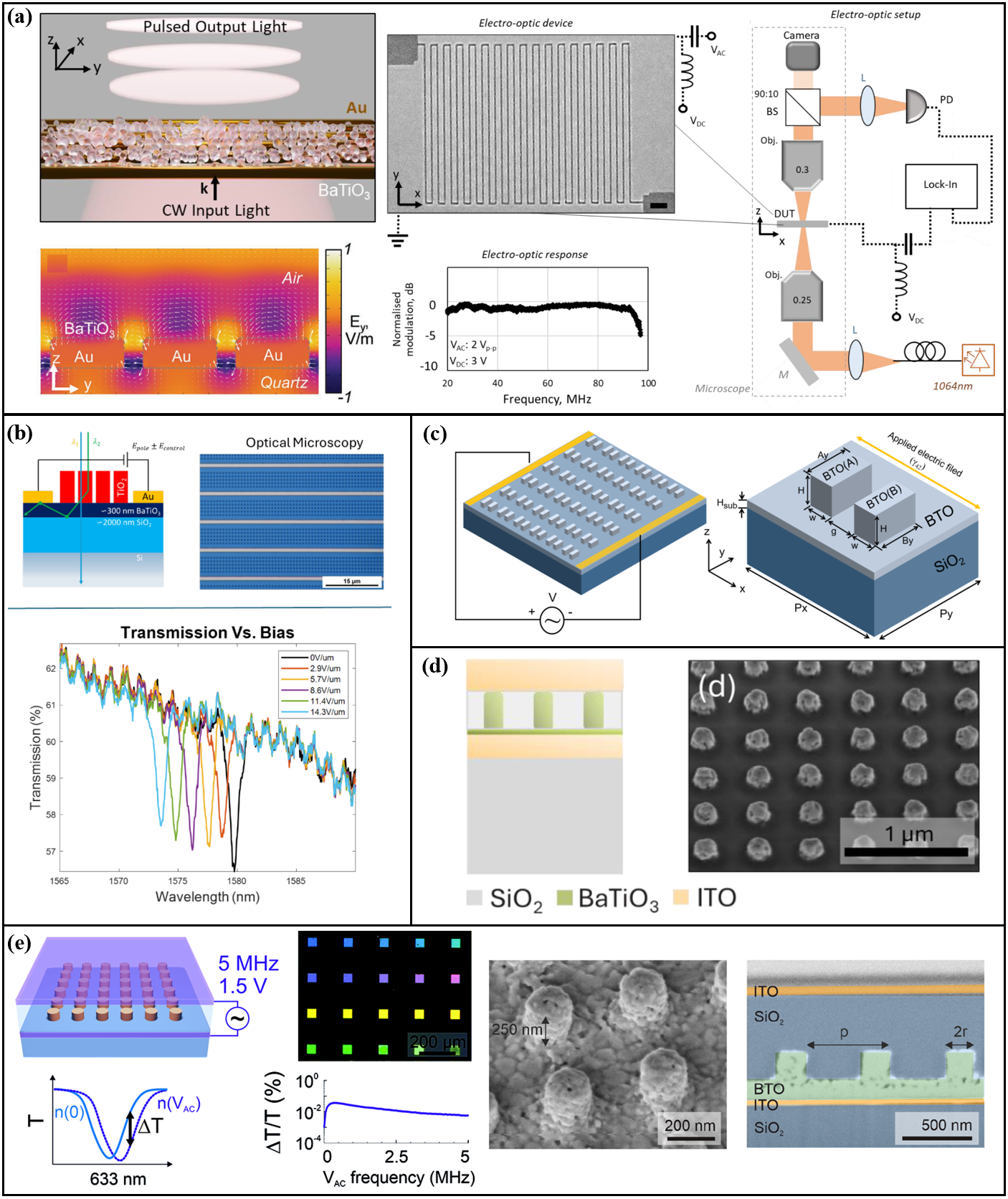}
    \caption{Metasurface free-space EO modulators, based on (a) a plasmonic/BTO nanocrystal structure, including the structural schematic diagram, scanning electron microscope (SEM) image of devices, simulated electric field distributions, EO characterization setups, and EO response curves. (b) all-dielectric configuration using TiO$_2$ cube array heterogeneously fabricated on thin-film BTO, including the structural schematic diagram, optical microscopy, and measured transmission spectra while varying the applied electric field. \cite{yousef2025batio3}, (c) \cite{shen2024ultra}, (d)  \cite{prountzou2025electric} and (e) \cite{weigand2024nanoimprinting} all-dielectric configuration using monolithic thin-film BTO, including the structural schematic diagram, SEM images, resonance shift curve and corresponding RF frequency response. (a)(b) Reproduced with permission from Yousef \textit{et al.},\copyright 2025 in CLEO: Science and Innovations (Optica Publishing Group) p.SS102-5. (c) Reproduced with permission from Shen \textit{et al.}, \copyright 2024 authors, licensed under a Creative Common Attribution (CC BY) license. (d) Reproduced with permission from Prountzou \textit{et al.},\copyright 2025 in CLEO/Europe-EQEC pp.1-1. (e) Reprinted with permission from Weigand \textit{et al.},\copyright 2024 American Chemical Society.}
    \label{fig:chap5-1}
\end{figure*}

Firstly, metasurface-based free-space EO modulators are fabricated by exploiting the EO property of BTO, enabling manipulation across optical degrees of freedom. An optical intensity modulator can be realized using a metasurface based on a hybrid metal–plasmonic resonance and BTO dielectric configuration \cite{karvounis202195,karvounis2020electro}, as shown in Fig. \ref{fig:chap5-1}(a). The device is fabricated by spin-coating thin-film BTO nanocrystals onto an array of paired gold nanowires. The Au nanowire array induces a plasmonic resonance, which enhances the applied electric field. This enhanced field, in turn, elicits a strong EO response from the BTO film, thereby modifying the plasmonic resonance condition and modulating the transmitted light intensity. The effective EO coefficient of the device is up to $\sim$63 pm/V, enabling modulation under a driving voltage of less than 3 V and achieving an EO bandwidth of $\sim$95 MHz \cite{karvounis202195}. Another implementation approach for intensity modulators with lower optical loss relies on an all-dielectric configuration, which utilizes dielectric pillar arrays to realize a guided-mode resonance (GMR) \cite{yousef2025batio3}. Using a TiO$_2$ cube array heterogeneously fabricated on thin-film BTO, this structure enables a high-quality ($Q$)-factor resonance, and an applied electric field modulates the refractive index via the EO effect of BTO, resulting in a spectral shift of the GMR resonance. Consequently, intensity modulation at specific wavelengths can be achieved, as shown in Fig. \ref{fig:chap5-1}(b). The $Q$ factor of the device reaches $\sim$4000, and the resonance wavelength modulation capability reaches $\sim$0.43 nm when the electric field intensity is 1 V/$\mu$m. In addition, monolithic thin-film BTO metasurface EO modulators with potential for achieving higher performance, theoretical simulations \cite{shen2024ultra} for enhancing the $Q$ factor through a pillar array design, and experimental exploration \cite{prountzou2025electric,weigand2024nanoimprinting} aimed at effectively confining the electric field within the BTO active region are respectively shown in Fig. \ref{fig:chap5-1}(c), (d) and (e). The BTO microstructures \cite{prountzou2025electric,weigand2024nanoimprinting} are then fabricated using a sol-gel-based BTO thin film via nanoimprinting. The device $Q$ factor has reached 200, and EO modulation has been achieved under a 1.5-V driving voltage, with a modulation frequency of up to 5 MHz \cite{weigand2024nanoimprinting}.

\begin{figure}
    \centering
    \includegraphics[width=1\linewidth]{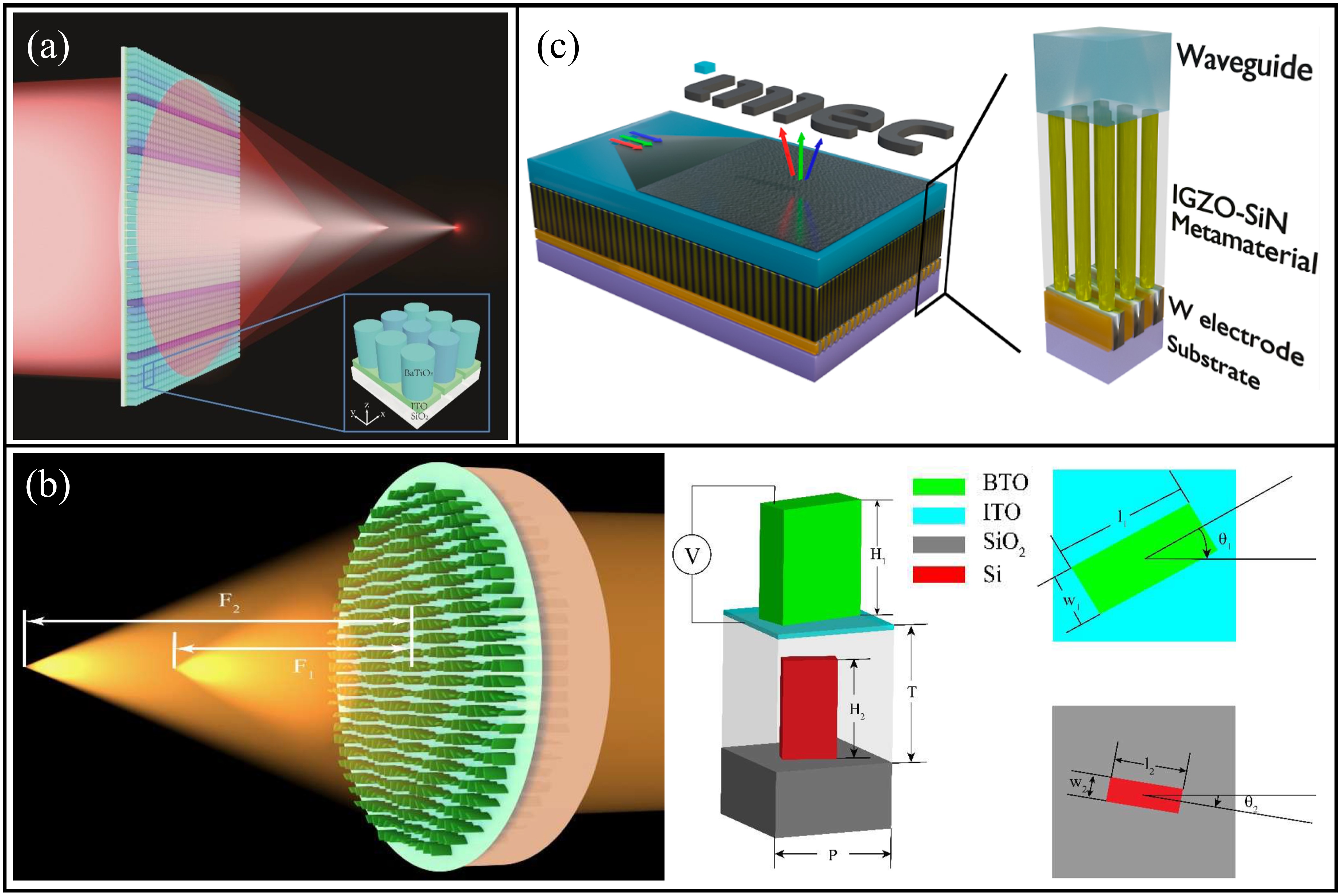}
    \caption{(a) Schematic of electrically driven zoom metalens. The illustration shows the arrangement of the BTO antennas where the author only changed the voltages. Reproduced with permission from Xu \textit{et al.}, Nanomaterials 11, 729 (2021). Copyright 2021 authors, licensed under a Creative Common Attribution (CC BY) license \cite{xu2021electrically}. (b) Schematic of an electrically modulated bifocal metalens and a unit cell. The metalens can focus at two different focal points at the same time. By applying different voltages, the intensity ratio of the two focal points changes. Reproduced with permission from Qin \textit{et al.}, Nanomaterials 11, 2023 (2021). Copyright 2021 authors, licensed under a Creative Common Attribution (CC BY) license\cite{qin2021active}. (c) Schematic of the waveguide–metasurface hybrid spatial light modulator and a unit cell. Reprinted with permission from Croes \textit{et al.}, ACS Photonics 11, 529–536 (2024). Copyright 2024 American Chemical Society\cite{croes2024subwavelength}.}
    \label{fig:chap5-2}
\end{figure}

Secondly, through metasurface microstructural engineering, it has become feasible to manipulate the wavefront of light, thereby enabling on-demand functionalities such as metalenses and spatial-light modulators \cite{neshev2018optical}. Electrically tunable metasurfaces, in particular, are generally desired over static metasurface devices due to their flexible reconfigurability \cite{ding2024electrically,jung2024rise}. As a material platform possessing exceptional EO properties, thin-film BTO demonstrates substantial potential and prospects for corresponding applications. Theoretical research has indicated that adjustment of a metalens focal length can be accomplished \cite{xu2021electrically} by applying a voltage to modulate the refractive index of BTO nano-column antennas in different regions of the metasurface, thereby enabling  dynamic tuning of the phase hologram of the entire metalens, as shown in Fig. \ref{fig:chap5-2}(a). Based on a more complex BTO-Si nanofin structure, a dual-focus design can be implemented \cite{qin2021active}. By applying a voltage to BTO, the light intensity distribution between the two foci can be controlled, which shows significant potential in applications such as dynamic holography and multi-imaging systems, as illustrated in Fig. \ref{fig:chap5-2}(b). Moreover, theoretical research proposed a spatial light modulator \cite{croes2024subwavelength}, in which any arbitrary desired grating pattern can be synthesized within a BTO waveguide by applying a modulating electric field, utilizing a novel waveguide–metasurface hybrid configuration, as illustrated in Fig. \ref{fig:chap5-2}(c).

\begin{figure}
    \centering
    \includegraphics[width=1\linewidth]{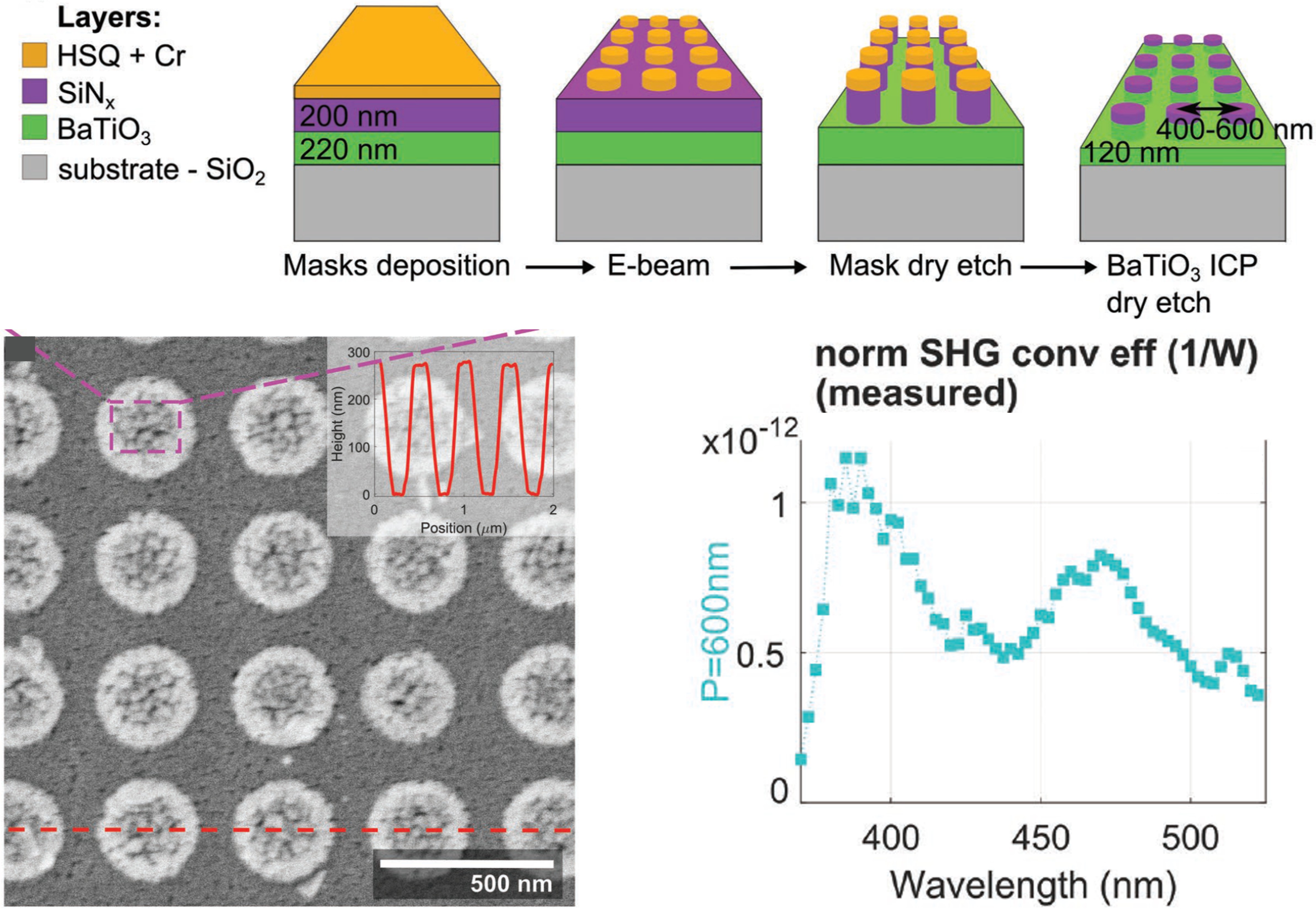}
    \caption{Schematic of thin-film BTO-based nonlinear optical metasurface, including the fabrication process flow, SEM image, and measured SHG results. Reproduced with permission from Timpu \textit{et al.},  Advanced optical materials 7, 1900936, \copyright 2019 Wiley-VCH GmbH\cite{timpu2019enhanced}.}
    \label{fig:chap5-3}
\end{figure}

In addition, by combining the optical resonance field enhancement of metasurface microstructures with the second-order nonlinear ($\chi^{(2)}$) optical characteristics of BTO materials, experimental work has demonstrated a BTO-based metasurface that exhibits enhanced second harmonic generation (SHG) across a broad spectral range from near-ultraviolet to visible light \cite{timpu2019enhanced}, as illustrated in Fig. \ref{fig:chap5-3}. This research stands out as one of the foundational studies in the field of nonlinear BTO metasurfaces, showcasing both the feasibility and considerable potential of integrating high-performance ferroelectric materials with metasurface nanophotonics to realize miniaturized, high-performance nonlinear optical devices.

\subsection{Optical sensing}
\begin{figure}
    \centering
    \includegraphics[width=1\linewidth]{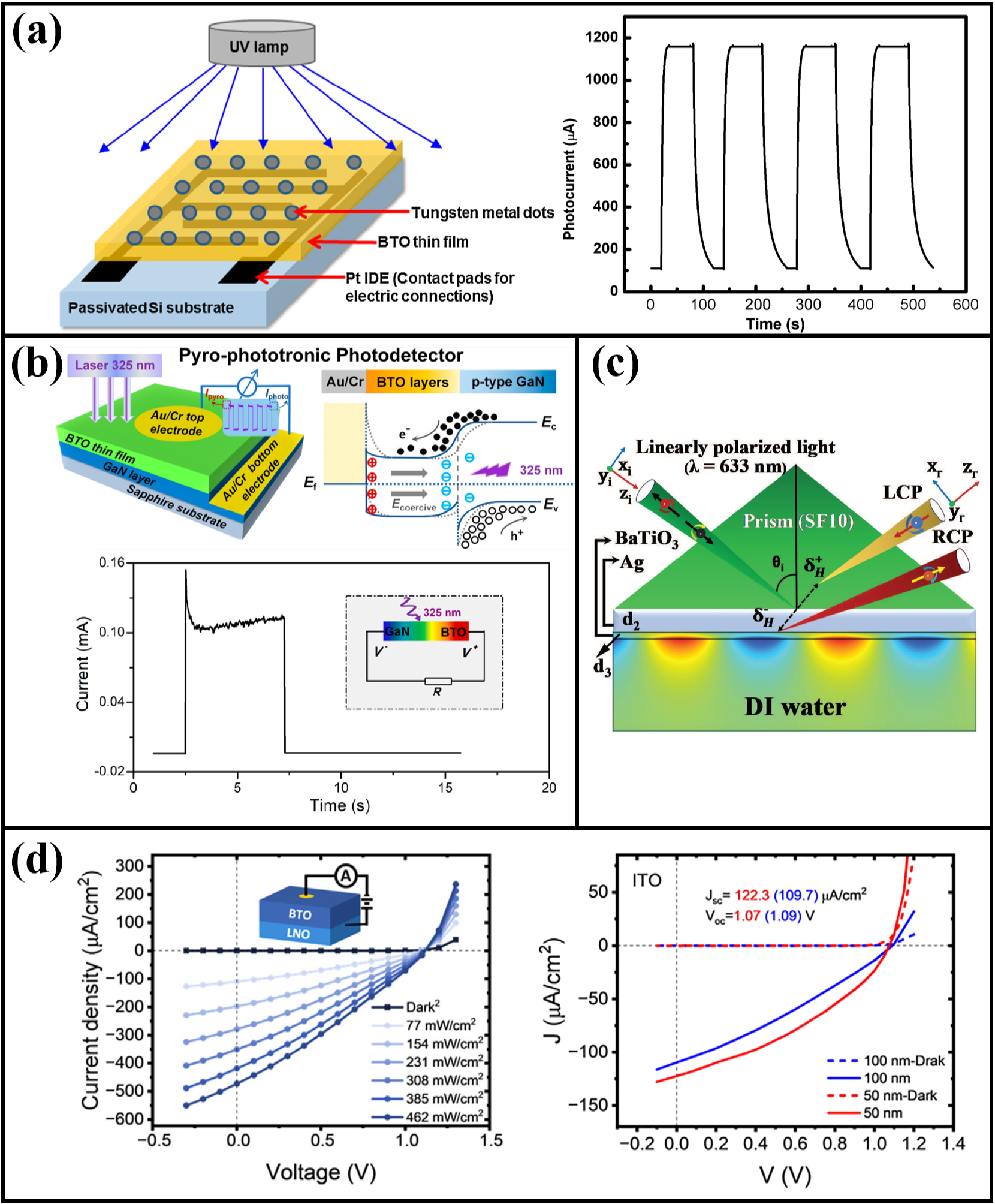}
    \caption{(a) Schematic of UV photodetector based on BTO thin film integrated with uniformly distributed circular dots of W modifier, with its photoelectric detection curve. Reproduced from Sharma \textit{et al.}, Sensors and Actuators A: Physical 230, 175–181 (2015), licensed under Creative Commons Attribution–NonCommercial–NoDerivatives (CC BY-NC-ND)\cite{sharma2015ultraviolet}. (b) Schematic illustration of the structure of the BTO/GaN photodetector, with its photoelectric detection curve. Reprinted with permission from Zhang \textit{et al.}, Nano Letters 21, 8808–8816. Copyright 2021 American Chemical Society\cite{zhang2021self}. (c) Systematic representation of the proposed structure for enhanced detection of photonic spin Hall effect. Reprinted with permission from IEEE Journal of Quantum Electronics. \copyright 2025 IEEE\cite{kumar2025high}. (d) Giant bulk photovoltaic effect in crystal orientation restructured quasi-epitaxial BaTiO$_3$ films on silicon substrates, with an open-circuit photovoltage exceeding 1.07 V \cite{ji2026giant}.}
    \label{fig:chap5-5}
\end{figure}

Owing to several distinct optical properties of BTO, it is also considered as a promising core component candidate in certain types of sensors. Firstly, as mentioned in Section \ref{Sec2.3}, BTO is transparent to photons with an energy below 3.2 eV (i.e., wavelengths longer than 387 nm), whereas photons exceeding this energy are absorbed within BTO, generating photocarriers. Driven by the built-in electric field induced by spontaneous polarization in BTO, these photocarriers produce a detectable photocurrent. Such photovoltaic effect of BTO is consequently utilized in ultraviolet (UV) photodetectors. As illustrated in Fig. \ref{fig:chap5-5}(a), BTO thin film modified with tungsten dots demonstrates photodetection with a photoconductive gain as high as $1.84\times10^4$ at a wavelength of 365 nm \cite{sharma2015ultraviolet}. In addition, as shown in Fig. \ref{fig:chap5-5}(b), a BTO thin film/GaN heterojunction enables high-performance ultraviolet detection at a wavelength of 325 nm by synergistically combining the photovoltaic and pyroelectric effects \cite{zhang2021self}. Secondly, by leveraging the enhancement effect of the high refractive index of BTO materials with surface plasmon resonances, a 28-fold amplification of the photonic quantum spin Hall effect, which is intrinsically difficult to observe directly, was achieved, thus enabling its experimental detection \cite{kumar2025high}, as shown in Fig. \ref{fig:chap5-5}(c). Moreover, Ji \textit{et al.} recently reported a strong bulk-photovoltaic-effect-based photoresponse in crystal-orientation-restructured quasi-epitaxial BTO films on silicon\cite{ji2026giant}. In this structure, the redistribution of $c$- and $a$-oriented domains across the film thickness contributed to enhanced photovoltaic response, with an open-circuit photovoltage exceeding 1.07 V in a 50 nm-thick BTO film, as shown in Fig. \ref{fig:chap5-5}(d). This result further indicates the potential of thin-film BTO for ferroelectric photovoltaic and photodetection-related applications. The micro-fabrication processes for these sensing devices are fully compatible with integrated optical platforms. In the future, by incorporating on-chip optical modulation components, we anticipate the realization of fully functional integrated photonic chips based on thin-film BTO.

\subsection{Nonlinear optical applications}
\begin{figure}
    \centering
    \includegraphics[width=1\linewidth]{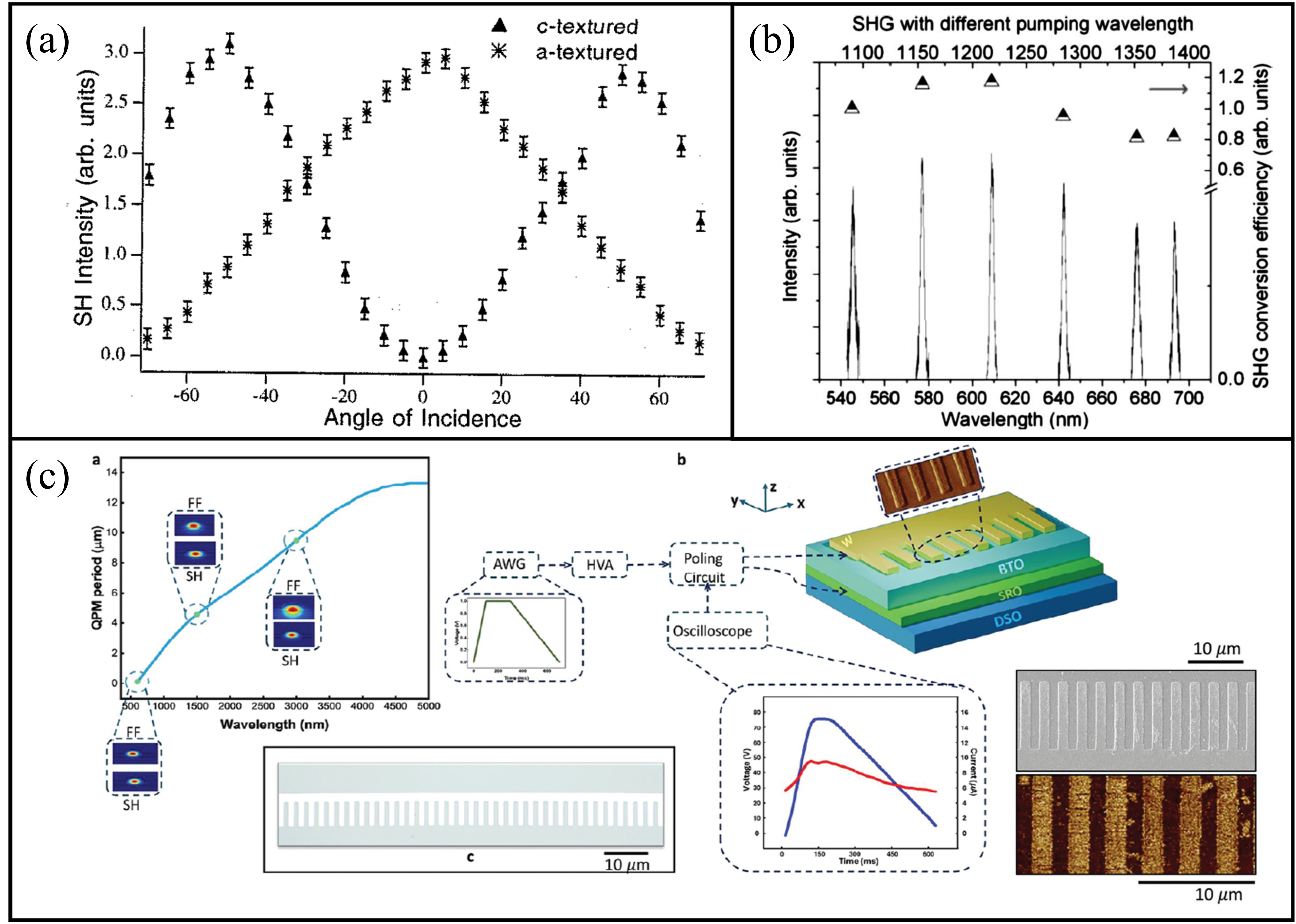}
    \caption{(a) SHG observed in $c$-textured and $a$-textured BTO films. Reproduced from Bihari \textit{et al.}, Journal of Applied Physics 76, 1169–1174 (1994), with the permission of AIP publishing  \cite{bihari1994investigation}. (b) SHG observed in thin-film BTO waveguides. Reproduced from Lin \textit{et al.}, Applied Physics Letters 92 (2008), with the permission of AIP publishing\cite{lin2008highly}. (c) Schematic of the periodically poled thin-film BTO, including the QPM period design, the poling setup, optical microscope images of the patterned electrodes, as well as SEM images and the corresponding piezo force microscopy (PFM) images of the poling pattern. Reproduced with permission from Aashna \textit{et al.}, Advanced Science 11, 2406248 (2024). Copyright 2024 authors, licensed under a Creative Common Attribution (CC BY) license \cite{aashna2024periodic}.}
    \label{fig:chap5-4}
\end{figure}

Although BTO shows excellent nonlinear optical properties (as illustrated in Section \ref{optical nonlinear}), its exceptionally prominent EO characteristics may somewhat overshadow its nonlinear optical properties, which appear to be comparatively underexplored. $\chi^{(2)}$ optical SHG has been observed using BTO thin films \cite{bihari1994investigation} and waveguides \cite{lin2008highly} in earlier work, as shown in Fig. \ref{fig:chap5-4}(a) and (b). Since BTO is a ferroelectric material, its polarization direction can be switched by applying a pulsed electric field. Consequently, single-crystalline BTO thin film can be periodically poled to enhance the nonlinear optical conversion efficiency based on the principle of quasi-phase matching (QPM). A recent experiment achieved stable periodic poling after preparing high-quality $c$-axis oriented single-crystal BTO films \cite{aashna2024periodic}, as shown in Fig. \ref{fig:chap5-4}(c). In the future, further exploration of chip-based nonlinear optics with BTO should be carried out. In addition, leveraging the strong third-order nonlinearity of BTO or metal-ion-doped BTO thin films introduced in Section \ref{chi3}, BTO holds significant potential for enabling applications such as high-precision optical clocks \cite{papp2014microresonator} and soliton-based optical frequency comb ranging \cite{suh2018soliton}.

\subsection{Neuromorphic optical computing}
\begin{figure}
    \centering
    \includegraphics[width=1\linewidth]{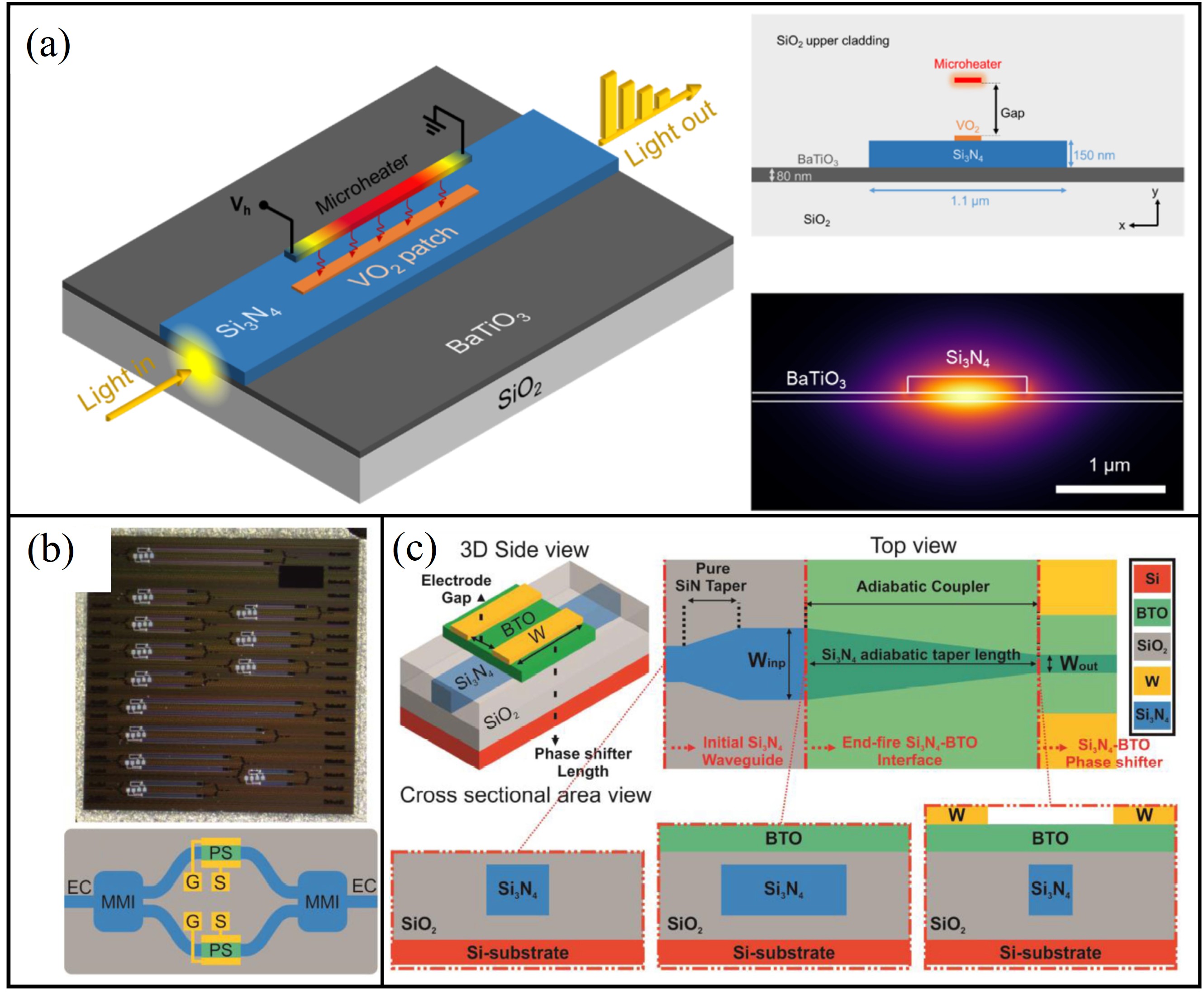}
    \caption{(a) Illustration and working principle of a proposed device for neuromorphic optical computing applications, with scalar multiplication functionality as an example, based on a hybrid VO$_2$/SiN/BTO waveguide, with optical mode distribution. Reproduced with permission from Seoane \textit{et al.}, Optical Materials Express 13, 3266–3276 (2023). Copyright 2023 authors, licensed under a Creative Common Attribution (CC BY) license\cite{seoane2023neuro}. (b) Schematic of SiN/BTO EO modulators serving as active weight elements for optical neural networks. Reprinted with permission from Journal of Lightwave Technology. \copyright 2025 IEEE.\cite{chrysostomidis2025ultra}.}
    \label{fig:chap5-6}
\end{figure}

Optical computing represents an emerging technological paradigm that utilizes light rather than electrons as information carriers for computational processing \cite{touch2017optical,mcmahon2023physics}. Leveraging the transmission and interference properties of light within integrated photonic chips, such a paradigm enables direct emulation of neural network computational processes, thereby realizing "in-propagation" neuromorphic optical computing \cite{shastri2021photonics,farmakidis2024integrated,li2025photonics}. Evidently, the thin-film BTO integrated photonic chip is regarded as a highly promising platform for this application \cite{wang2025neuro} due to its exceptional EO properties, which impart outstanding light modulation capabilities. Recently, a hybrid BTO/SiN platform augmented with VO$_2$ has been proposed as a building block for large-scale neuromorphic photonic systems, in which BTO can provide low-loss, high-speed phase modulation, while VO$_2$ can supply a intensity coding \cite{seoane2023neuro}, as shown in Fig. \ref{fig:chap5-6}(a). Moreover, ultra-efficient SiN Mach-Zehnder interferometers employing BTO as an EO active material have been experimentally demonstrated as the active weight element in optical neural networks \cite{chrysostomidis2025ultra} as illustrated in Fig. \ref{fig:chap5-6}(b), emphasizing a low tuning energy down to only 121 nW, a high extinction ratio surpassing 27 dB for precise weight control, a low insertion loss of 4.05 dB, and CMOS-compatible drive voltages of a few volts. In the future, the integration of BTO high-speed EO modulators with BTO-based ferroelectric domain-controlled memory devices \cite{pal2025switch,sun2022nonvolatile} is expected to enable the realization of computing and storage integration on a single BTO-based chip. Finally, it should be mentioned that BTO has potential use in memristors\cite{youngblood2023integrated}, indicating that there exist interesting use cases beyond what we have discussed. A key challenge for the field is to move from isolated device demonstrations to integrated systems in which BTO’s various features are co-designed—rather than treated in isolation—to implement complete optical computing primitives such as weighted summation, activation, and memory within a unified material and fabrication platform.

\section{Outlook and Conclusion}
This review has surveyed recent progress in thin-film BTO for integrated photonics, including thin-film material growth, crystallographic orientation control, and EO device performance. We emphasize the critical roles of film orientation, ferroelectric domain structure, and device geometry to determine the accessible EO coefficients and modulation mechanisms. By comparing monolithic and hybrid platforms, film-level and device-level characterization approaches, and both $a$- and $c$-axis-oriented films, this review presented both an overview as well as an up-to-date summary of recent advances in thin-film BTO growth, fabrication, and applications. 

One of the most attractive advantages of thin-film BTO is its exceptionally large Pockels coefficients, particularly the $r_{42}$ or $r_{51}$ coefficients, which in principle ought to enable strong EO modulation and low-voltage operation. Owning to its significantly larger EO coefficients compared to other widely used EO materials, such as LN, BTO has the potential to surpass performance of current state-of-the-art EO devices. In addition, recent demonstration of low-loss monolithic BTO ridge waveguides, enabled by optimized dry-etching methods, establish the BTOI platform as a route toward highly efficiency EO modulation. This platform is increasingly competitive with well-established lithium niobate on-insulator (LNOI) and lithium tantalate on-insulator (LTOI) technologies. Furthermore, recent progress in wafer-scale integration, BEOL compatibility, and hybrid integration with mature Si and SiN photonics indicates that BTO is not limited to laboratory-scale demonstrations but has practical potential for scalable and manufacturable EO photonic systems. 
For instance, Veeco and Imec recently announced the development of a 300 mm-compatible epitaxial BTO integration process for silicon photonics, highlighting the growing industrial interest in large-scale BTO photonic platforms\cite{veeco2026BTO}.
These attributes make BTO a promising EO material for applications ranging from high-speed optical modulation and low-power optical tuning to emerging areas such as quantum photonics and large-scale integrated photonic circuits.

Despite this progress, several challenges remain before thin-film BTO EO devices can fully reach their potential. Device performance remains highly dependent on film quality, crystallographic orientation, and ferroelectric domain configuration. Currently, BTO thin films on insulators are not yet commercially standardized or as widely accessible as platforms such as SOI, LNOI, or LTOI, and many reported films exhibits mixed $a$-/$c$-axis orientations or polycrystalline structures. These factors limit reproducibility and often prevent devices from consistently achieving bulk-like EO coefficients. In the widely used in-plane multi-domain $a$-axis-oriented films, although promising EO performance has been demonstrated, the effective refractive index and EO response depend highly on domain structure, electric-field alignment, and poling processes. As a result, careful electrode orientation, pre-poling or applying constant bias voltage is typically required to activate the maximum effective EO response. Further advances in domain engineering and understanding the domain structure and its relationship with poling stability and EO response would enable more robust and repeatable utilization of these films.

For $c$-axis-oriented BTO films, the single-domain nature provides direct access to the intrinsic $r_{42}$ coefficient without degradation from the in-plane domain disorder. However, EO modulation driven purely by $r_{42}$ inherently induces both refractive-index changes and polarization rotation, complicating their use in devices that require only phase modulation. While this limits their applicability as conventional phase shifters or MZI interferometers, it also opens opportunities for alternative EO device concepts that intentionally exploit polarization rotation, including active polarization controller or mode converters. At the same time, although EO modulation in BTO offers advantages of low optical loss and and potentially ultra-high speed, the large dielectric permittivity of BTO can make group-velocity and impedance matching of optical and RF modes challenging. Addressing these RF-optical design challenges will be critical for realizing higher-speed BTO-based EO devices.

Another important challenge for BTO photonics is the development of reliable and scalable etching processes. The etching of BTO has been one of the major obstacles for integrated photonic device fabrication. Although hybrid platforms have enabled various devices with exceptional performance, their inherent drawback is the relatively low optical mode overlap with the BTO material, which subsequently reduces the EO overlap factor and limits the efficient utilization of BTO's large EO coefficients. Therefore, monolithic BTO platforms are generally more desirable. However, their performance has been largely limited by the difficulty of etching BTO. Although recent reports have successfully demonstrated monolithic BTO ridge waveguides using dry-etching methods, achieving both low propagation loss and high fabrication quality remains challenging, and such performance has only been demonstrated in a limited number of reports. In addition, wet-etching methods provide an alternative route toward compact BTO photonic devices, offering advantages such as high etch selectivity, relatively deep etch depths, and simpler fabrication processes. However, wet etching typically results in sidewall slopes of approximately 45\degree\ due to its isotropic nature. Although the propagation losses reported for wet-etched BTO waveguides remain higher than those achieved using optimized dry-etching approaches, further improvements in BTO crystalline quality, etching chemistry, and etching condition may provide a pathway toward low-loss monolithic BTO waveguides. Continued advances in BTO etching processes will be critical for realizing low-loss monolithic photonic devices and fully exploiting the advantages of the BTO-on-insulator platform.

Furthermore, with respect to practical applications, thin-film BTO has demonstrated considerable potential across a range of domains - including high-speed optical communications, meta surface-based free-space light modulation, and neuromorphic optical computing - owing to its exceptional EO properties. With continued advances in EO device fabrication, as discussed above, BTO-based integrated photonic platforms are positioned as a promising candidates for next-generation technologies in these fields. In addition, nonlinear optical applications based on BTO materials represent another promising direction that merits further dedicated investigation. By integrating passive optical components such as low-loss waveguides, the BTO-based integrated photonic platform holds the potential for realizing a toolbox capable of comprehensive optical manipulation functionalities on a single chip.

In summary, thin-film BTO has emerged as one of the most promising EO materials for integrated photonics due to its strong intrinsic EO response with a growing list of demonstrated compatibilities suitable for scalable fabrication platforms. Continued progress in film growth, film transfer techniques, etching processes, domain engineering, and device-level design are expected to further improve performance, reproducibility, and bandwidth. With these developments, BTO-based EO devices are well positioned to play an important role in next-generation low-power, high-speed, and multifunctional integrated photonic systems in years to come.

\section*{Acknowledgments}
The authors acknowledge funding support from the A*STAR White Space Fund - National Semiconductor Translation and Innovation Centre (NSTIC) of Singapore and the Ministry of Education, Singapore, under its Academic Research Fund (Tier 2) Programme Award T2EP50224-0037.

\section*{Author Declarations}
\textbf{Conflict of Interest}

The authors have no conflicts to disclose.

\textbf{Author Contributions}

Hong-Lin Lin and Minghao Shang contributed equally to this work.

\textbf{Hong-Lin Lin:} Conceptualization (equal); Formal Analysis (lead); Investigation (lead); Visualization (lead); Writing – original draft (equal); Writing – review \& editing (equal). \textbf{Minghao Shang:} Conceptualization (equal); Formal Analysis (supporting); Investigation (supporting); Visualization (supporting); Writing original draft (equal); Writing – review \& editing (supporting). \textbf{Yuhui Yin:} Investigation (supporting); Formal Analysis (lead); Writing – original draft (supporting). \textbf{Wujie Fu:} Investigation (supporting); Writing – original draft (supporting). \textbf{Luo Qi:} Conceptualization (supporting); Investigation (supporting). \textbf{Aaron J. Danner:} Conceptualization (equal); Project administration (lead); Supervision (lead); Writing – review \& editing (equal).

\section*{Data Availability Statement}
The data that support the findings of this study are available within the article.

\nocite{*}
\bibliography{aipsamp}

\end{document}